\documentclass[twocolumn,trackchanges,twocolappendix]{aastex701} 

\usepackage{amsmath}

\begin{document}

\title{Probing Multi-formation Channels of Dark Matter-Deficient Galaxies in a Simulated Universe}

\author[orcid=0009-0008-0644-9630,sname='Han']{Sanghyeon Han}
\affiliation{Department of Physics and Astronomy, Sejong University, 209 Neungdong-ro, Gwangjin-gu, Seoul 05006, Republic of Korea}
\affiliation{Department of Physics and Astronomy, Seoul National University, Gwanak-gu, Seoul 08826, Republic of Korea}
\email[show]{chzg1@sejong.ac.kr}  

\author[orcid=0000-0003-3428-7612, sname='Hwang']{Ho Seong Hwang} 
\affiliation{Department of Physics and Astronomy, Seoul National University, Gwanak-gu, Seoul 08826, Republic of Korea}
\affiliation{SNU Astronomy Research Center, Seoul National University, Seoul 08826, Republic of Korea}
\affiliation{Institute for Data Innovation in Science, Seoul National University, Seoul 08826, Korea}
\email[show]{hhwang@astro.snu.ac.kr}

\author[orcid=0000-0002-6810-1778, sname='Lee']{Jaehyun Lee}
\affiliation{Korea Astronomy and Space Science Institute, 776, Daedeokdae-ro, Yuseong-gu, Daejeon 34055, Republic of Korea}
\email{jaehyun@kasi.re.kr}


\begin{abstract}

We investigate the formation and evolution of dark matter--deficient galaxies (DMDGs) using {the IllustrisTNG100} cosmological hydrodynamical simulation. 
We define DMDGs as systems with a dark matter--to--stellar mass ratio {below unity, and find two formation channels}. 
In the first, galaxies that initially form within {massive} dark matter halos lose the majority of their dark matter and transition into DMDGs {(acquired DMDGs)}. 
In the second, galaxies form with little or no dark matter from the {beginning} {(inborn DMDGs)}. 
The two populations show distinct physical properties. 
{Acquired DMDGs} exhibit significant temporal variations in mass content, size, and kinematics, whereas {inborn DMDGs} show mild evolution. 
The former population tends to form closer to their nearest neighbor galaxies in large--scale overdense regions than the latter population does. 
{We find that most stars in the inborn DMDGs form in situ. Their precursor gas is already metal--enriched and associated with other galaxies before formation. 
Tidal interactions and filamentary gas flows may contribute to the assembly of compact gas clumps, suggesting multiple formation pathways for inborn DMDGs.}
Our findings support the idea that there are at least two distinct channels to make galaxies dark matter deficient, which is important for understanding the formation mechanisms of these extreme systems.

\end{abstract}



\section{Introduction} \label{sec:intro}

Dark matter is a fundamental component in structure formation in the $\Lambda$ {cold dark matter ($\Lambda$CDM)} paradigm. 
In the hierarchical bottom--up picture, cosmic structures originate from small--scale density perturbations by gravitational instability \citep{Peebles1982}. 
Dark matter first collapses into small halos, which then grow through accretion and merger. 
{Baryonic matter} falls into the potential well generated by dark matter clustering, where gas cooling leads to star formation \citep{WhiteRees1978}. 
{In this standard picture of galaxy formation, galaxies are expected to form within the dark matter halo.} 

{Even in the framework of $\Lambda$CDM, however, some galaxies may have low fractions of dark matter.}
During interactions such as mergers {and} close encounters, material can be tidally stripped to form tails and bridges \citep{Toomre1972}. \citet{Zwicky1956} proposed that star formation within such debris can form new galaxies. 
These systems, called tidal dwarf galaxies (TDGs), have been identified observationally and reproduced in simulations \citep[e.g.,][]{Mirabel1992,Barnes1992}. 
TDGs are expected to retain little dark matter because of the high velocity dispersion of dark matter {and the shallow potential wells preventing TDGs from capturing a significant amount of dark matter} \citep[see, e.g.,][and references therein]{Braine2000,Duc2004,Ploeckinger2018}.

\setcounter{footnote}{0}
Recent studies have reported galaxies that appear to lack dark matter but are not associated with TDGs. \citet{vanDokkum2018,vanDokkum2019} and \citet{Keim2026} found the dwarf galaxies NGC 1052-DF2, NGC 1052-DF4, and {NGC 1052-DF9}\footnote{DF2 and DF4 are also classified as ultra diffuse galaxies (UDGs), characterized by large sizes ($R_{e}>1.5~\mathrm{kpc}$) and very low surface brightness ($\mu_{g}>24~\mathrm{mag~arcsec^{-2}}$) \citep{vanDokkum2015}. {DF9 does not meet the definition of the UDG \citep{Gannon2023}.}}, which have been interpreted as dark matter--deficient galaxy (DMDG) candidates. 
Their inferred halo--to--stellar mass ratios are below unity \citep{vanDokkum2018,vanDokkum2019,Shen2023}, in contrast to expectations for typical galaxies at similar stellar mass \citep[$M_{\rm halo}/M_{\star} > 100$,][]{Behroozi2010,Moster2013}. 
{One could expect that DF2 and DF4 are TDGs.} 
However, typical TDGs tend to be metal--rich outliers above the mass--metallicity relation \citep[e.g.,][]{Hunter2000,Duc2007,Duc2014,Sweet2014}, whereas DF2 and DF4 exhibit low metallicities, consistent with the standard mass--metallicity relation \citep{Fensch2019,Buzzo2022}.

Several studies have raised concerns about the reliability of the inferred properties of DF2 and DF4. 
For example, \citet{Martin2018} and \citet{Laporte2019} challenged the mass estimates, noting the limitations of the available data. 
\citet{Trujillo2019} and \citet{Monelli2019ApJ} argued that the distances to DF2 and DF4 are $\sim13\texttt{-}14~\mathrm{Mpc}$ rather than $\sim20~\mathrm{Mpc}$, which imply that these systems are not extremely dark matter--deficient. 
More recent analyses, however, continue to support low dark matter content in these objects \citep{Danieli2020,Shen2021,Shen2023}.

Motivated by these discoveries, follow--up observations have expanded the sample of DMDGs. \citet{Guo2020} {reported} 19 additional candidates, five in the vicinity of galaxy groups or clusters and fourteen in relatively isolated environments. 
{AGC~114905 is a gas--rich UDG that is located at $76\pm5~\mathrm{Mpc}$ and is identified as a dark matter--deficient galaxy \citep{ManceraPina2022}.} 
In contrast to DF2 and DF4, NGC~1277 is a massive early--type galaxy ($M_{\star}=1.3\times10^{11}~\mathrm{M}_{\odot}$) with a very low dark matter fraction out to $\sim5$ times the effective radius \citep{Comeron2023}. 
Notably, FCC~224 hosts an unusually bright globular cluster system akin to those of DF2 and DF4 \citep{Romanowsky2024,Tang2025}. 
\citet{Buzzo2025} {classified} FCC~224 as a dark matter--deficient galaxy, {highlighted} its similarities to DF2 and DF4, and {proposed} a framework to identify additional DMDGs.

There are several mechanisms proposed to explain the origin of DMDGs within 
the $\Lambda$CDM paradigm. 
In one scenario, galaxies form within {massive} dark matter halos but subsequently experience severe tidal stripping, leaving behind predominantly baryonic remnants \citep{Ogiya2018,Nusser2020}. 
This mechanism has been realized in a variety of numerical studies \citep[e.g.,][]{Jing2019,Niemiec2019,Carleton2019,Maccio2021,Jackson2021,Ogiya2022,Montero-Dorta2024,Katayama2024,Contreras-Santos2024}. 
{{In particular, \citet{Moreno2022} showed that galaxies resembling DF2 and DF4 can arise in cosmological hydrodynamical simulations. 
Hereafter, we refer to dark matter--deficient galaxies with stellar masses, sizes, velocity dispersions, and morphologies comparable to those of DF2 and DF4 as ``DF--like" systems.}} 
Although \citet{Saulder2020} {argued} that some isolated, massive DMDGs are numerical artifacts associated with unphysical processes near the periodic boundaries of simulation boxes, \citet{Mitrasinovic2023} {demonstrated} that physically plausible isolated DMDG candidates can exist and are not solely attributable to numerical effects.

A second scenario, proposed by \citet{Silk2019}, is a ``mini--Bullet Cluster" event in which high--velocity collisions between gas--rich galaxies produce baryonic condensations (gas clouds) that form dwarf galaxies while retaining little or no dark matter. 
This mechanism has been reproduced in high resolution simulations \citep{Shin2020,LeeJoo2021,Otaki2023}. \citet{vanDokkum2022} further {suggested} that DF2 and DF4 may lie along a trail of DMDGs generated by such collisions. 
{\cite{LeeJoo2024} used} hydrodynamical simulations to validate the formation of multiple remnants in this scenario and {assessed} its occurrence under conservative criteria in cosmological {simulations}.
{{\cite{Buzzo2026} identified the FCC~224--FCC~240 pair in the Fornax cluster as possible analogues of DF2 and DF4. 
These galaxies share several properties with DF2 and DF4, including low internal velocity dispersions, unusually luminous globular clusters, and similar stellar populations and structures.
These similarities suggest that FCC~224 and FCC~240 may have formed through a high velocity collision similar to that proposed for DF2 and DF4.}}

{As mentioned earlier, TDGs are expected to contain little or no dark matter.
\citet{Hammer2026} proposed an alternative interpretation in which a head-on 1:1 major merger associated with the formation of present-day NGC~1052 produces long-lived tidal tails and TDGs.
Their model produces tidal features that remain visible 8.4 Gyr after the merger.
It also produces TDGs whose spatial configuration, sizes, and some kinematic properties broadly resemble those of the observed DM--deficient dwarfs. 
These results renew interest in the TDG origin for these systems.
}

Observations and simulations now support the existence of DMDGs. 
{However, it is unclear which pathway is dominant. It is also uncertain whether all DMDGs, including DF2 and DF4, share similar physical properties.} 
{Therefore, a single formation mechanism is unlikely to account for these kinds of objects.} 
Previous studies have primarily demonstrated individual mechanisms and reproduced specific analogs (e.g., DF2--like or NGC 1277--like objects). 
{It remains poorly constrained whether DMDG populations exhibit systematic differences in their internal and external physical properties, either among themselves or relative to normal galaxies.} 
The goal of this study is to elucidate the formation and evolution of DMDGs and to investigate their physical properties. 
We use the IllustrisTNG100 cosmological hydrodynamical simulations, identify DMDGs at $z=0$, and trace their progenitors to determine formation epochs and evolutionary pathways across cosmic time.

This paper is organized as follows. In Section~\ref{sec:data}, we provide a brief description of the IllustrisTNG100 simulations and our DMDG selection procedure. 
In Section~\ref{sec:results}, we explore the formation channels of DMDGs. 
We derive the internal physical properties of DMDGs and investigate them with respect to the formation channel. We also investigate the environmental effects on DMDGs. 
{In Section~\ref{sec:discuss}, we discuss the implications of} our findings. 
We summarize the key findings and offer concluding remarks in Section~\ref{sec:conclusion}. 
{Several appendices provide additional validation and robustness analyses that supplement the main text.}
{Appendix~\ref{sec:app_a} examines the reliability of the merger trees for selected DMDGs. 
Appendix~\ref{sec:app_b} examines numerical artifacts associated with the periodic boundary conditions in the TNG simulations.
Appendix~\ref{sec:app_c} shows the spatial distributions of two DMDGs that are outliers from the stellar mass--size relation of DMDGs.}

\section{Data} \label{sec:data}

\subsection{IllustrisTNG} \label{subsec:tng}
The IllustrisTNG project \citep{Nelson2019a,Pillepich2018a,Springel2018,Nelson2018a,Naiman2018,Marinacci2018} is a suite of cosmological hydrodynamical simulations performed in three volumes: TNG50, TNG100, and TNG300. 
These simulations differ in their comoving box size, as well as the mass and spatial resolutions of their dark matter particles and baryonic components. 
All runs adopt the Planck 2015 cosmology \citep{PlanckColl2016} with $\Omega_{\Lambda}=0.6911$, $\Omega_{\rm m}=0.3089$, $\Omega_{\rm b}=0.0486$, $\sigma_{8}=0.8159$, and $H_{0}=67.74~\mathrm{km~s^{-1}~Mpc^{-1}}$. 
The numerical implementations of the galaxy formation model are detailed in \citet{Weinberger2017} and \citet{Pillepich2018b}.

The TNG project provides catalogs of galaxies, groups, and clusters with various physical information at different redshifts. 
For dark matter structures, two catalogs are available: halo and subhalo catalogs. 
Halos are identified with a friends--of--friends (FoF) algorithm \citep{Davis1985,Audit1998}, and subhalos are identified in the FoF halos using the \texttt{SUBFIND} algorithm \citep{Springel2001,Dolag2009}.

The TNG100-1 simulation, the highest--resolution realization of the TNG100 suite, contains $1820^3$ dark matter particles and an equal number of initial gas cells within a comoving volume of $(106.5~\mathrm{cMpc})^3$. 
The baryonic (dark matter) mass resolution is $1.4~(7.5)\times10^6~\mathrm{M}_{\odot}$. 
The gravitational softening length for dark matter and stellar particles is $0.74~\mathrm{kpc}$ at $z=0$. 
While TNG50 achieves higher resolution, its smaller volume yields fewer galaxies and limits statistical power. 
In contrast, TNG300 spans a larger volume {at the cost of mass and spatial resolution, limiting our ability to study low mass galaxies.} 
We therefore adopt TNG100-1, which provides a proper balance between resolution and sample size for our study.

\begin{figure}[!t]
\centering
\includegraphics[width=0.47\textwidth]{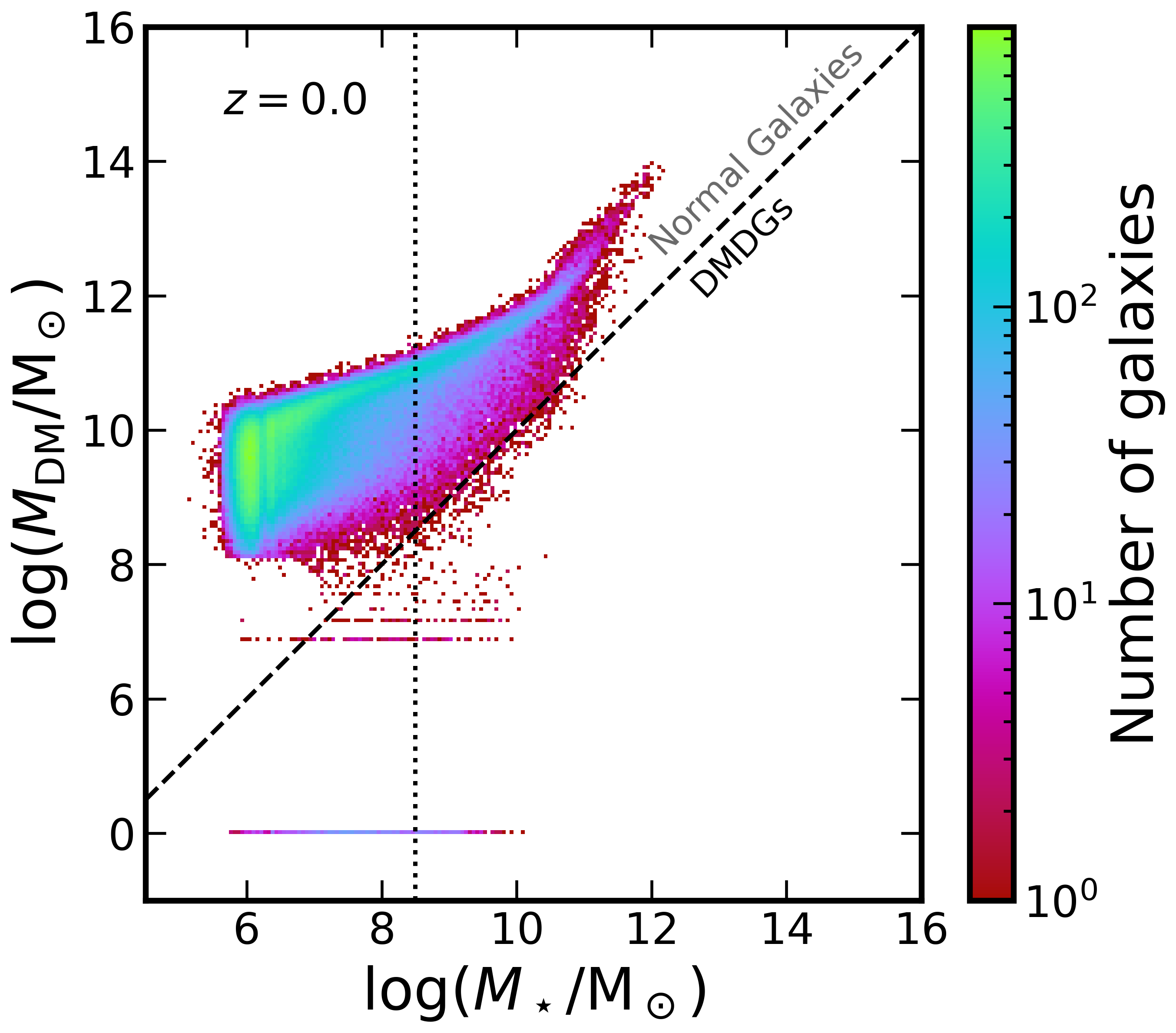}
\caption{
Two--dimensional histogram of stellar mass versus dark matter mass for galaxies at $z=0$. Colors indicate the number of galaxies per bin. 
The vertical dotted line marks the stellar mass cut at $M_{\star} = 10^{8.5}~\rm M_{\odot}$. 
The black dashed line shows the DMDGs threshold ($M_{\rm DM}=M_{\star}$); galaxies below this line satisfy $M_{\rm DM}/M_{\star}<1$.
}
\label{fig:Mstar_Mdm}
\end{figure}

\subsection{Sample Selection} \label{subsec:sample}
\begin{figure*}[!t]
\centering
\includegraphics[width=\textwidth]{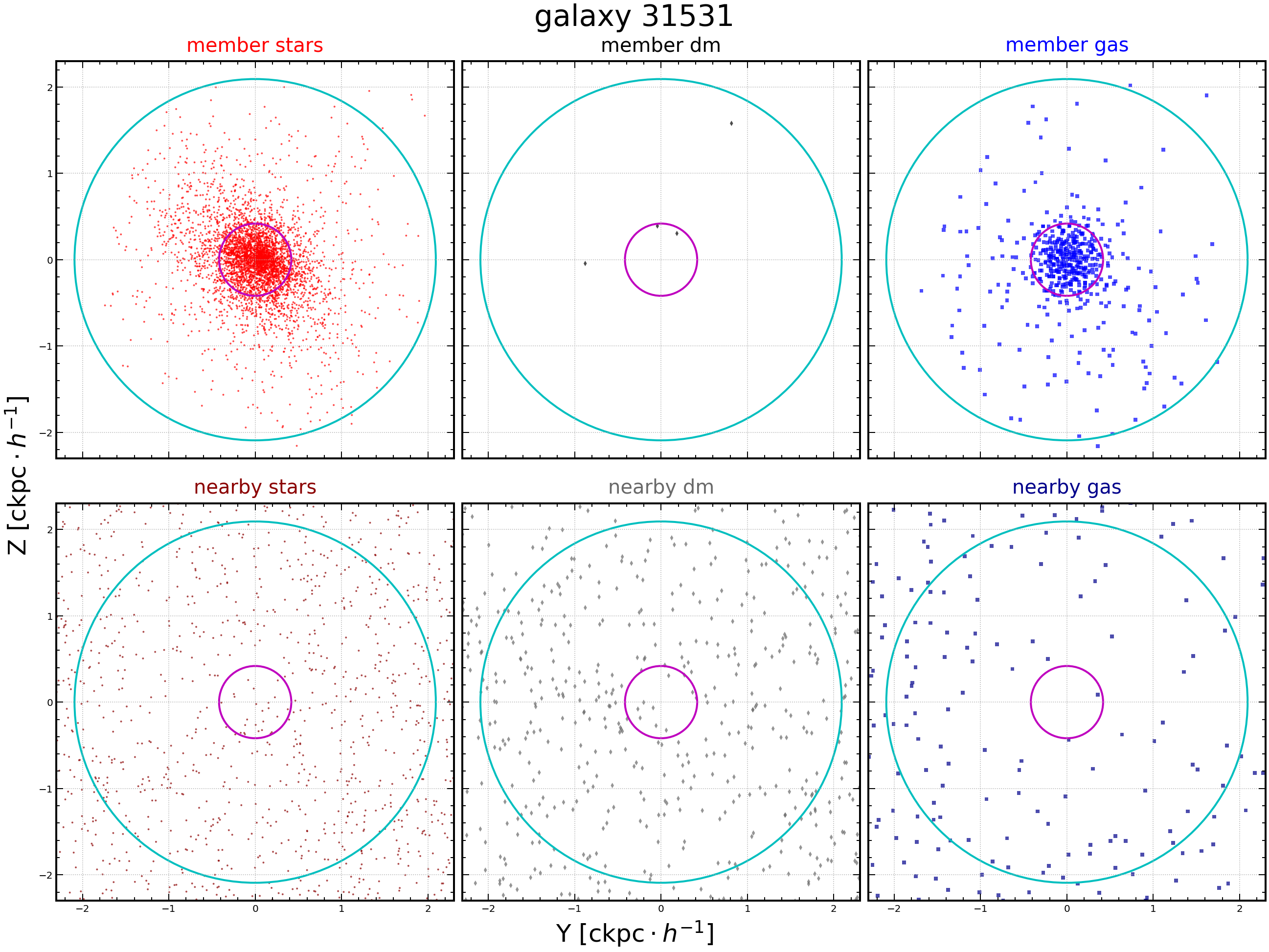}
\caption{
Projected spatial distributions of particles in the $y$–$z$ plane, shown relative to the galaxy center. 
The top row displays member particles of the target galaxy. 
{The bottom row shows particles within a box of $20\ \mathrm{ckpc \cdot} h^{-1}$ centered on the target galaxy, excluding particles bound to the target galaxy and including particles bound to other galaxies as well as unbound particles.}
From left to right, the columns correspond to stars, dark matter, and gas. 
The target galaxy is spatially well defined and not significantly contaminated. 
In all panels, the magenta circle marks the half--stellar mass radius, and the cyan circle marks five times this radius.
}
\label{fig:spatial_good}
\end{figure*}

We obtain the {stellar and dark matter masses} of galaxies from the subhalo catalog. 
We restrict the sample to ${M_{\star}\ge10^{8.5}~\mathrm{M}_{\odot}}$, {which corresponds to about {300} stellar particles, a conservative number for the statistically reliable measurement of internal properties of galaxies.} 
Accordingly, we exclude DF--like systems with $M_{\star}\sim10^{8}~\mathrm{M}_{\odot}$. 
We exclude subhalos which are transient or non-physical fragments, based on the \texttt{SubhaloFlag}\footnote{{\texttt{SubhaloFlag} indicates whether objects identified by \texttt{SUBFIND} are likely to have formed through the process of gravitational collapse and structure formation. 
Some satellite subhalos correspond to baryon-driven fragments or clumps (e.g., disk instabilities) within already formed galaxies, and \texttt{SUBFIND} cannot, a priori, distinguish these from genuine subhalos \citep[for more detail, see][]{Nelson2019a}.}}. 
In total, {33,423} galaxies satisfy these criteria. Galaxies are classified by their dark matter--to--stellar mass ratios $(M_{\rm DM}/M_{\star})$ at $z=0$. 
We define DMDGs as objects satisfying $M_{\rm DM}/M_{\star}<1$. 
Figure~\ref{fig:Mstar_Mdm} summarizes the classification in the $M_{\star}\texttt{-}M_{\rm DM}$ plane at $z=0$. 
{Our selection criteria yield {733} DMDGs and {32,690} normal galaxies, respectively.}

\begin{figure*}[!t]
\centering
\includegraphics[width=\textwidth]{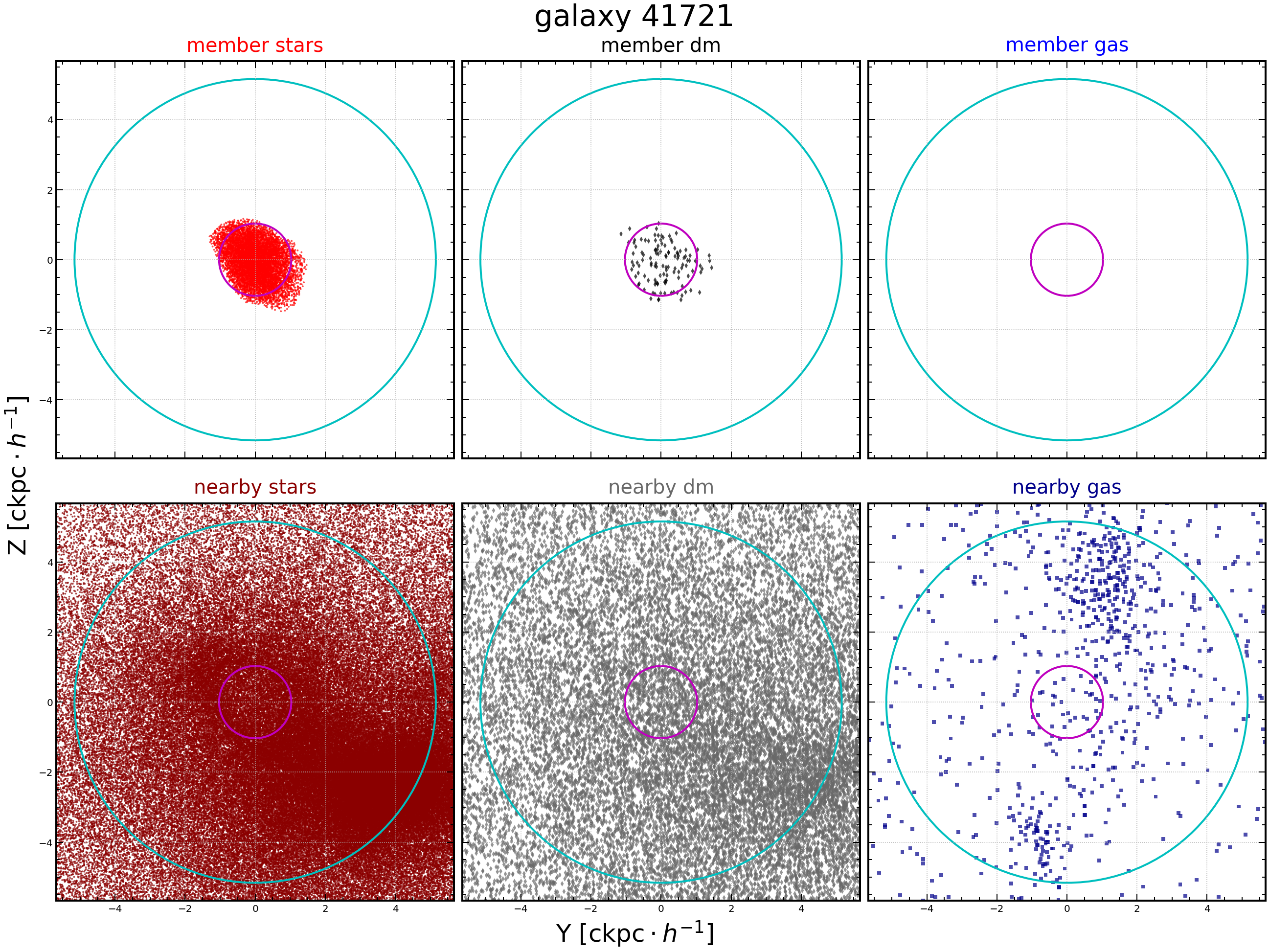}
\caption{
Similar to Figure~\ref{fig:spatial_good}, but the target galaxy is spatially embedded within the envelope of a larger galaxy, indicating substantial contamination.
}
\label{fig:spatial_bad}
\end{figure*}

Because of limitations in the galaxy identification scheme, {an object that is not gravitationally self-bound may be misidentified as a real galaxy.} 
To exclude such systems, we first conduct a visual inspection of projected spatial distributions of the stellar, dark matter, and gaseous components for each DMDG. 
Several subhalos are found to be contaminated by interloper particles. 
These contaminants consist of two populations: particles gravitationally bound to neighboring galaxies and stray particles not bound to any galaxy. 
Figures~\ref{fig:spatial_good} and \ref{fig:spatial_bad} show two examples. 
For each system, all member particles, together with any interloping particles, are projected onto the $y$--$z$ plane.
In Figure~\ref{fig:spatial_good}, the galaxy is clearly isolated in space with negligible contamination by non-member particles. 
We classify these systems as the spatially clean sample. 
{In contrast, Figure~\ref{fig:spatial_bad} shows a case where a galaxy is entirely embedded within the boundary of a larger halo, making boundary detection difficult and complicating the evaluation of virialization.}

To quantify and verify the visual classification, we compute a contamination fraction for each galaxy by counting stellar particles within boundaries defined in phase space as follows. 

We define the velocity boundary as
\begin{equation}\label{eq:vcut}
v_{\rm cutoff} = 1.5v_{\rm max},
\end{equation}
where $v_{\rm max}$ is the maximum velocity among stellar particle members of the galaxy. 
Here, we empirically increase the velocity boundary by 50\% to include interlopers unbound to the galaxy.

\begin{figure}[!t]
\centering
\includegraphics[width=0.47\textwidth]{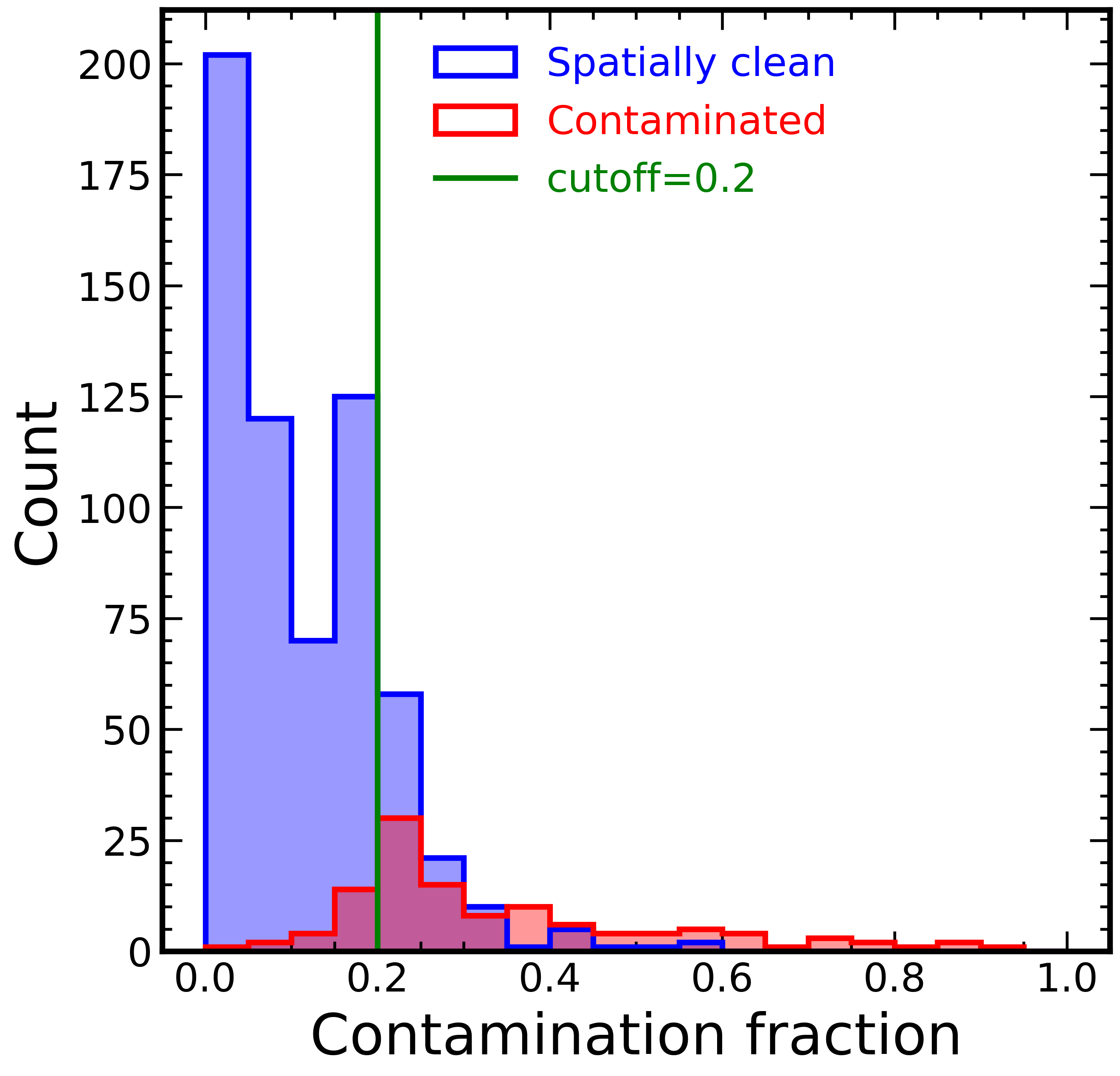}
\caption{
Histogram of the contamination fraction for the {733} DMDG candidates. 
The contamination fraction is computed within the adopted boundary. 
Colors encode the visual classification (spatially clean vs. contaminated). 
A vertical line marks the adopted cutoff, chosen to maximize sample size while minimizing contamination.
}
\label{fig:contamination}
\end{figure}

{For the spatial boundary, we restrict the particles to those clustered around the main body of the galaxy.} 
{To identify the boundary separating member particles from neighboring particles, we perform the following procedure.} 
Considering member particles with $r>R_{\star,1/2}$, let $r_k$ denote the distance of the $k$-th nearest member (ordered by increasing $r$), and define the outermost member radius as $r_{\rm last}$. 
With definitions of $\Delta r \equiv r_{k+1}-r_k$ and $S = \{i \mid \Delta r_i > R_{\star,1/2}\}$, we introduce the spatial cutoff radius as
\begin{equation}
    r_{\mathrm{cutoff}}
     = \begin{cases}
       r_{\mathrm{last}}, & \text{if } S=\varnothing,\\
       r_k \text{ for some } k \in \operatorname*{arg\,max}_{i\in S}\Delta r_i,
         & \text{if } S\neq\varnothing.
     \end{cases}
\end{equation}
{In other words, the cutoff radius is set by the radius of the outermost concentric spherical shell that contains no member particle, subject to two {conditions; one} is that the shell's thickness must exceed the half--stellar mass radius of the galaxy. 
The other is that the shell's outer radius must be less than the maximum distance reached by the member particles. 
If no such shell satisfies these criteria, the cutoff radius is instead set equal to the maximum distance of the member particles. 
We then count stellar member $(N_{\textrm{member}})$ and non-member $(N_{\textrm{non-member}})$ particles which satisfy both the velocity and distance cutoff criteria as shown above.}
{The contamination fraction is then defined as}
\begin{equation}\label{eq:fcont}
f_{\rm cont} \equiv \frac{N_{\rm non\text{-}member}}{N_{\rm member}+N_{\rm non\text{-}member}}.
\end{equation}
Figure~\ref{fig:contamination} shows the distribution of $f_{\rm cont}$ for two subsamples classified by the visual classification. 
We adopt a cutoff threshold of $f_{\rm cont}=0.2$ to maximize the reliability of the sample while keeping contamination low, allowing us to select DMDG candidates in an objective manner. 
This yields {538} DMDGs in our sample.

\section{Results} \label{sec:results}

In this section, we present our results on DMDGs, beginning with their formation epochs. 
To this end, we use galaxy merger trees constructed with SubLink \citep{Rodriguez-Gomez2015}. 
{The TNGs provide two types of trees}: a dark matter--based tree and a baryonic--based tree constructed from star forming gas cells and stellar particles. 
Because DMDGs often contain very few dark matter particles, we adopt the baryonic--based tree, SubLink\texttt{\char`_}gal.

\begin{figure}[!t]
\centering
\includegraphics[width=0.47\textwidth]{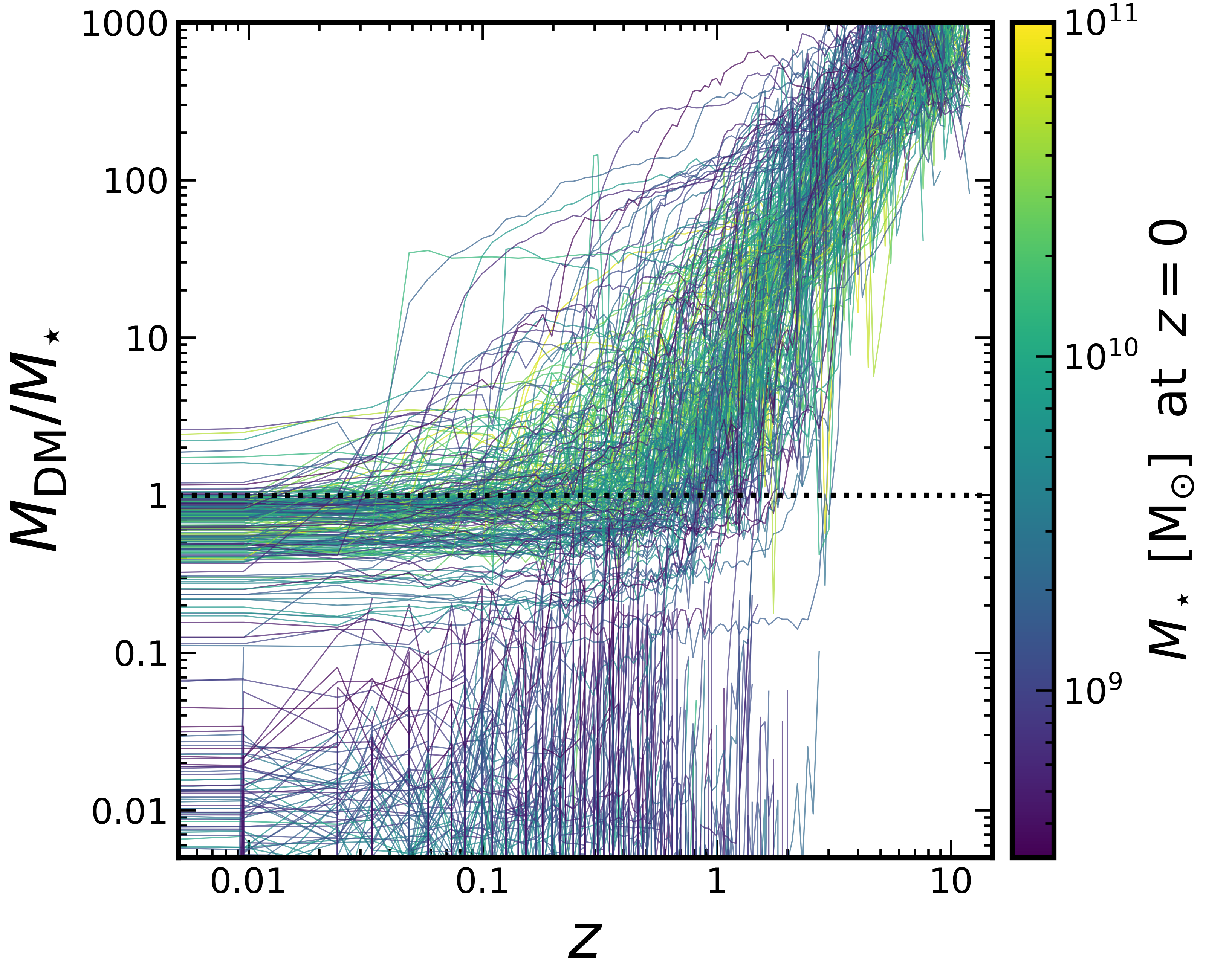}
\caption{
Dark matter--to--stellar mass ratio ($M_{\rm DM}/M_{\star}$) of DMDGs as a function of redshift. 
The gray lines represent the evolution of mass ratio of DMDGs. 
The mass ratio evolution displays two branches; some decline from high values to below unity. 
Others reside below one at birth. 
The dotted line indicates the mass ratio criterion used to classify DMDGs. 
Note that the galaxies with $M_{\rm DM}/M_{\star} \ll 1$ are not illustrated in this figure.
{{Colors indicate stellar mass at $z=0$.}}
}
\label{fig:ratio0}
\end{figure}

\subsection{Diverse evolutionary histories of DMDGs} \label{subsec:tree}

To investigate the formation epochs of DMDGs, we trace their progenitors through the merger tree.
For each of the {538} DMDGs, we identify the earliest progenitor and examine the dark matter--to--stellar mass ratios as a function of redshift. 
Figure~\ref{fig:ratio0} shows the redshift evolution of the mass ratio of DMDGs. 
We identify two branches; some DMDGs are born within massive dark matter halos and subsequently lose dark matter over cosmic time, whereas others form with little or no dark matter.
{We therefore classify DMDGs into two types based on their dark matter--to--stellar mass ratio at birth: ``acquired DMDGs", in which the ratio exceeds unity, and ``inborn DMDGs", in which it does not.}
In our sample, {229} galaxies are acquired DMDGs and {309} are inborn DMDGs. 

{{We note that all the galaxies in Figure~\ref{fig:ratio0} satisfy the DMDG criterion described in Section~\ref{subsec:sample} at $z=0$. 
Because the $x$-axis is logarithmic, $z=0$ is not shown. 
Some galaxies became DMDGs only between $z=0.01$ and $z=0$.
Their dark matter--to--stellar mass ratios remain above unity at the lowest redshifts shown in the figure.}}

\begin{figure*}[!t]
\centering
\includegraphics[width=0.98\textwidth]{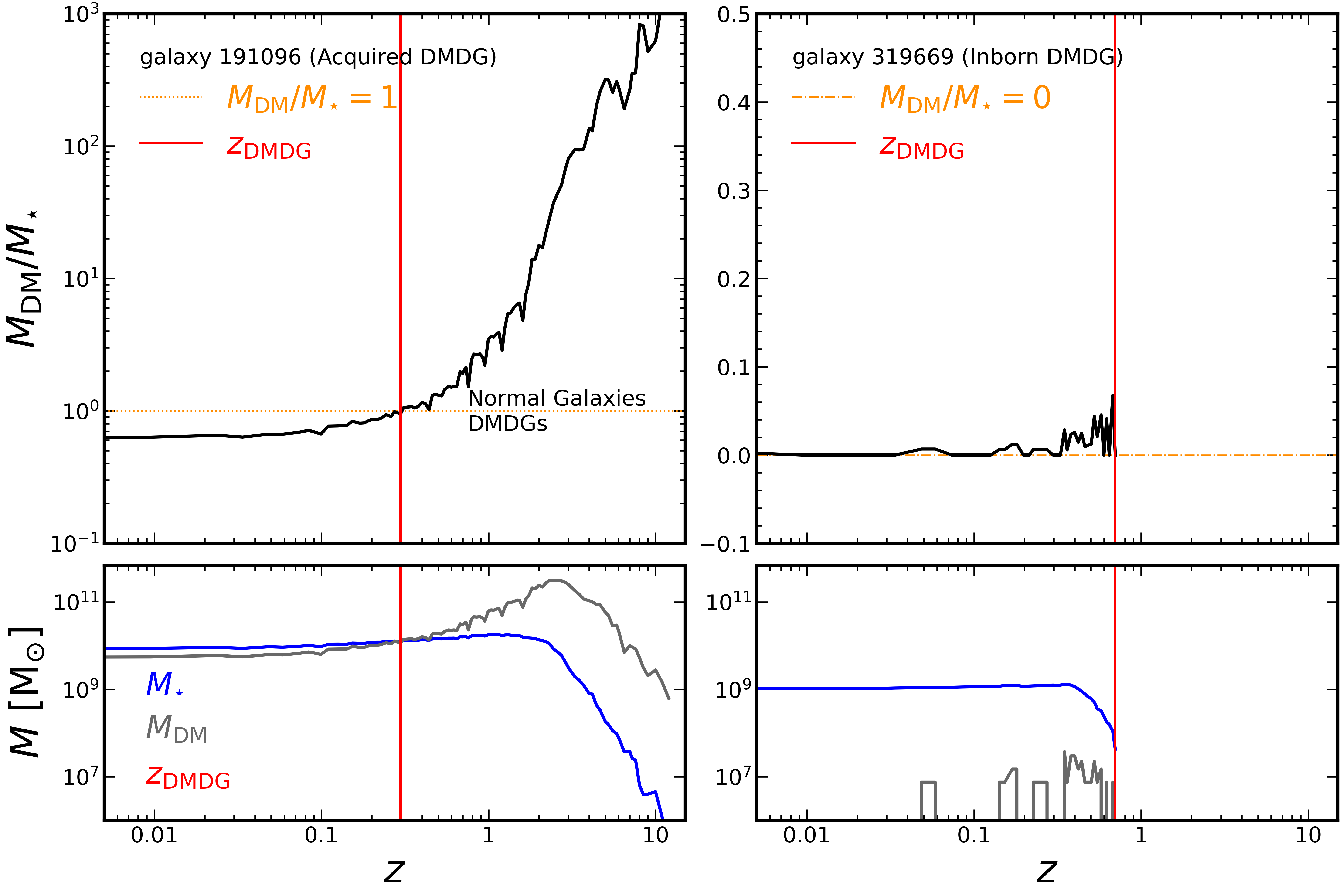}
\caption{
{Examples of dark matter--to--stellar mass ratio, stellar and dark matter masses of two DMDGs with different channels as a function of redshift. 
\textit{Left}: The acquired DMDG case (formed within a massive dark matter halo but late dark matter loss). 
\textit{Right}: The inborn DMDG case (formed with little/no dark matter). 
\textit{Top}: Dark matter--to--stellar mass ratio. The horizontal dotted (dot-dash) line indicates $M_{\rm DM}/M_{\star}=1~(0)$.
\textit{Bottom}: Stellar and dark matter masses of galaxies. The blue and gray lines show stellar and dark matter masses, respectively.
In the left (right) panels, the red vertical line marks the DMDG transition (formation) epoch.
}
}
\label{fig:ratio}
\end{figure*}

{We find that, for 15 inborn DMDGs, the earliest progenitors identified from the merger tree are inconsistent with their star formation histories.}
{In these galaxies, the merger tree assigns the first progenitor at an epoch later than the formation of their oldest stellar populations.}
{To examine this issue, we trace the particles back in time and search for earlier progenitor candidates that were missed by the merger tree.}
{We identify such candidates for all 15 galaxies.}
{However, in 14 cases, these candidates do not satisfy the TNG \texttt{SubhaloFlag} criterion and therefore cannot be treated as reliable progenitors.}
{We therefore retain only the one galaxy whose progenitor can be robustly reconstructed and exclude the remaining 14 galaxies from the sample because their formation epochs are not well defined.}
{As a result, our final sample includes 295 inborn DMDGs.}
{Further details are provided in Appendix~\ref{sec:app_a}.}

{\citet{Mitrasinovic2025} recently showed that the numerical artifacts associated with periodic boundary conditions (PBC) can contaminate DMDG samples in the IllustrisTNG simulations. 
We examine the positions, boundary crossing events, and mass evolution of DMDGs.
We identify two acquired DMDGs that undergo abrupt mass loss during boundary crossings and first satisfy the DMDG criterion at nearly the same epochs.
We exclude these systems from the fiducial sample and present the detailed tests in Appendix~\ref{sec:app_b}.}

Figure~\ref{fig:ratio} shows representative tracks of $M_{\rm DM}/M_{\star}(z)$ for two examples. 
In the acquired DMDG case, the dark matter mass grows after the initial formation, but declines steadily after $z \sim 2.5$, entering the DMDG regime. 
In the inborn DMDG case, the galaxy forms with very low dark matter content. 
{The bottom panels show} that the stellar masses tend to remain approximately constant over time after transition or formation epochs regardless of formation channel. 
{This suggests that continuous star formation is suppressed in dark matter--deficient environments.}

\begin{figure*}[!t]
\centering
\includegraphics[width=\textwidth]{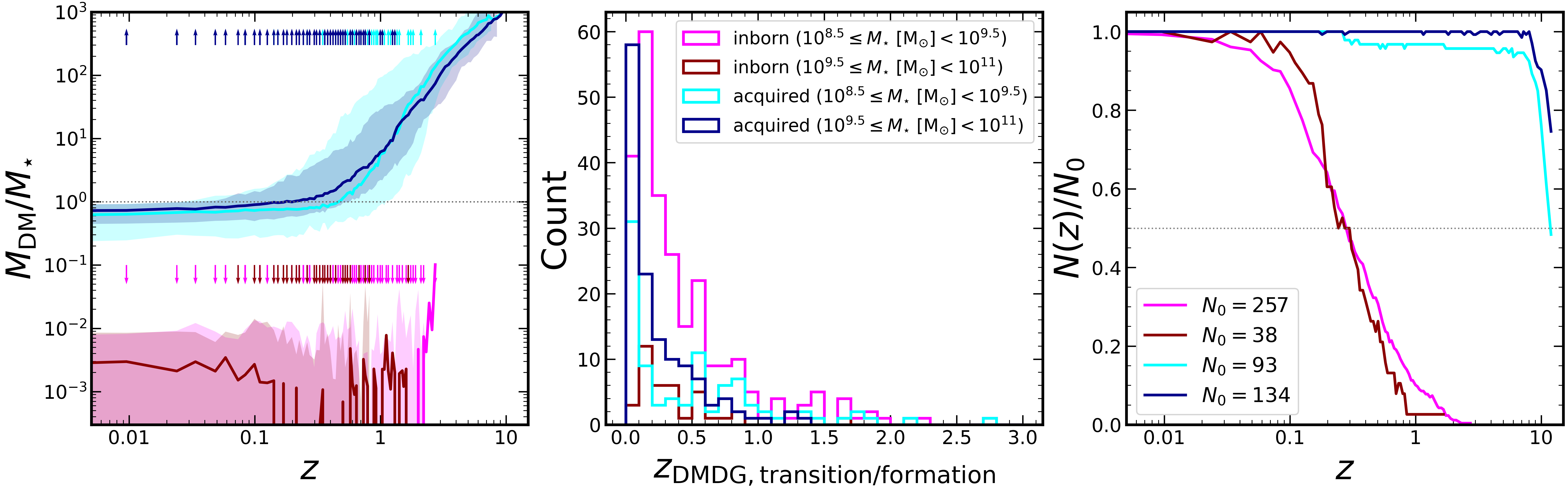}
\caption{
\textit{Left}: Similar to Figure~\ref{fig:ratio0}, evolutionary histories of the dark matter--to--stellar mass ratio for DMDGs differentiated by formation channels and stellar mass. 
{The curves indicate the median. 
The shaded bands show the $1\sigma$ scatter.} 
The arrows mark the transition (acquired) or formation (inborn) redshifts. 
\textit{Middle}: Distributions of DMDG transition/formation redshifts by channel. 
{{\textit{Right}: The number of DMDGs at each redshift, normalized by the total sample size at $z = 0$.
The horizontal dotted line marks $N(z)/N_{0} = 0.5$.}} 
The cyan and magenta colors denote the formation channels: acquired and inborn, respectively. 
{In all panels, the lines with bright colors correspond to less--massive $(M_{\star}=10^{8.5-9.5}\ \rm M_\odot)$ galaxies, while the lines with dark colors indicate the more--massive $(M_{\star}=10^{9.5-11}\ \rm M_\odot)$ galaxies.}
}
\label{fig:ratio_evol_zdmdg}
\end{figure*}

\begin{figure*}[!t]
\centering
\includegraphics[width=1.\textwidth]{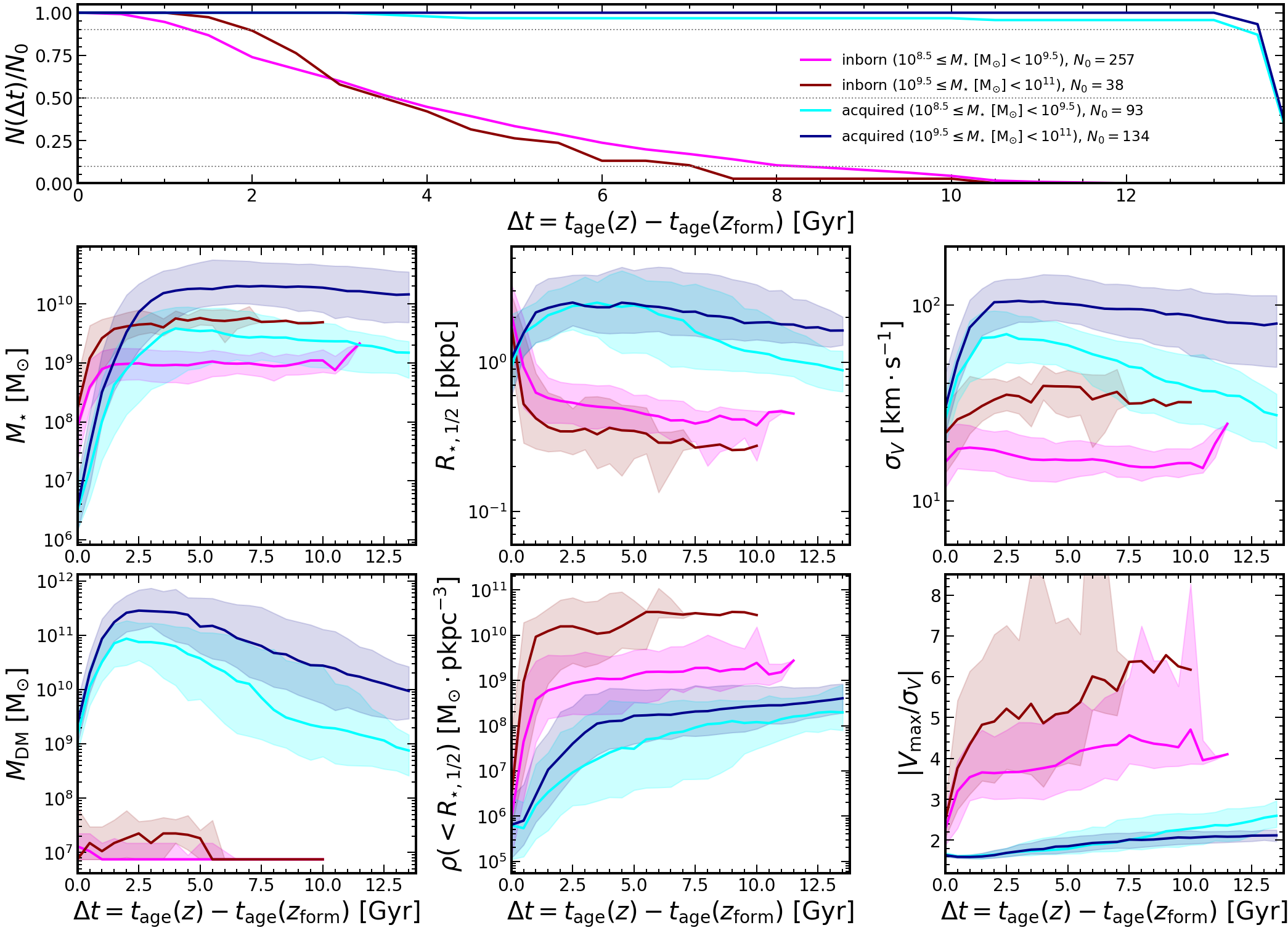}
\caption{
Evolutionary histories of the internal properties of DMDGs. 
The line colors follow Figure~\ref{fig:ratio_evol_zdmdg}, encoding formation channel (acquired vs. inborn) and the $z=0$ stellar mass bin.
{{\textit{Top panel}: Fraction of galaxies contributing to the median tracks at each elapsed time.
The horizontal dotted lines mark fractions of 0.1, 0.5, and 0.9.}}
\textit{Upper row (left to right)}: stellar mass, half--stellar mass radius {in the proper scale}, and one-dimensional velocity dispersion, measured from all bound particles/cells. 
\textit{Lower row (left to right)}: dark matter mass, central stellar mass density (stellar mass enclosed within the half--stellar mass radius {in the proper scale}, divided by the corresponding volume), and kinematic morphology, defined as the ratio between the maximum rotation velocity and velocity dispersion $|V_{\rm max}/\sigma_V|$, where $V_{\max}$ is the maximum of the spherically averaged rotation curve.
}
\label{fig:prop_evol}
\end{figure*}

The stellar mass range of our sample is $M_{\star}=10^{8.5-11}~\rm M_\odot$. 
Considering the sample size and the stellar mass range, we divide our sample into two stellar mass bins: less--massive galaxies with $10^{8.5} \leq M_{\star}(z=0)~[\rm M_\odot] < 10^{9.5}$ and more--massive galaxies with $10^{9.5} \leq M_{\star}(z=0)~[\rm M_\odot] < 10^{11}$. 
Table~\ref{tab:sample_size} summarizes the size of subsamples separated by formation channel and stellar mass. 
The left panel in Figure~\ref{fig:ratio_evol_zdmdg} shows the evolutionary histories of $M_{\rm DM}/M_{\star}$ for the subsamples of DMDGs depending on the class defined above and on the stellar mass. 
The curves show the median, and shaded regions indicate the 16th--84th percentiles ($\approx1\sigma$). 
Some of the less--massive galaxies exhibit a tendency to become DMDG earlier than more--massive ones. 
However, this trend is not statistically significant, given the $1\sigma$ scatter.
{The {middle} panel} represents {the redshift distribution of the transition or formation epochs for DMDGs, $z_{\rm DMDG}$.} 
Most DMDGs appear at comparatively late times ($z<1$). 
{The majority of more--massive DMDGs regardless of type form at $z<0.5$, whereas less--massive systems tend to form at higher redshifts.} 
{{The right panel represents the number of DMDGs at each redshift, which is normalized by the total sample size at $z = 0$, $N_0$.
While the acquired subsamples maintain $N(z)/N_{0}\approx 1$ across nearly all redshifts shown, the inborn subsamples drop below 50\% already at $z\sim 0.3$.}}
\begin{deluxetable}{lcccc}[!h]
\tabletypesize{\scriptsize}
\tablewidth{1\textwidth} 
\renewcommand{\arraystretch}{1.5}
\tablecaption{{The number of DMDGs separated by formation channel and stellar mass} \label{tab:sample_size}}
\tablehead{
\colhead{Stellar mass at $z=0$} & & \colhead{Acquired DMDGs} & & \colhead{Inborn DMDGs}
} 
\startdata 
 $10^{8.5-9.5}~\rm{M_{\odot}}$  &   & $93$ &   &   $257$   \\
 $10^{9.5-11}~\rm{M_{\odot}}$  &   & $134$ &   &   $38$    \\
\hline
 Total  &   & $227$ &   &   $295$
\enddata
\end{deluxetable}
\subsection{Internal Properties} \label{subsec:int_prop}

Now we turn our attention to the evolution of physical properties of DMDGs. 
{{As shown in Figure~\ref{fig:ratio_evol_zdmdg}, the fraction of inborn DMDGs contributing to the median tracks decreases rapidly with redshift. 
The redshift--binned medians do not trace a fixed population. 
At a given redshift, the sample combines galaxies that have evolved for several Gyr after formation with those that formed recently. 
We account for this difference by aligning each galaxy by its formation epoch and examining its evolution as a function of time since formation.}}
{{Figure~\ref{fig:prop_evol} shows the post--formation history of the DMDGs.
The panels represent the fraction of galaxies contributing to the median tracks (top panel), stellar mass (upper left), dark matter mass (lower left), half--stellar mass radius (upper center), central density (lower center), velocity dispersion (upper right), and kinematic morphology (lower right).}}

{{We calculate median trends using a window of 0.5 Gyr.
At each elapsed time, we collect data points within the window and assign one representative value to each galaxy by taking the median of its points in the window. 
We compute the median and the 16th--84th percentile range across galaxies. 
The adopted window width is approximately twice the maximum snapshot spacing in TNG100-1 ($\sim 0.24$ Gyr).
This width is chosen to balance irregular time sampling against the need to resolve the rapid evolution during the first $\sim 1$ Gyr. 
Most galaxies with available data contribute at least one snapshot to each window without excessive smoothing.
Using window widths of 0.3--1.0 Gyr does not change the qualitative trends.}}

{{The top panel shows the fraction of galaxies contributing at each elapsed time, $N(\Delta t)/N_0$, where $N(\Delta t)$ is the number of galaxies with at least one measurement within the window centered on $\Delta t$, and $N_0$ is the size of each subsample (See Table~\ref{tab:sample_size}). 
This fraction for inborn DMDGs drops below 0.5 after $\sim 3$ Gyr. 
We interpret the trends of inborn DMDGs at larger elapsed times with caution, because they are calculated from less than half of the corresponding $z=0$ subsample.}}

{{Acquired DMDGs form in the early Universe and their stellar masses increase steadily during the first $\sim 4$ Gyr after formation. 
The stellar mass of the less--massive acquired DMDGs (cyan) grows more slowly than that of more--massive ones (darkblue) during this period. 
At later times, less--massive systems show a gradual decrease in stellar mass, while more--massive systems maintain nearly constant stellar masses. 
Stellar masses of the inborn DMDGs increase rapidly during the first $\sim 1$ Gyr after formation, and then remain nearly constant. 
The more--massive inborn DMDGs (darkred) reach higher stellar masses than the less--massive ones (magenta), but both mass bins show broadly similar behavior.}}

{{The lower left panel shows the evolution of dark matter mass. 
acquired DMDGs, the early-time growth of acquired DMDGs is broadly similar to their stellar mass increase. 
After $\sim 2.5$ Gyr, their dark matter mass declines with time, with a stronger decrease for the less--massive subsample. 
Less--massive inborn DMDGs form with very little dark matter and remain dark matter poor, whereas the more--massive subsample shows a weak increase. 
However, given the dark matter particle mass and the low dark matter content of these systems, this apparent fluctuation appears to be driven by the stochastic accretion of a few particles and thus it is not statistically significant.}}

{{The upper center panel shows the evolution of the half--stellar mass radius.
The acquired DMDGs grow in size during the first $\sim 2.5$ Gyr and subsequently shrink, with a more rapid decrease in the less--massive subsample.
The inborn DMDGs shrink rapidly during the first $\sim 1$ Gyr and change little thereafter. 
The more--massive inborn DMDGs have the smallest sizes among the four subsamples.
While the acquired DMDGs show different size evolution depending on stellar mass, the inborn DMDGs show similar size evolution in the two stellar mass bins.}}

{{The lower center panel shows the central stellar mass density, defined as the stellar mass enclosed within the half--stellar mass radius divided by the corresponding spherical volume. 
The two acquired subsamples have similar central densities, indicating that their mass and size evolution largely compensate for each other.
Among the inborn DMDGs, the more--massive subsample has substantially higher central densities, consistent with their larger stellar masses and smaller sizes.}}

{{The upper right panel shows the one-dimensional velocity dispersion, $\sigma_V$, computed from all bound particles and cells. 
The less--massive acquired DMDGs show a steep decline in $\sigma_V$ at late times, broadly following their dark matter loss, whereas the more--massive systems show little evolution.
The inborn DMDGs have systematically lower velocity dispersions than the acquired systems.}}

\begin{figure*}[!t]
\centering
\includegraphics[width=1.\textwidth]{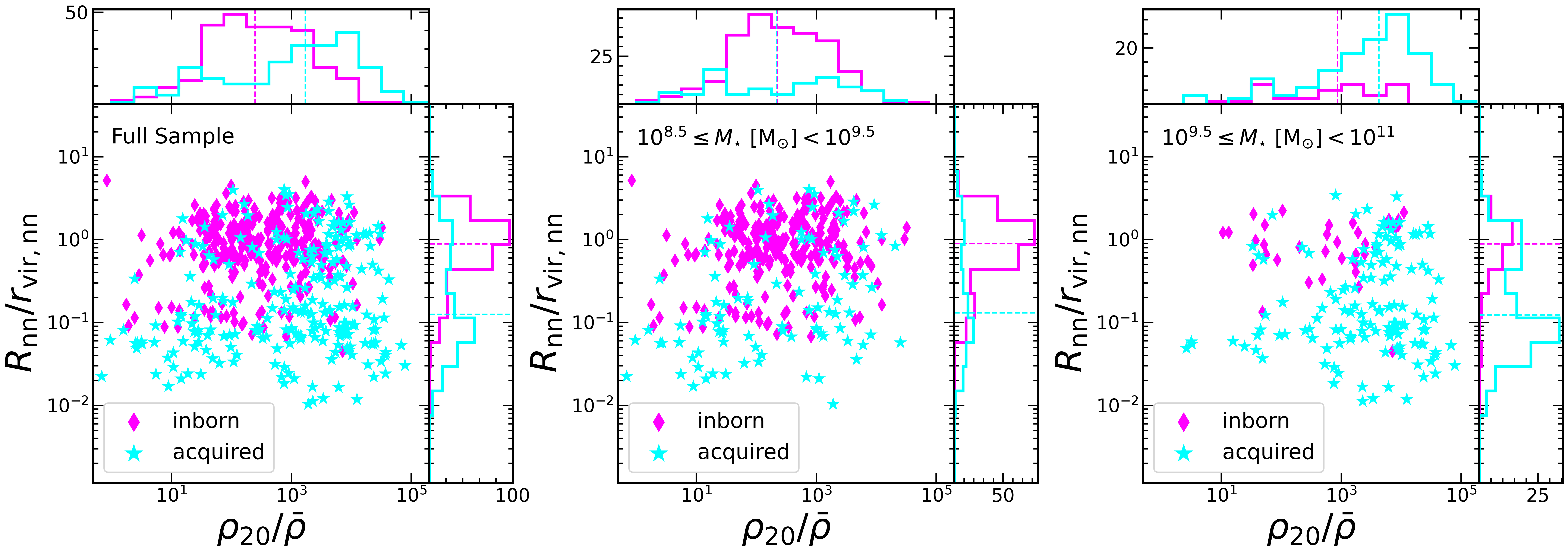}
\caption{
Distribution of DMDGs in the plane of large--scale density versus the nearest neighbor distance, where distance is normalized by the virial radius of the nearest neighbor. 
Both quantities are evaluated at the epoch of DMDG transition (acquired) or formation (inborn). 
{The diamond and star symbols represent the inborn and the acquired DMDGs, respectively.} 
\textit{Left}: full DMDG sample. \textit{Middle}: less--massive DMDGs. 
\textit{Right}: more--massive DMDGs. For each panel, the top and right subpanels show the histograms of the large--scale density and the normalized distance, respectively. 
The dashed lines indicate the medians.
}
\label{fig:env}
\end{figure*}

{{The lower right panel shows the kinematic morphology, defined as $|V_{\rm max}/\sigma_V|$, where $V_{\rm max}$ is the maximum of the spherically averaged rotation curve.
The acquired DMDGs are predominantly dispersion-supported, whereas the inborn DMDGs are more rotation-supported. 
Among the acquired DMDGs, the less--massive subsample has slightly higher $|V_{\rm max}/\sigma_V|$ than the more--massive subsample. 
The inborn population shows the opposite mass dependence, with higher values in the more--massive subsample despite higher velocity dispersions.}}

\subsection{Environmental Effects} \label{subsec:env}

The influence of environment on galaxy formation and evolution has been well established through the observed morphology--density relation \citep[e.g., ][]{Dressler1980,Postman1984}. 
Numerous studies have explored the environmental dependence of galaxy properties from the local universe \citep[e.g.,][]{Park2007, Park2009, Thomas2010, Hwang2010, Hwang2012} to intermediate redshift \citep[e.g.,][]{Postman2005, Smith2005, Hwang2009}. 
This section investigates whether the formation and evolution of DMDGs are linked to specific environmental conditions. 
We adopt two environmental parameters for this purpose. 
The first one is the large--scale background density on scales of a few Mpc \citep{Park2007, Muldrew2012}. 
For a given galaxy position, we measure the local density using the $n$-nearest dark matter particles as
\begin{equation}
    \rho_{n}/\bar{\rho} = \sum_{i=1}^{n}W(|r_i|, h)/\bar{\rho},
\end{equation}
where $W(|r_i|, h)$ is a smoothing kernel, $r_i$ is the distance to the $i$-th dark matter particle, and $h$ is a scaling factor normalized by the distance to the $n$-th nearest particle \citep[see Section 3 in][for details]{Muldrew2012}. 
We normalize the local density by the mean density of dark matter over the simulation volume. 
{We adopt $n=20$, which is often used as a proxy for the large--scale environment in previous studies \citep{Park2007, GLee2024, Jeong2025}. 
Because a snapshot of TNG100 contains too many particles $(N_{\rm DM} \sim 6\times10^9)$, we randomly down-sample the data to a subset containing $7.5 \times 10^{-6}$ of the total dark matter particles. 
With this sampling ratio, $\rho_{20}$ represents the environment on a few Mpc scale, beyond the typical size of a galaxy cluster.}

{The second one is the small--scale environment parameterized by the distance to the nearest galaxy normalized by its virial radius \citep{Park2008, Park2009b}.} 
The virial radius is defined by the following equation:
\begin{equation}
    r_{\rm vir} = \left(\frac{3}{4\pi}\frac{M_{\rm tot}}{200\rho_c}\right)^{1/3},
\end{equation}
where $M_{\rm tot}$ is the total mass of the nearest neighbor and $\rho_c$ is the critical density.
The nearest neighbor of each target galaxy is the closest system that satisfies a stellar mass criterion motivated by the major merger threshold, $M_{\star, \rm nei} > 0.2~M_{\star, \rm tar}$, where $M_{\star, \rm nei}$ is the stellar mass of the neighbor, and $M_{\star, \rm tar}$ is that of the target galaxy {at the formation epoch}. 
{We vary the stellar mass criterion for the nearest neighbor from 0.2 to 2 times the stellar mass of the target galaxy and confirm that the qualitative distinction between the two formation channels remains consistent over this range.}
We note that the virial radius estimated here is based on the assumptions that the galaxy is virialized and that its total mass corresponds to the virial mass. 
It should therefore be regarded as a pseudo--virial radius.

In this analysis, we exclude targets whose neighbors have total masses below $\sim 1.842\times10^{8}~\rm{M_\odot}$, which is the mass limit coming from the incomplete mass function of the simulation \citep{Nelson2019a}.
In our sample, 14 inborn DMDGs have neighbors with total masses below this threshold.
Figure~\ref{fig:env} shows the joint distribution of large--scale density and the nearest neighbor distance for the DMDGs. Both environmental quantities are evaluated at their transition (acquired) or formation (inborn) epochs. 
{The left panel shows the total sample while the middle and right panels exhibit the subsamples split by $z=0$ stellar mass.} 
In the left panel, the inborn DMDGs preferentially occupy relatively underdense regions that indicates low large--scale density and low influence from the nearest neighbor, whereas the acquired DMDGs tend to lie toward the lower--right side. 
Thus, the acquired DMDGs form in denser large--scale environments and experience stronger influence from the nearest neighbor. 
{In the middle panel, the median values of $\rho_{20}$ are similar. 
However, {the shapes of histograms differ}, and a Kolmogorov--Smirnov (KS) test comparing the $\rho_{20}$ distributions yields $p$--value $<0.001$, indicating a statistically significant difference.} 
{Comparing the middle and right panels, the more--massive acquired DMDGs form in denser environments than the less--massive ones, whereas the inborn subsamples show little difference.} 

{According to \cite{Park2009b}, there are two characteristic scales for the distance to the nearest neighbor. 
The first is the virial radius of the nearest neighbor galaxy $r_{\rm vir,nn}$, and the second is the separation where galaxies start to merge, approximately $0.05r_{\rm vir,nn}$. 
We conduct the KS test for the DMDG subsamples separated at the first scale. 
For $R_{\rm nn} > r_{\rm vir, nn}$, differences among the less--massive DMDG subsamples are not statistically significant in the background density distributions $(p=0.054)$. 
The more--massive DMDGs also show similar distributions $(p=0.060)$. 
For $R_{\rm nn} < r_{\rm vir, nn}$, the background density distributions of DMDGs in each stellar mass bin differ, with $p$--value of 0.001 and 0.002, respectively. 
We do not pursue further analysis for DMDGs with $R_{\rm nn}<0.05r_{\rm vir,nn}$, given that only two inborn DMDGs fall within this scale.} 
Overall, the inborn DMDGs {reside in underdense regions probed by both small and large--scale environment measures} than the acquired DMDGs.

\section{Discussion} \label{sec:discuss}

\subsection{Formation Channels} \label{subsec:formchan}

In this section, we {discuss} the origin and evolution of DMDGs based on our findings. 
We find that {there are at least two ways for producing DMDGs}, and that the resulting internal properties depend on the formation channel and on stellar mass. 
The scenarios discussed in the literature include {(i) tidal stripping, and (ii) mini--Bullet Cluster event, and (iii) tidal debris environment}. 
All three scenarios can occur {in cosmological simulations} \citep[e.g.,][]{Ploeckinger2018, Haslbauer2019, LeeJoo2024, Contreras-Santos2024}. 

{Tidal Stripping:} The evolution of dark matter mass and {the dark matter--to--stellar} mass ratio in our results {could be understood in the context of tidal stripping.} 
As shown in Figures~\ref{fig:ratio_evol_zdmdg} and \ref{fig:prop_evol}, the acquired DMDGs lose dark matter over time. 
These galaxies reside in dense large--scale environments and in close proximity {to massive neighbors}. 
Such environments promote strong tidal interactions that can remove dark matter from their halos. 
Furthermore, the subsequent increase in the nearest neighbor distance suggests reduced opportunities for re-accretion, reinforcing the persistence of low dark matter content. 
According to \cite{Lovell2025}, {the stripping of dark matter depends on the dynamical mass of infalling galaxies} and the mass of the host halos, yet {it is consistently observed} across the IllustrisTNG suite. 
This suggests that although the degree and time of stripping can be influenced by numerical effects, the phenomenon itself is observed regardless of the simulation resolution. 

{{We compare our results with previous simulation studies of DMDG formation. 
The acquired DMDGs initially have ordinary dark matter contents but subsequently become dark matter deficient. 
Previous studies identify tidal stripping as central to this evolutionary channel. 
Cosmological simulations indicate that satellites can become dark matter deficient through close interactions with massive central or companion galaxies \citep[e.g.,][]{Jing2019,Jackson2021}. 
In particular, \citet{Moreno2022} found that repeated close encounters with massive neighbors remove much more dark matter than stellar mass, with the mass-loss episodes occurring near pericentric passages. 
Idealized $N$-body simulations similarly show that dark matter is stripped primarily during successive pericentric passages, when the tidal field of the host is strongest \citep[e.g.,][]{Ogiya2022}. 
These studies consistently show that dark matter is stripped more efficiently than stars, producing a rapid decline in dark matter mass accompanied by a much weaker or delayed decrease in stellar mass.
Using TNG50, \cite{deAlmeida2026} showed that dark matter--deficient compact satellites lose most of their dark matter after entering their hosts.
These satellites also lose a substantial fraction of their stellar mass. 
The tidal removal of their outer stellar envelopes reduces their sizes. 
This coupled mass and size evolution is consistent with that of acquired DMDGs. 
Our environmental analysis examines these systems when they first satisfy the DMDG criterion.
At this epoch, acquired DMDGs reside in environments conducive to strong tidal interactions with more massive neighbors. 
Taken together, these results support tidal stripping as the primary mechanism driving their transition to dark matter deficiency.}}

{We also note that the oscillating feature in the dark matter mass evolution shown in Figure~\ref{fig:ratio} may arise from a limitation of the galaxy-finding algorithm, which identifies only the densest part of the structure of a subhalo above the background density field, especially when the subhalo passes near pericenter in a larger halo \citep{Springel2021,Lovell2025}.}

{Mini--Bullet Cluster Event:} {Our results indicate that DMDGs can arise through more than one formation channel. 
We now consider the inborn population in our sample. 
Because these systems are already dark matter deficient when first identified, they may at first be qualitatively consistent with the mini--bullet scenario. 
However, we could not directly compare our results to collision--induced DMDGs formed within this mechanism.}
As pointed out by \cite{Shin2020}, the resolution of TNG100 is insufficient to confirm whether DMDGs form in such an event. 
Nevertheless, \cite{LeeJoo2024} provided suggestive evidence for their occurrence and for environmental signatures that can be applicable partly to inborn DMDGs. 
{In the same simulation (TNG100), collision events are identified with significant probability, yielding between 11 and 115 occurrences under stringent conditions over $0.01 < z < 3$.}
{In contrast to the tidal stripping scenario, DMDGs formed through these events are born with very little dark matter, consistent with the properties of inborn DMDGs \citep{Silk2019,Shin2020,LeeJoo2021,LeeJoo2024}.}

{Tidal Debris Environment:} {{Observed TDG candidates span a broad range in stellar mass, $10^{6\texttt{-}9}\ \rm M_{\odot}$ \citep{Sheen2009, Wen2012, Zaragoza-Cardiel2024}. 
This range partially overlaps with that of the inborn DMDGs.
Resolved stellar populations in TDGs indicate continuous star formation over the preceding 200 Myr \citep{Makarova2002}.
Chemodynamical simulations find star formation lasting more than 300 Myr and, in later models, throughout 3 Gyr of simulated evolution \citep{Recchi2007, Ploeckinger2015}.
The old TDG formed approximately 4 Gyr ago and continues to form stars \citep{Duc2014}.
We examine the cumulative star formation histories of the inborn DMDGs and quantify the duration of their star formation (see Section~\ref{subsubsec:sfh}). 
Most inborn DMDGs have intervals shorter than 5 Gyr, broadly consistent with the range reported for TDGs.}}

{{As mentioned earlier, TDGs tend to be more metal--rich than normal dwarf galaxies of similar luminosity because they form from pre--enriched material stripped from the outer disks of massive galaxies \citep[e.g.,][]{Weilbacher2003,Duc2012,Fensch2019b}.
The precursor gas of the inborn DMDGs is already enriched to approximately half the solar metallicity about 0.8 Gyr before the formation epoch (Section~\ref{subsubsec:inborngas}).
Spatial tracking further shows that, in some systems, this gas assembles through tidal interactions between extended gaseous structures. 
These results suggest that some inborn DMDGs form through a channel similar to that of TDGs. 
However, TDGs occupy the upper locus of the dwarf galaxy size--mass relation, with unusually large effective radii at fixed stellar mass \citep{Duc2014}. 
By contrast, the inborn DMDGs tend to have smaller effective radii at fixed stellar mass. 
Further investigation is needed to determine why systems with apparently similar formation channels exhibit contrasting structural properties.}}

{{We further discuss the possible origins of the inborn DMDGs.
The high fractions of the in situ star formation of the inborn DMDGs indicate that nearly all their stars form along their main progenitor branches (Section~\ref{subsubsec:sfh}).
Before formation, nearly all their precursor gas is bound to another galaxy and is already metal enriched (Section~\ref{subsubsec:inborngas}). 
These findings indicate that the inborn DMDGs form from pre--enriched material previously associated with other systems rather than from pristine IGM accreted by their own pre--existing massive dark matter halos. 
The spatial analysis reveals diverse gas assembly configurations, including tidal interactions, filamentary flows, mergers involving gaseous structures, and clumps detaching from disk structures (Section~\ref{subsubsec:inborngas}). 
We interpret tidal debris environment as one of several pathways for the inborn DMDGs, rather than as the origin of the entire population. 
Although the precursor gas becomes more compact, kinematically colder, and increasingly star forming, our analysis does not resolve the gravitational processes responsible for this assembly.
Identifying these processes will require higher resolution simulations and a detailed analysis of the gravitational interactions within the gas and between the gas and its surroundings.}}


\subsubsection{STAR FORMATION HISTORY OF THE INBORN DMDGS} \label{subsubsec:sfh}
{{In this section, we construct the cumulative star formation history of each inborn DMDG from the formation times and initial masses of its member stellar particles identified at $z=0$ \citep[e.g.,][]{Simha2014,Diemer2017,Joshi2021}.
We define $t_{10}$ and $t_{90}$ as the elapsed times at which the cumulative fraction reaches 0.1 and 0.9, respectively. 
We  characterize the star formation duration as
}}
\begin{equation}
    \Delta t_{\rm SF} = t_{90} - t_{10}.
\end{equation}

\begin{figure*}[!t]
\centering
\includegraphics[width=0.98\textwidth]{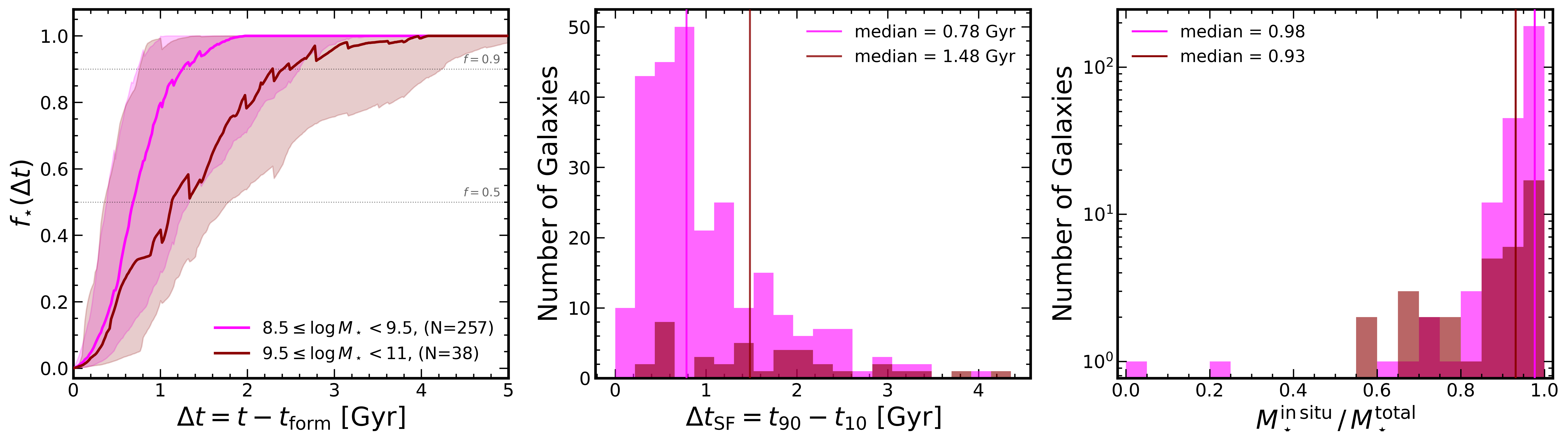}
\caption{
{{Post identification stellar mass assembly of the inborn DMDGs. 
\textit{Left}: Cumulative star formation history. 
Both the numerator and the denominator include only star particles formed after galaxy formation, weighted by their initial mass, so that each curve reaches unity by construction.
Solid curves show the median across galaxies at each $\Delta t$, and shaded regions span 16th--84th percentiles.
Dotted horizontal lines mark $f_\star = 0.5$ and 0.9. 
\textit{Middle}: Distribution of the star formation duration, $\Delta t_{\rm SF} = t_{90} - t_{10}$, the interval over which the cumulative stellar mass fraction increase from 10\% to 90\%. 
\textit{Right}: Distribution of the in situ stellar mass fraction at $z=0$, where in situ refers to stars formed within the galaxy’s own main progenitor branch, so that the ex situ fraction is the complement.
}}
}
\label{fig:sfh}
\end{figure*}

{{The cumulative history based on all $z=0$ member particles may include stars formed in other progenitors that subsequently merged with the galaxy \citep{Diemer2017}. 
We examine the stellar mass assembly of the inborn DMDGs to determine whether post--identification portions of these histories are dominated by in situ star formation or by the accretion of ex situ stars. 
We use the TNG stellar assembly catalog, which decomposes stellar mass growth into in situ and ex situ components \citep{Rodriguez-Gomez2015,Rodriguez-Gomez2016,Rodriguez-Gomez2017}. 
The catalog further classifies the ex situ stars according to whether they formed before or after infall and whether they were accreted through major mergers, minor mergers, or flyby encounters.
}}

{{Figure~\ref{fig:sfh} shows the cumulative star formation histories (left panel), distributions of the star formation duration (middle panel), and in situ stellar mass fractions (right panel). 
The median durations are 0.78 and 1.48 Gyr for the less-- and more--massive subsamples, respectively.
The duration is shorter than 1 Gyr for 160 less--massive inborn DMDGs (62\%) and 11 more--massive inborn DMDGs (29\%). 
Thus, the less--massive inborn DMDGs form their stars over shorter timescales than the more--massive subsample, although both subsamples assemble most of their stellar mass on sub-Gyr to Gyr timescales.}}

{{The in situ stellar fraction, $f_{\rm in\ situ}$, is the fraction of the $z=0$ stellar mass formed along the main progenitor branch. 
Ex situ stars formed in other galaxies and were subsequently accreted. 
In situ stars dominate the stellar mass of the inborn DMDGs, with median $f_{\rm in\ situ}$ values of 0.98 and 0.93 for the less-- and more--massive subsamples, respectively. 
Only two inborn DMDGs have $f_{\rm in\ situ}<0.5$. 
One system (galaxy 289450) has $f_{\rm in\ situ}=0.217$ because a single major merger with a mass ratio greater than 1:4 contributes most of its ex situ mass. 
The merger tree of the other system (galaxy 108506) fails to recover the true main progenitor branch, causing the stellar assembly catalog to assign $f_{\rm in\ situ}=0$. 
We reconstruct its progenitor branch separately throughout this study (Appendix~\ref{sec:app_a}).}}

{{As discussed above, the in situ classification is defined relative to the main progenitor branch, which begins when each inborn DMDG is first identified.
The high in situ fractions show that most stars form along this branch rather than being accreted from other galaxies. 
Thus, most inborn DMDGs do not inherit a substantial pre-existing stellar population. 
However, the in situ classification does not demonstrate that the gas was gravitationally self-bound or that its self-gravity drove star formation.
We examine the dynamical state of the gas separately in Section~\ref{subsubsec:inborngas}.}}

\subsubsection{GAS ASSEMBLY AND FORMATION PATHWAYS OF THE INBORN DMDGS} \label{subsubsec:inborngas}
\begin{figure*}[!t]
\centering
\includegraphics[width=0.98\textwidth]{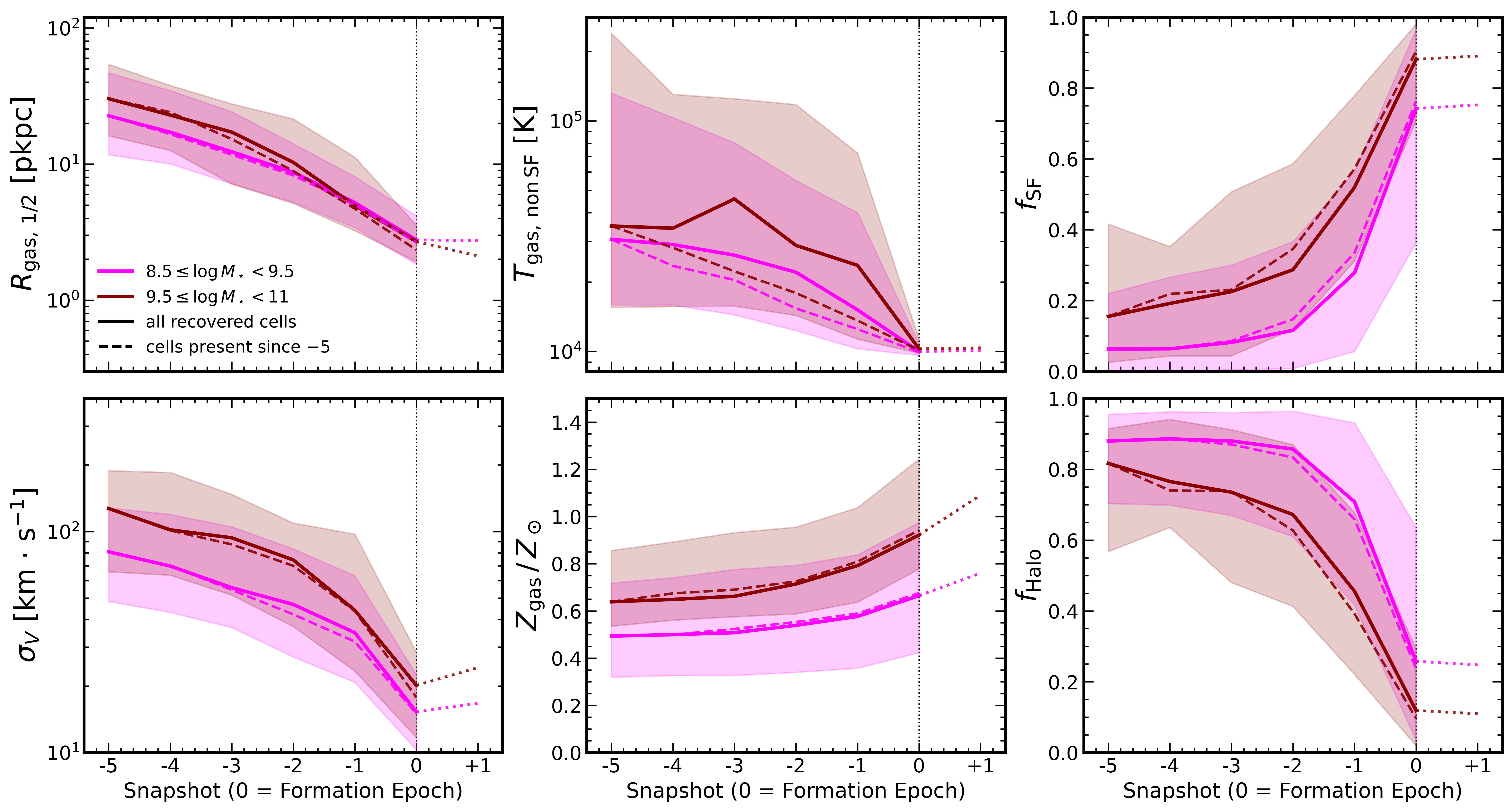}
\caption{
{{Evolution of the gas properties of inborn DMDGs around the first-identification epoch.
The $x$-axis shows time, with the formation epoch denoted as zero. 
\textit{Upper left}: half--gas mass radius, $R_{\rm gas,1/2}$, shown for both mass bins.
\textit{Lower left}: one-dimensional velocity dispersion, $\sigma_{V}$, measured in the center of mass frame of the tracked gas. 
\textit{Upper center}: mass--weighted temperature of non-star-forming gas, $T_{\rm gas,\,non\,SF}$.
\textit{Lower center}: mass--weighted metallicity, $Z_{\rm gas}$, in solar units $(Z_{\odot}=0.0127)$.
\textit{Upper right}: star forming gas fraction, $f_{\rm SF}$.
\textit{Lower right}: halo phase gas fraction, $f_{\rm Halo}$.
Solid curves denote the median across galaxies at each stage and shaded bands the 16th--84th percentiles. 
Dashed curves show the same quantities computed using only those gas cells that are already recovered at offset $-5$, so that an identical cell set is followed at every stage. 
The agreement between solid and dashed curves indicates that the trends are not driven by the changing number of recovered cells. 
Dotted curves at positive offsets are shown for reference only; after first identification we read all gas cells bound to the descendant subhalo at each snapshot.
}}
}
\label{fig:gasprops}
\end{figure*}

{{In this section, we investigate how gas assembles into the inborn DMDGs, which contain little dark matter at formation. 
We identify member gas cells of each galaxy at the formation epoch and search for the same cell IDs in the preceding five snapshots, extending back by approximately 0.8 Gyr. 
We measure the half--gas mass radius $(R_{\rm gas,1/2})$, one-dimensional velocity dispersion $(\sigma_{V})$, temperature $(T)$, hydrogen number density $(n_{\rm H})$, and mean metallicity $(Z_{\rm gas}/Z_{\odot})$.
Following \citet{Martizzi2019}, we classify the gas into six phases: star forming gas, halo gas, the diffuse intergalactic medium (IGM), the warm-hot intergalactic medium (WHIM), the warm circumgalactic medium (WCGM), and the hot medium. 
We calculate the fraction in each phase (e.g., $f_{\rm SF}$).
The temperature of star forming gas is set by the subgrid equation of state and does not represent a physical gas temperature\footnote{{{See the TNG data release documentation: \url{https://www.tng-project.org/data/docs/specifications}.}}}.
Therefore, we calculate the temperature evolution using only the non-star-forming gas.
}}

{{Figure~\ref{fig:gasprops} shows the time evolution of $R_{\rm gas,1/2}$, $\sigma_{V}$, $T$, $Z_{\rm gas}/Z_{\odot}$, $f_{\rm SF}$, and $f_{\rm Halo}$.
The simultaneous decreases in $R_{\rm gas,1/2}$ and $\sigma_{V}$ indicate that the precursor gas becomes more compact and kinematically colder toward the formation epoch. 
Its mean metallicity is already 0.5--0.64 $Z_{\odot}$ well before galaxy formation and increases only mildly thereafter. 
This pre--enrichment is inconsistent with an origin in chemically pristine IGM. 
The non-star-forming gas cools from temperatures above $10^4\ \rm K$ toward $\sim 10^4\ \rm K$. 
This temperature scale is characteristic of photoionized gas, and the decline is consistent with radiative cooling of enriched gas exposed to the metagalactic ultraviolet background \citep{Hui1997, Wiersma2009}. 
The star forming gas fraction increases toward formation, whereas the halo gas fraction decreases. 
The WCGM accounts for less than 10\% of the gas mass at early times and declines further toward formation. 
The diffuse IGM, WHIM, and hot medium make negligible contributions throughout the tracked period.}}

\begin{figure*}[!t]
\centering
\includegraphics[width=\textwidth]{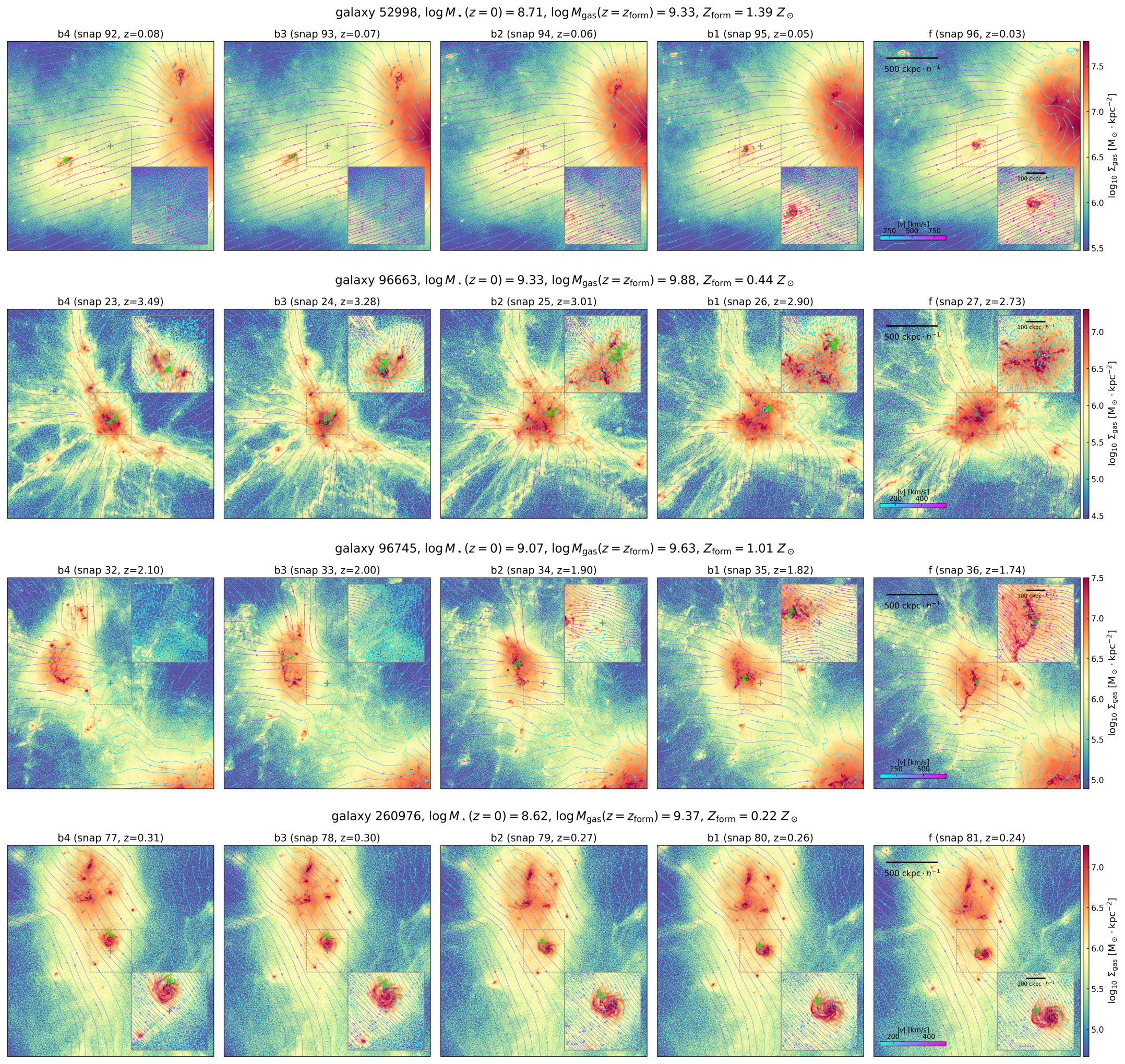}
\caption{
{{Four representative examples of the gas distribution projected onto the $xy$-plane within a cubic region of side length $2\ {\rm cMpc}\cdot h^{-1}$, centered on the formation-epoch position of each inborn DMDG. 
Each row shows one galaxy, labeled by galaxy ID at $z=0$, stellar mass at $z=0$, gas mass at the formation redshift, and the metallicity of the precursor gas at formation. 
Columns show, from left to right, five snapshots before formation through the formation epoch itself. 
The color scale shows the projected gas surface density, $\log{\Sigma_{\rm gas}}$. 
Each panel includes an inset showing the $400\ {\rm ckpc}\cdot h^{-1}$ cubic region enclosed by the dashed square. 
Green contours mark the distribution of the precursor gas. 
Arrows show the peculiar velocity field of the gas, colored by speed in $\rm km\cdot s^{-1}$.
In each row, all panels share the same color scale including subpanels.
In all four examples, the precursor gas develops into a distinct clump toward the formation epoch. 
The surrounding environments differ substantially among the systems though.
}}
}
\label{fig:gasenv}
\end{figure*}

{{Because AREPO cells can be refined or derefined and exchange mass across cell boundaries, gas cell IDs do not provide exact Lagrangian trajectories \citep{Springel2010,Genel2013}.
Consequently, only a fraction of the gas cells identified at the formation epoch can be recovered at earlier times. 
As a robustness test, we repeat the analysis using only gas cells whose IDs are recovered at every tracked snapshot.
This selection retains a fixed set of cell IDs throughout the tracked period, although it does not account for mass exchange across cell boundaries.
The evolutionary trends in $R_{\rm gas,1/2}$, $\sigma_V$, $Z_{\rm gas}/Z_{\odot}$, and the gas phase fractions derived from this fixed-ID sample closely agree with those obtained using all gas cells recovered at each snapshot.
The mean temperature of the non-star-forming gas differs by up to a factor of approximately 1.6 at intermediate stages, with the fixed cell sample yielding lower temperatures. 
Nevertheless, both samples show declining temperatures toward formation.}}

{{We also examine whether the precursor gas was bound to other galaxies, resided in the diffuse intragroup medium, or was gravitationally unbound before formation. 
Nearly all the precursor gas is bound to other galaxies throughout the pre--formation interval and becomes associated with the newly identified inborn DMDG only at the formation epoch. 
The diffuse intragroup and unbound components make negligible contributions.
This trend is consistent across both stellar mass bins.}}


{{We inspect the spatial distribution of the precursor gas within a cube of side length $2\ {\rm cMpc}\cdot h^{-1}$ centered on each galaxy at its formation epoch to examine possible formation pathways. 
Figure~\ref{fig:gasenv} shows four representative examples, with arrows indicating the peculiar velocity field of the gas at each epoch. 
These examples exhibit several gas assembly configurations.
A gas clump forms when a smaller halo falls into a more massive halo (galaxy 52998). 
Gas flowing along the large--scale filaments accumulates through interactions with the surrounding structures and forms a compact clump (galaxy 96663). 
A clump develops within a merging massive gaseous halo, possibly through an internal instability (galaxy 96745). 
A tidal interaction appears to destabilize an extended gas distribution and produce a distinct condensation (galaxy 260976).}}

{{These examples indicate that inborn DMDGs form in diverse environments rather than through a single pathway. 
In each case, the gas develops a compact clump and begins forming stars, but the assembly process differs. 
This diversity contrasts with the predominantly stripping--driven evolution of the acquired DMDGs.
Across the inborn population, however, the precursor gas shows common evolution; it becomes more compact and kinematically colder, and an increasing fraction enters the star forming phase. 
These trends indicate that the assembling gas increasingly satisfies the conditions imposed by the TNG star-formation prescription before the system is identified as a galaxy.}}

{{The onset and efficiency of star formation in the inborn systems may depend on the adopted subgrid ISM model. 
In TNG, the unresolved multiphase ISM is treated with an effective equation of state, and gas above \(n_{\rm H}\simeq0.1\,{\rm cm}^{-3}\) becomes eligible for stochastic star formation according to an empirical Kennicutt--Schmidt relation \citep{Springel2003,Pillepich2018b}. 
The increasing star forming gas fraction indicates that more precursor gas crosses the density threshold imposed by the TNG model. 
However, it does not establish that the same condensations would collapse and form stars with comparable efficiency if cloud scale dynamics were resolved or additional local collapse criteria were imposed. 
Higher resolution simulations that resolve these dynamics or implement such criteria are needed to determine how sensitively the predicted formation and abundance of inborn DMDGs depend on the star formation prescription.}}

\subsection{How special are DMDGs in comparison with normal galaxies?} \label{subsec:comp}

\begin{figure*}[!t]
\centering
\includegraphics[width=.98\textwidth]{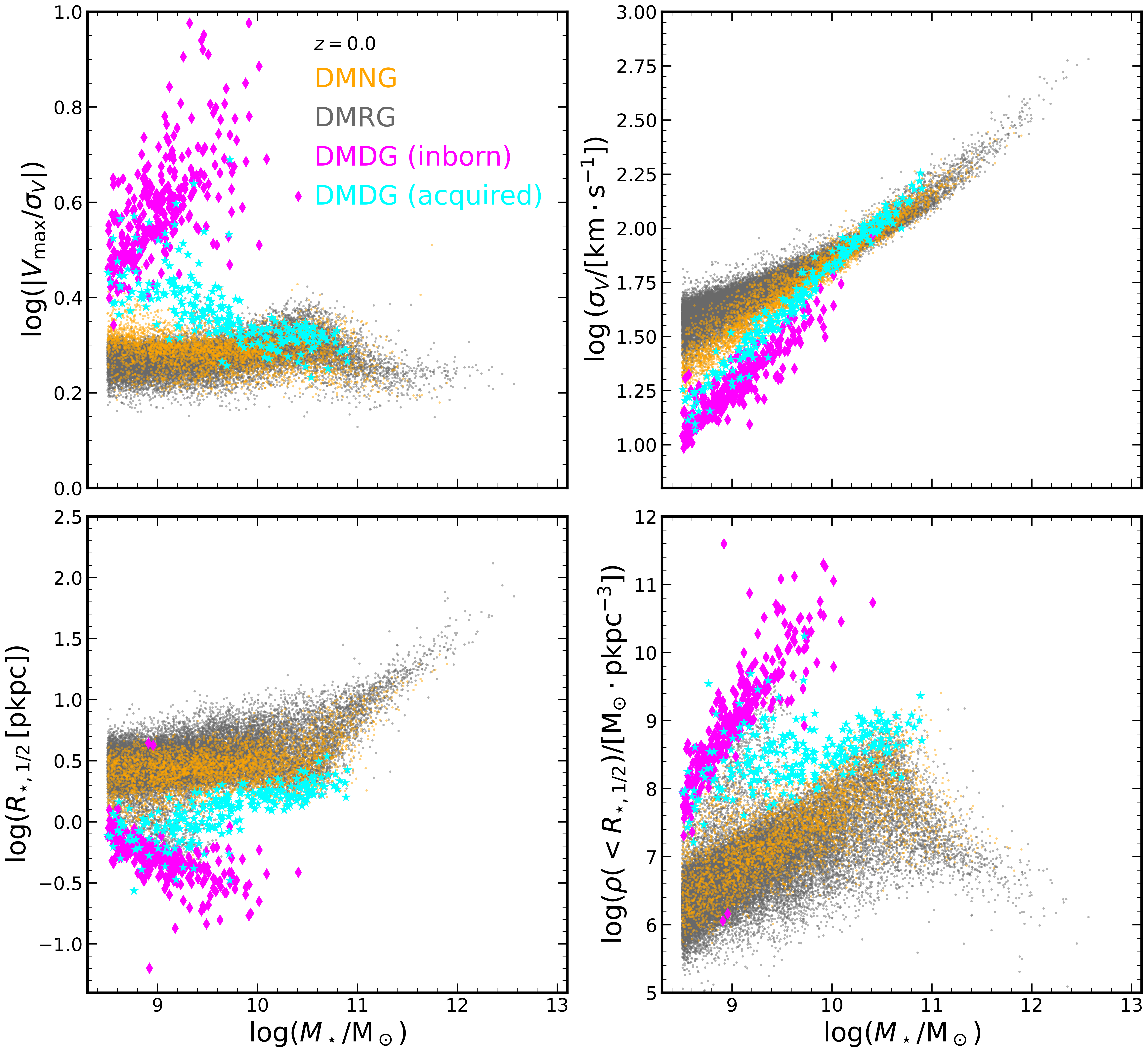}
\caption{
Internal properties of galaxies as a function of stellar mass at $z=0$. 
DMNGs and DMRGs are selected based on the mass ratio criteria satisfying $1\leq M_{\rm DM}/M_{\star}<10$ and $10 \leq M_{\rm DM}/M_{\star}$, respectively. 
{For all panels, the magenta diamond symbol represents the inborn DMDGs. The cyan star marker is the acquired DMDGs.} The orange dot indicates DMNGs. The gray dot marks DMRGs. 
Each panel displays the kinematic morphology, velocity dispersion, half--stellar mass radius, and central density, respectively.
}
\label{fig:prop_z0}
\end{figure*}

We compare the internal properties of DMDGs with those of other galaxy populations and discuss this. 
We define dark matter--normal galaxies (DMNGs) as systems with dark matter--to--stellar mass ratios between 1 and 10, and dark matter--rich galaxies (DMRGs) as those with mass ratios larger than 10. 
We identify {4,205} DMNGs, and {28,485} DMRGs. 
Figure~\ref{fig:prop_z0} shows the physical properties of galaxies as a function of the stellar mass at $z=0$. 

Most inborn DMDGs have stellar masses smaller than $10^{10}\ \rm M_{\odot}$, whereas the acquired DMDGs span a wider range of $10^{8.5-11}\ \rm M_{\odot}$. 
{At a given stellar mass, DMNGs and DMRGs largely overlap in their structural and kinematic properties, while DMDGs occupy a distinct locus.}
The inborn DMDGs are typically smaller, more centrally concentrated, and more rotationally supported than the acquired DMDGs. 
As shown in Figure~\ref{fig:prop_evol}, the inborn DMDGs tend to become increasingly compact and dense with increasing stellar mass, while the acquired DMDGs follow trends more similar to those of DMNGs and DMRGs. 

{The observed decrease in size with increasing stellar mass is unusual in the context of the galaxy stellar mass-size relation \citep[e.g.,][]{vanderWel2014,vanDokkum2015b}, where galaxy size is generally expected to increase with stellar mass.}
{\citet{Cappellari2013} shows a decrease in the size over stellar mass range $2 \times 10^{9} \lesssim M_{\star} \lesssim 3 \times 10^{10}\ \rm M_{\odot}$.}
{This decrease is linked to the in-situ star formation and reflects the evolutionary sequence in which star-forming spiral galaxies evolve into fast rotators through gas accretion and bulge growth, followed by quenching \citep{Cappellari2016}.}
{The upper left panel of Figure~\ref{fig:gasprops} shows the evolution of the half--gas mass radii of the inborn DMDGs during their formation.}
{The gas components are confined within a few pkpc.}
{This compact gas distribution suggests that gas accretion drives the growth of the dense central regions of the inborn DMDGs.}
{However, the spirals-to-fast-rotators sequence has mainly been discussed for galaxies more massive than dwarf galaxies.}
{Galaxies in that stellar mass range are typically embedded in massive dark matter halos, which can promote the gas accretion.}
{In contrast, the inborn DMDGs do not reside in the massive dark matter halos during their formation and evolution.}
{We discuss the possible origins of the inborn DMDGs in Section~\ref{subsec:formchan}, but their formation is unlikely to be associated with direct gas accretion driven by the massive dark matter halo.}
{Therefore, the distinctive structural and kinematic properties of the inborn DMDGs may originate from a mechanism different from that proposed for the spirals-to-fast-rotators sequence.}

We also note that more--massive acquired DMDGs overlap with the dark matter--dominated populations {in the stellar mass--kinematics relation}. 
These galaxies are typically identified near the $M_{\rm DM} \simeq M_{\star}$ locus and still contain $\gtrsim10^{3}$ dark matter particles, which can influence their dynamics and sizes, thereby reducing the differences between DMDGs and DMNGs at high $M_{\star}$.

{Two inborn DMDGs show unusually large half--stellar mass radii and low central densities compared with the rest of the inborn population.}
{We investigate their spatial distribution as a function of redshift in Appendix~\ref{sec:app_c}.}
{These galaxies undergo mergers after very recent encounters with neighboring galaxies at $z<0.03$.}
{This ongoing merger process likely makes these galaxies larger than the other inborn DMDGs.}

Considering the different dark matter fractions between normal galaxies and DMDGs, it is clear that their internal properties differ. 
{Although the baryonic processes responsible for building the stellar component may be similar, the presence or absence of dark matter halos can substantially influence their kinematic properties.} 
For example, both observations and simulations have confirmed that the rotation curves in the outer regions of galaxies vary with the dark matter fraction \citep{Yoon2021, Jeong2025}. 
In contrast to DMDG, {there exist dark matter–dominated galaxies;} \citet{GLee2024} shows the influence of dark matter on internal properties by comparing dark galaxies lacking stellar components with normal galaxies \citep[see also][]{Kwon2025}. 
Numerous studies have also investigated how dark matter affects the baryonic components of galaxies \citep[e.g.,][]{Ubler2021, Farahi2022, Puglisi2023, Shankar2025}. 
These results suggest that the differences between DMDGs and normal galaxies are primarily driven by the presence or absence of dark matter halos. 
In particular, the inborn and acquired DMDGs are further separated from each other, consistent with distinct formation channels that imprint different evolutionary paths and present-day properties.

\begin{figure}[!t]
\centering
\includegraphics[width=0.47\textwidth]{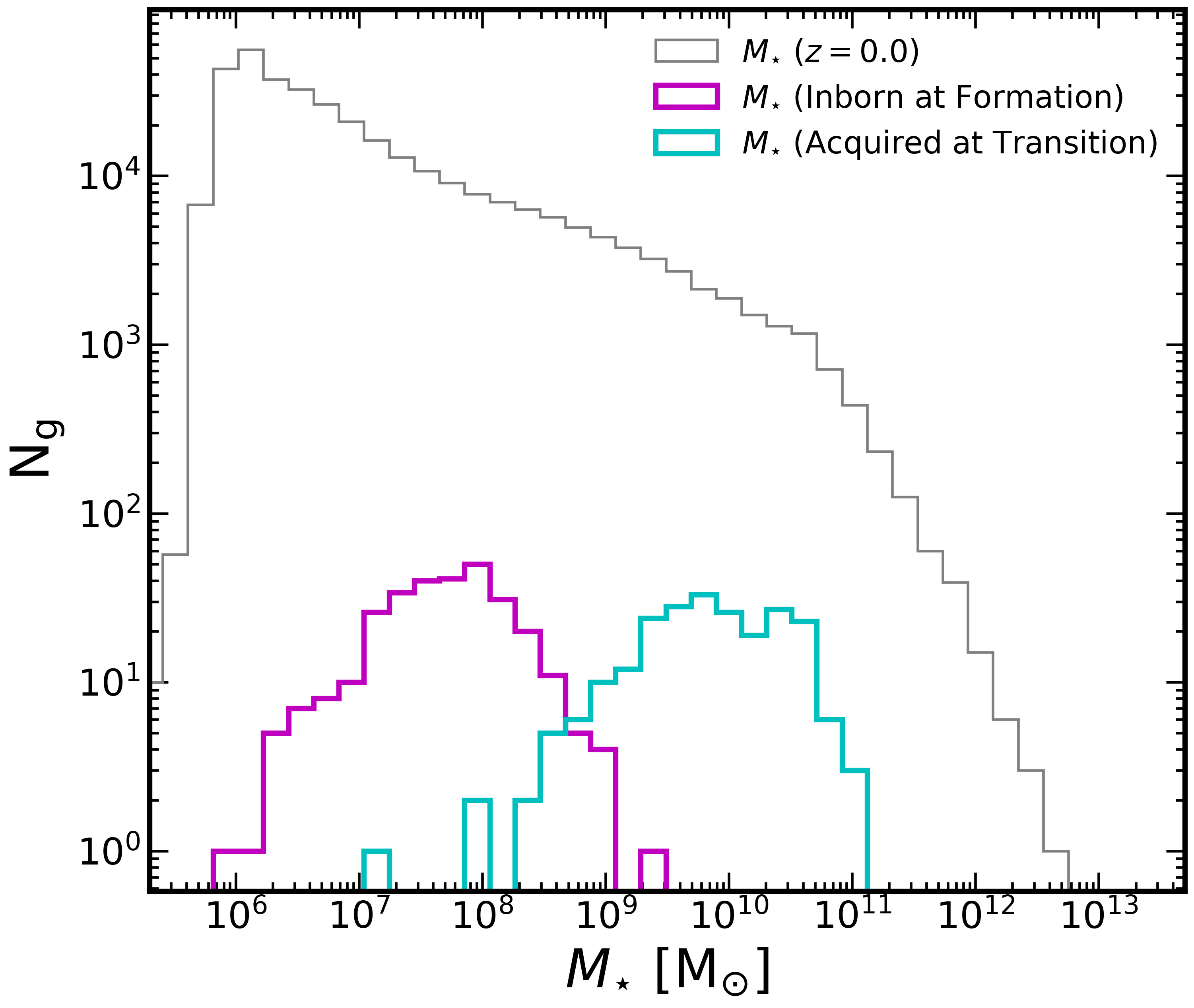}
\caption{
Stellar mass function of DMDGs at their transition/formation epochs. 
The gray histogram shows the galaxy stellar mass function at $z=0$.
}
\label{fig:massfunction}
\end{figure}

\subsection{Caveats} \label{subsec:caveat}
In this study, we investigate the internal properties and the environments of DMDGs at their transition/formation epochs. 
Figure~\ref{fig:massfunction} further shows that their {stellar} mass functions at the transition/formation epochs are distinct. 
This difference is important when selecting the nearest neighbors because the selection is based on the stellar masses at those epochs.

\begin{figure}[!t]
\centering
\includegraphics[width=0.47\textwidth]{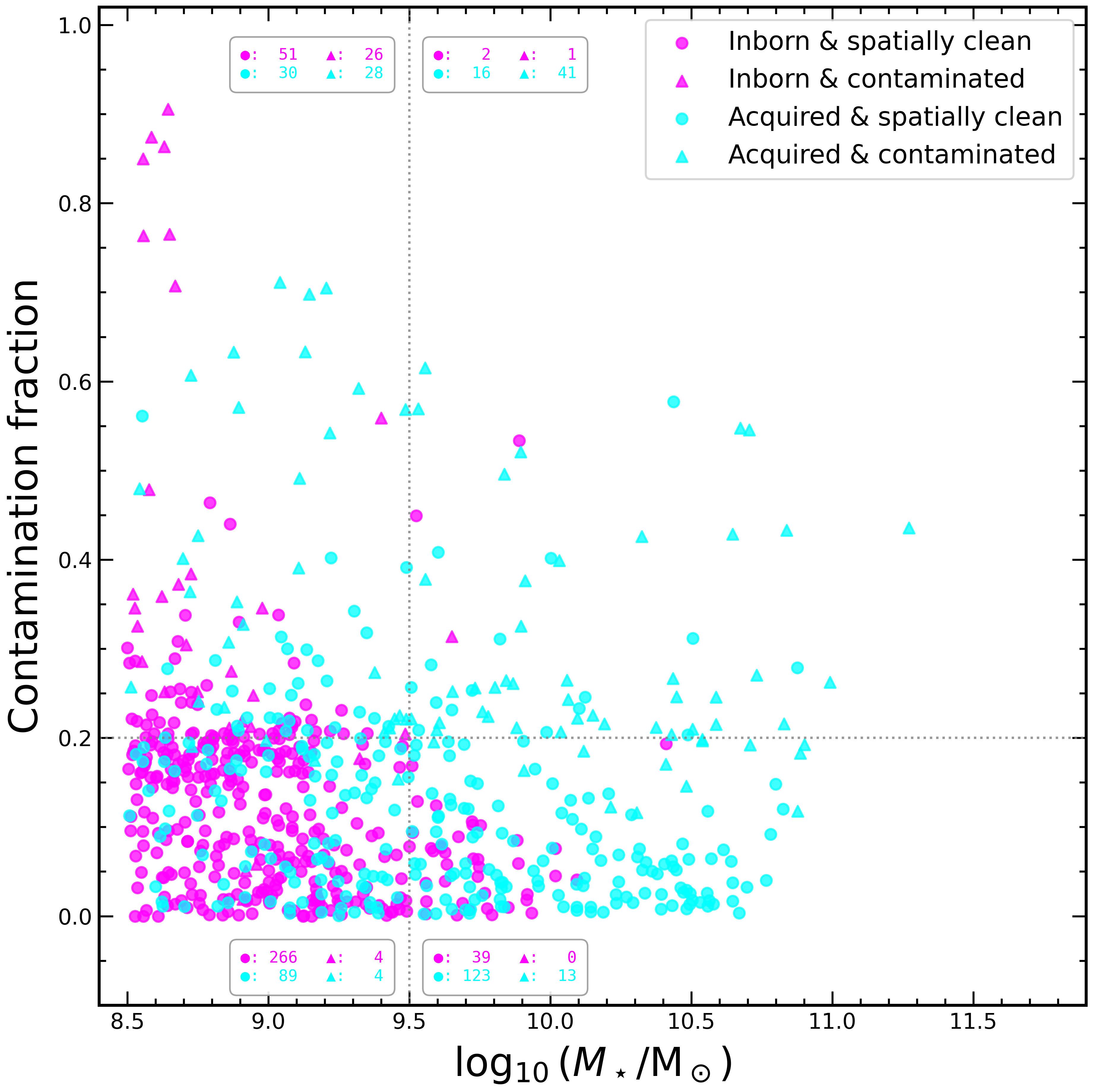}
\caption{
{Distribution of DMDGs in the stellar mass--contamination fraction plane.
Circles and triangles denote galaxies classified through visual inspection as spatially clean and contaminated, respectively. 
Magenta and cyan colors represent inborn and acquired channels, respectively. 
Vertical and horizontal dotted lines mark the stellar mass division at $\log_{10}(M_\star/{\rm M_\odot})=9.5$ and the contamination fraction threshold at $f_{\rm cont}=0.2$, respectively.
The small box in each quadrant gives the number of galaxies for each combination of visual classification and formation channel.}
}
\label{fig:contemmassdep}
\end{figure}

{{In Section~\ref{sec:data}, we reduced contamination arising from limitations of the galaxy identification algorithm by applying the $f_{\rm cont}$ criterion (Figure~\ref{fig:contamination}). 
Here, we examine how $f_{\rm cont}$ varies with stellar mass and formation channel.
Figure~\ref{fig:contemmassdep} shows $f_{\rm cont}$ as a function of stellar mass, with colors distinguishing inborn and acquired candidates.
Table~\ref{tab:contamination_fraction} lists the number of galaxies with $f_{\rm cont} >0.2$ in each stellar mass bin and formation channel.
For the full candidate sample, the fractions with $f_{\rm cont}>0.2$ are similar in the less-- and more--massive bins: 27\% and 26\%, respectively.
This overall similarity masks clear differences between the formation channels.
In both mass bins, acquired candidates exceed the threshold more often than inborn candidates.
Within each formation channel, less--massive candidates exceed the threshold more often than more--massive candidates. 
The $f_{\rm cont} \leq 0.2$ criterion removes different fractions of candidates from the stellar mass and formation channel subsamples.
Nevertheless, the fraction of visually contaminated candidates among those with $f_{\rm cont}\leq0.2$ remains low in all subsamples.
Therefore, we retain $f_{\rm cont}\leq0.2$ as a uniform criterion for selecting the low contamination sample.}}
\begin{deluxetable}{lccc}[!h]
\tabletypesize{\scriptsize}
\tablewidth{1\textwidth}
\renewcommand{\arraystretch}{1.}
\tablecaption{{Number and fraction of DMDGs with $f_{\rm cont} > 0.2$, split by stellar mass and formation channel.}
\label{tab:contamination_fraction}}
\tablehead{
\colhead{} & \colhead{All} & \colhead{Inborn} & \colhead{Acquired}
}
\startdata
Less-massive  & 135/498 (27\%) & 77/347 (22\%) & 58/151 (38\%) \\
More-massive  & {60}/235 (26\%)  & 3/42 (7\%)    & {57}/193 (30\%) \\
\hline
All mass bins & {195}/733 (27\%) & 80/389 (21\%) & {115}/344 ({33}\%)
\enddata
\end{deluxetable}

\begin{figure}[!t]
\centering
\includegraphics[width=0.47\textwidth]{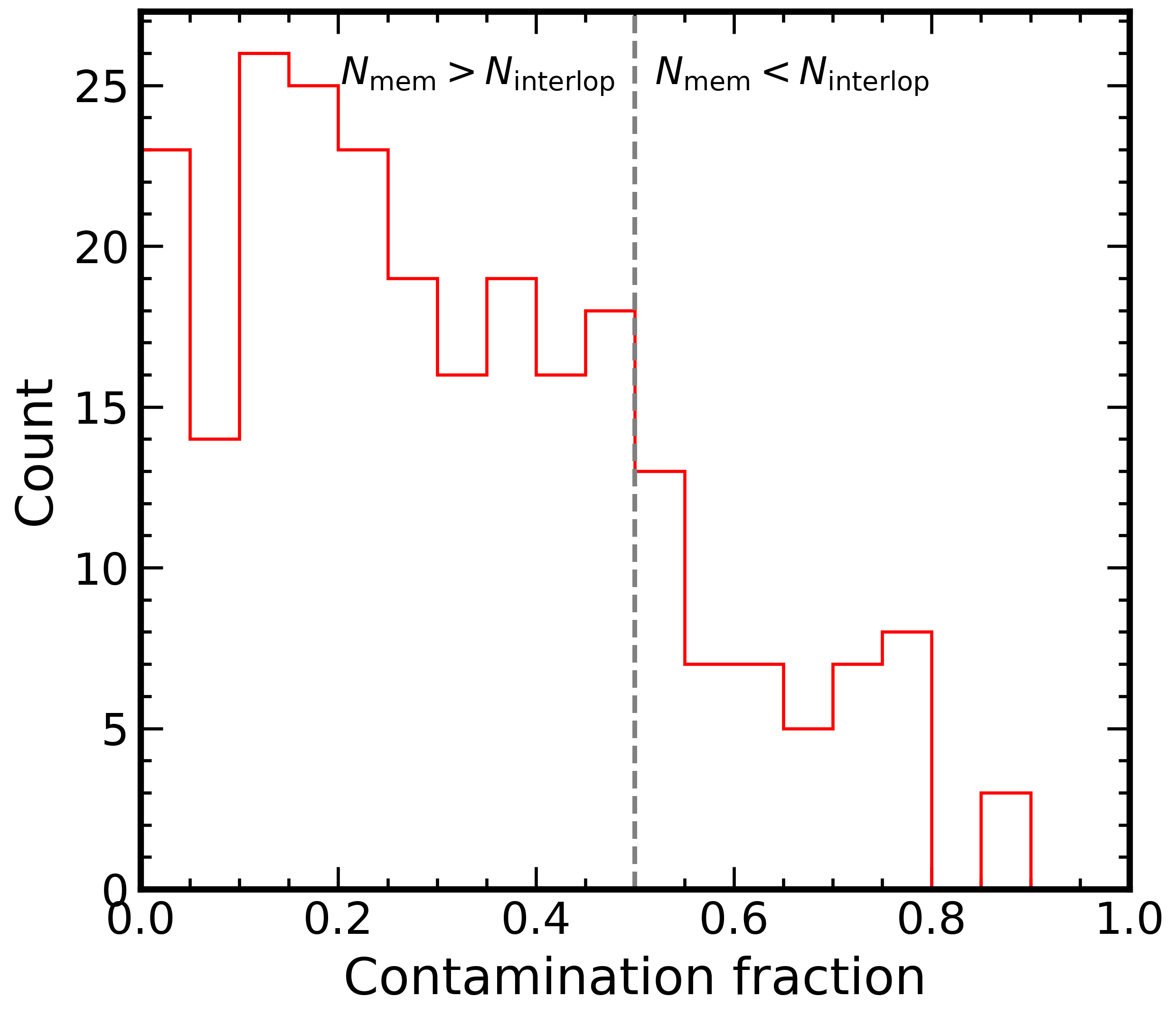}
\caption{
Histogram of the contamination fraction for the inborn DMDGs at their formation epochs, similar to Figure~\ref{fig:contamination}. The gray line indicates $N_{\rm{member}}=N_{\rm{interloper}}$.
}
\label{fig:inborncontam}
\end{figure}

The inborn DMDGs form without a massive dark matter halo from the beginning. 
Figure~\ref{fig:env} shows that they are relatively isolated at birth. 
However, within the standard galaxy formation framework of the $\Lambda$CDM paradigm, it is difficult to understand the formation of clumpy gaseous or stellar structures without an underlying gravitational potential well. 
{{To examine this issue, we inspect the spatial distribution of stars, gas and dark matter before and after the formation of the inborn DMDGs, and also estimate the contamination fraction at their formation epochs. 
Figure~\ref{fig:inborncontam} shows the contamination fraction of the inborn DMDGs at formation\footnote{{At formation epochs, we set $r_{\rm{cutoff}}=r_{\rm{last}}$ and $v_{\rm{cutoff}}=v_{\rm{max}}$ because the boundary procedure in Section~\ref{sec:data} is not robust for systems with few member stellar particles.}}.}}

{{The $f_{\rm cont}=0.2$ criterion in Figure~\ref{fig:contamination} was adopted for the DMDG candidates at $z=0$.
We do not apply this criterion at the formation epochs of the inborn DMDGs.
Most inborn DMDGs contain fewer than 300 member stellar particles at formation, and 50 inborn DMDGs contain fewer than ten stellar particles. 
Thus, a few interloping particles can lead to $f_{\rm cont}>0.2$.
Figure~\ref{fig:inborncontam} is used to assess possible contamination at the formation epochs, rather than to accept or reject the inborn DMDGs.}}

{{Many inborn DMDGs have low contamination fractions, whereas some have high contamination fractions.
The latter systems require caution when interpreting the formation of inborn DMDGs because they may be affected by limitations of the galaxy identification scheme or numerical effects.}}

\section{Conclusions} \label{sec:conclusion}

In this study, we investigate the formation and evolution of dark matter--deficient galaxies using the IllustrisTNG100 simulation. 
We define DMDGs as systems with low dark matter content relative to their stellar mass. 
We refine our sample through visual inspection and {by quantifying} the contamination fraction of nonmember stellar particles.
We analyze {the evolution} of DMDGs and the environments in which DMDGs form. 
We also compare our sample to dark matter--dominated galaxies briefly.

The key findings of our work are summarized as follows:

\begin{enumerate}
    \item DMDGs form through two different channels. 
    One kind refers to galaxies that initially formed inside a dark matter halo but lost their dark matter content over time. 
    The other kind comprises the galaxies that form with intrinsically little or no dark matter. 
    We classify the former as acquired DMDGs and the latter as inborn DMDGs. 
    \item Regardless of the formation channel, most DMDGs appear at $z<1$. 
    Some less--massive systems appear at $0.5<z<1$, whereas the more massive DMDGs predominantly appear at $z<0.5$. 
    \item The evolution of the internal properties of DMDGs depends on both stellar mass and formation channel. 
    The acquired DMDGs exhibit pronounced changes across the transition epoch. 
    {After formation, the inborn DMDGs rapidly build their stellar mass during the first $\sim1$ Gyr. 
    Their physical properties change only mildly thereafter.}
    \item At their formation epochs, the acquired and the inborn DMDGs occupy distinct environments on both small and large scales. 
    The acquired DMDGs tend to lie close enough to their nearest neighbor to be affected by tidal interaction and reside in denser large scale regions. 
    The inborn DMDGs are relatively isolated from their nearest neighbor and inhabit underdense region.
    \item {The acquired DMDGs are consistent with a tidal stripping scenario.
    The inborn DMDGs form from gas clumps that assemble without pre-existing massive dark matter halos of their own.
    In contrast to the stripping-driven evolution of the acquired population, the assembly of these clumps is not limited to a single mechanism.}
    \item At the present epoch, DMDGs exhibit distinctive structural and kinematic properties relative to dark matter--dominated galaxies. 
    DMDGs are typically smaller, have higher central stellar densities, and are more rotation--supported. 
    The inborn DMDGs are a more extreme subset, showing the strongest compactness and rotational support.
\end{enumerate}

Our analysis favors the multi--channel origin for DMDGs rather than a single dominant mechanism. 
Within the stellar mass range considered here, DMDGs from different formation channels exhibit distinct structural and kinematic properties, both from one another and from normal galaxies. 
{However, it should be noted that we do not include DF--like, much less--massive systems in our analysis. 
The inclusion of such systems could be done in future studies when the mass resolution of cosmological simulations is improved.} 
Nevertheless, our findings provide qualitative diagnostics and trends that can inform future observational studies.

\begin{acknowledgments}
We thank the referee for constructive comments
that helped us to improve the manuscript.
We also thank the IllustrisTNG collaboration for making their simulation data publicly available and providing a JupyterLab workspace. 
We thank Juhan Kim for priceless comments on the manuscript. 
HSH acknowledges support from the National Research Foundation of Korea (NRF) funded by the Korea government (MSIT; RS-2026-25482692) and the Global-LAMP Program funded by the Ministry of Education (RS-2023-00301976). 
SH acknowledges a support from the National Research Foundation of Korea (NRF) grant funded by the Korea government (Ministry of Science and ICT: MSIT) (No. RS-2026-25469119).
\end{acknowledgments}

\appendix
\section{EXAMINATION OF THE PROGENITOR OF THE SUSPICIOUS INBORN DMDGS} \label{sec:app_a}
\renewcommand\thefigure{\thesection.\arabic{figure}} 
\setcounter{figure}{0}
In this appendix, we examine the merger trees of the inborn DMDGs. 
To assess the reliability of their formation, we inspect the star formation histories of these systems. 
{Among the 295 inborn DMDGs, we identify 15 galaxies for which the detection time of the earliest progenitor in the merger tree is inconsistent with the onset of star formation inferred from their star formation histories.} 
The top panels in Figure~\ref{fig:app_a} show the age of the stellar particles in the suspicious inborn DMDGs. 
Their star formation histories indicate that the member stars were formed much earlier than {the time when the galaxy first appears in the merger tree.} 

\begin{figure*}[!t]
\centering
\includegraphics[width=0.98\textwidth]{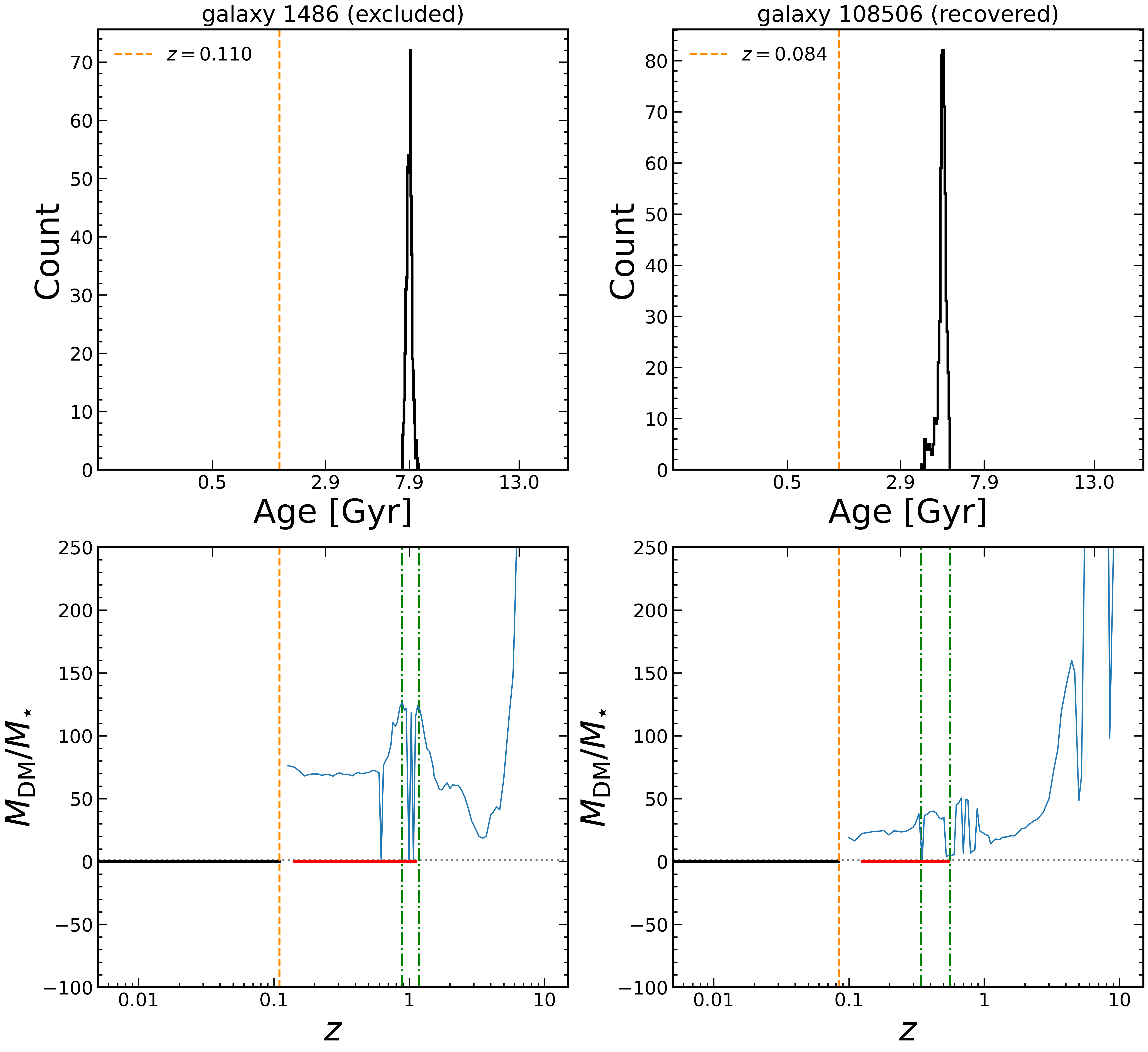}
\caption{
Examples of suspicious DMDGs.
\textit{Top}: Histograms of stellar ages of the member particles.
\textit{Bottom}: Dark matter--to--stellar mass ratio.
The upper $x$--axes show the stellar ages corresponding to the \textit{Top} panels.
The black curves show the mass ratios of the progenitors identified in the merger tree.
The orange dashed lines mark the epochs when the first progenitors are detected in the merger tree.
The green dot-dashed lines on the right and left mark the formation times of the first and last stars, respectively.
The red curves show the progenitors that contain the majority of the member particles. 
The blue curves indicate systems that temporarily contain the member particles during the disconnected phase between the red and black curves.
The gray dotted lines represent $M_{\rm DM}/M_{\star}=0$.
\textit{Left}: Example of a galaxy for which the true progenitor does not satisfy the cosmological origin criterion (\texttt{SubhaloFlag}).
\textit{Right}: The galaxy whose true progenitor is successfully recovered.
}
\label{fig:app_a}
\end{figure*}

To identify the true progenitors, we trace the member particles back to the epoch when the first particles formed. 
The bottom panels in Figure~\ref{fig:app_a} show the dark matter--to--stellar mass ratio of these galaxies. 
In most cases, the merger tree loses the true progenitor. 
This is because these galaxies form close to massive systems, and \texttt{SUBFIND} identifies them as the fragments produced by baryonic processes. 
As a result, most of these progenitors do not satisfy the cosmological origin criterion represented by \texttt{SubhaloFlag}, with only one exception. 
Therefore, we exclude 14 suspicious galaxies and include one galaxy in our sample.

\section{EXAMINATION OF THE NUMERICAL EFFECTS RELATED TO PERIODIC BOUNDARY CONDITIONS} \label{sec:app_b}
\renewcommand\thefigure{\thesection.\arabic{figure}} 
\setcounter{figure}{0}
\renewcommand\thetable{\thesection.\arabic{table}} 
\setcounter{table}{0}
{{In this appendix,  we examine whether our DMDG sample is affected by numerical artifacts associated with PBC, following the sanity checks recommended by \cite{Mitrasinovic2025}.}}

{{We first identify DMDGs located close to the simulation box boundaries. 
We define the boundary buffer as the region within a fixed distance of any box boundary. 
Using the most bound particle to define each galaxy position, we find that the  $2\ {\rm cMpc}\cdot h^{-1}$ buffer occupies 15.2\% of the simulation volume and contains 123 of the 524 DMDGs (23.5\%) in the original cleaned sample at $z=0$.
These comprise 76 inborn and 47 acquired DMDGs. 
Table~\ref{tab:app_tab_1} summarizes the number and fraction of DMDGs within buffers of different sizes. 
We also determine whether each DMDG experiences periodic boundary crossings along its main progenitor branch. 
Table~\ref{tab:app_tab_2} reports the number of boundary crossing systems using two position definitions: the most bound particle position (Pos) and the center of mass position (CM).
Table~\ref{tab:app_tab_3} further divides these systems according to whether they are located within the boundary buffer at $z=0$.}}
\begin{deluxetable}{lcccc}[!h]
\tabletypesize{\scriptsize}
\tablewidth{0pt}
\renewcommand{\arraystretch}{1.}
\tablecaption{{{Number and fraction of DMDGs located within the boundary buffer at $z=0$ for different buffer sizes.
The total fraction is computed with respect to all 524 DMDGs in the original cleaned sample.
The percentage in parentheses after each buffer size denotes the fraction of the total simulation volume occupied by the corresponding boundary buffer.
The fractions for inborn and acquired DMDGs are computed with respect to the corresponding formation channel subsamples.}}
\label{tab:app_tab_1}}
\tablehead{
\colhead{Buffer Size} & \colhead{Category} & \colhead{$N$} & \colhead{/\, Total} & \colhead{/\, Channel} \\
\colhead{$[{\rm cMpc}\cdot h^{-1}]$} & \colhead{} & \colhead{}& \colhead{} & \colhead{} 
}
\startdata
{           } & Total    & 16  & 3.1\%  & ---    \\
{0.5 (3.9\%)} & Inborn   & 9   & 1.7\%  & 3.1\%  \\
{           } & Acquired & 7   & 1.3\%  & 3.1\%  \\
\hline
{            } & Total    & 36  & 6.9\%  & ---    \\
{0.75 (5.9\%)} & Inborn   & 17  & 3.2\%  & 5.8\%  \\
{            } & Acquired & 19  & 3.6\%  & 8.3\%  \\
\hline
{           } & Total    & 72  & 13.7\% & ---    \\
{1.0 (7.8\%)} & Inborn   & 41  & 7.8\%  & 13.9\% \\
{           } & Acquired & 31  & 5.9\%  & 13.5\% \\
\hline
{            } & Total    & 104 & 19.8\% & ---    \\
{1.5 (11.5\%)} & Inborn   & 62  & 11.8\% & 21.0\% \\
{            } & Acquired & 42  & 8.0\%  & 18.3\% \\
\hline
{            } & Total    & 123 & 23.5\% & ---    \\
{2.0 (15.2\%)} & Inborn   & 76  & 14.5\% & 25.8\% \\
{            } & Acquired & 47  & 9.0\%  & 20.5\% \\
\hline
{            } & Total    & 136 & 26.0\% & ---    \\
{3.0 (22.1\%)} & Inborn   & 83  & 15.8\% & 28.1\% \\
{            } & Acquired & 53  & 10.1\% & 23.1\%
\enddata
\end{deluxetable}

\begin{deluxetable}{lcccc}[!h]
\tabletypesize{\scriptsize}
\tablewidth{1\textwidth}
\renewcommand{\arraystretch}{1.}
\tablecaption{{{Number and fraction of DMDGs that experience periodic boundary crossings along their progenitor histories.
We identify boundary crossings using the most bound particle position (Pos) and the center of mass position (CM).}} 
\label{tab:app_tab_2}}
\tablehead{
\colhead{Coordinate} & \colhead{Category} & \colhead{$N$} & \colhead{/\, Total} & \colhead{/\, Channel} 
}
\startdata
    & Total    & 44 & 8.4\% & ---    \\
Pos & Inborn   & 10 & 1.9\% & 3.4\%  \\
    & Acquired & 34 & 6.5\% & 14.8\% \\
\hline
   & Total    & 46 & 8.8\% & ---    \\
CM & Inborn   & 10 & 1.9\% & 3.4\%  \\
   & Acquired & 36 & 6.9\% & 15.7\%
\enddata
\end{deluxetable}

\begin{deluxetable}{llcc}[!h]
\tabletypesize{\scriptsize}
\tablewidth{1\textwidth}
\renewcommand{\arraystretch}{1.}
\tablecaption{{{Classification of DMDGs by periodic boundary crossing and $z=0$ buffer zone membership.
The buffer zone is defined as the region within $2\ {\rm cMpc}\cdot h^{-1}$ of a simulation box boundary.
``Crossing'' indicates at least one periodic boundary crossing along the merger tree history.
``Buffer Zone'' indicates that the DMDG is located within the buffer zone at $z=0$. 
``Safe Zone'' indicates that the DMDG is located outside the boundary buffer at $z=0$.}}
\label{tab:app_tab_3}}
\tablehead{
\colhead{Coordinate} & \colhead{Category} & \colhead{$N$} & \colhead{/\, Total} 
}
\startdata
    & Crossing $\cap$ Buffer Zone & 38  & 7.3\%  \\
Pos & Crossing $\cap$ Safe Zone   &  6  & 1.1\%  \\
    & No crossing, Buffer Zone    & 85  & 16.2\% \\
    & No crossing, Safe Zone      & 395 & 75.4\% \\
\hline
   & Crossing $\cap$ Buffer Zone & 39  & 7.4\%  \\
CM & Crossing $\cap$ Safe Zone   &  7  & 1.3\%  \\
   & No crossing, Buffer Zone    & 84  & 16.0\% \\
   & No crossing, Safe Zone      & 394 & 75.2\% 
\enddata
\end{deluxetable}

{{Present day proximity to the simulation box boundary does not by itself establish that a galaxy is a numerical artifact. 
In TNG100, DMDGs near the box boundaries may include both PBC related artifacts and genuine physical systems \citep{Mitrasinovic2025}. 
We do not remove DMDGs solely because they lie within the boundary buffer at $z=0$. 
Instead, we examine their position and mass histories for temporal associations between boundary crossings and their transition to or formation as DMDGs.}}

{{For galaxies with at least one boundary crossing, we define a PBC related artifact candidate using three criteria:
(i) the galaxy crosses a boundary according to either Pos or CM, (ii) the crossing occurs within 1 Gyr of its DMDG transition or formation epoch, and (iii) an abrupt mass change of more than 1 dex occurs within 1 Gyr of the crossing.
These criteria identify systems in which the DMDG classification is temporally associated with the boundary crossing and the unphysical mass loss.}}
\begin{figure*}[!t]
\centering
\includegraphics[width=\textwidth]{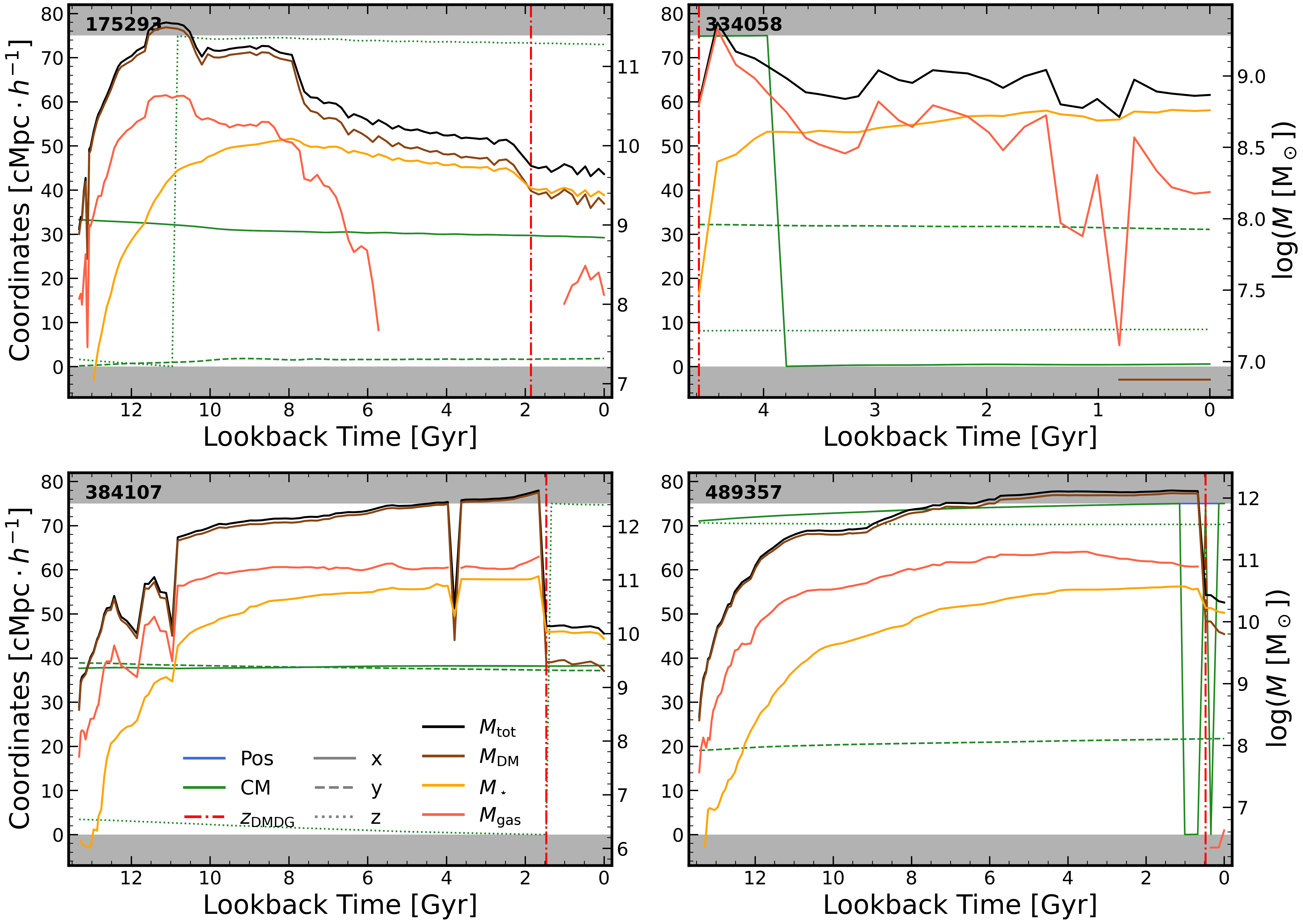}
\caption{
{{Examples of the orbital and mass histories of DMDGs that experience boundary related events.
The left axis in each panel shows the coordinate, and the right axis shows the mass.
We show both the position of the most bound particle (Pos) and the center of mass coordinate (CM).
For each coordinate definition, $x$, $y$, and $z$ components are shown separately.
The black, brown, yellow, and orange curves show the total, dark matter, stellar, and gas masses, respectively.
The vertical dot--dashed line marks the DMDG transition or formation epoch.
The gray shaded regions represent coordinate values outside the simulation box.
The top row shows galaxies that cross a periodic boundary but do not show an associated unphysical mass loss event.
The bottom row shows two acquired DMDGs that satisfy our PBC artifact criteria.
These two systems exhibit abrupt mass loss near the boundary related event and become DMDGs at nearly the same epoch.
We exclude these two systems from the fiducial sample.}}
}
\label{fig:app_b_1}
\end{figure*}

{{Figure~\ref{fig:app_b_1} shows examples of the position and mass histories of DMDGs with boundary crossing events. 
The top row shows galaxies that cross the periodic boundary without an associated unphysical mass loss. 
The bottom row shows two acquired DMDGs that satisfy all three criteria. 
Similar to the example shown in Figure~1 of \cite{Mitrasinovic2025}, both galaxies exhibit abrupt mass loss near the boundary crossing and become DMDGs at nearly the same epochs. 
This temporal coincidence indicates that their identification as DMDGs likely results from PBC related numerical artifacts rather than physical tidal processes.}}

\begin{figure*}[!t]
\centering
\includegraphics[width=0.98\textwidth]{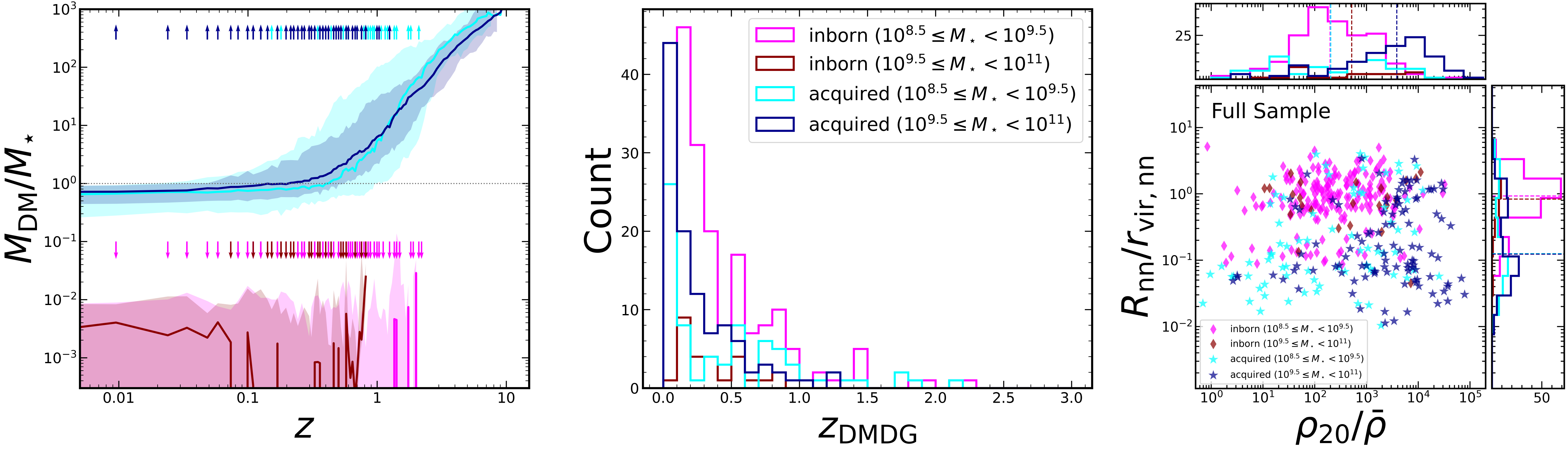}
\caption{
{{Mass ratio evolution, DMDG transition or formation redshift distribution, and environment of DMDGs in the safe zone.
The safe zone is defined as the region outside the $2\ {\rm cMpc}\cdot h^{-1}$ boundary buffer at $z=0$.
The format is similar to Figures~\ref{fig:ratio_evol_zdmdg} and~\ref{fig:env}.}}
}
\label{fig:app_b_2}
\end{figure*}

{{We exclude these two systems from the fiducial sample, leaving 522 DMDGs: 227 acquired and 295 inborn systems. 
As a conservative robustness test, we also repeat the main analyses after excluding all DMDGs within $2\ {\rm cMpc}\cdot h^{-1}$ of the box boundary at $z=0$. 
Figure~\ref{fig:app_b_2} shows the mass ratio evolution, redshift distribution, and environments of DMDGs outside the buffer, which we define as the safe zone. 
The qualitative differences between the inborn and acquired populations remain unchanged, indicating that boundary buffer systems do not drive our main conclusions.}}

\section{SPATIAL DISTRIBUTIONS OF TWO OUTLYING INBORN DMDGS: GALAXIES 193016 AND 231763} \label{sec:app_c}
\renewcommand\thefigure{\thesection.\arabic{figure}} 
\setcounter{figure}{0}
In this appendix, we investigate the spatial distributions of two inborn DMDGs that have larger half--stellar mass radii and lower central densities than those of the other inborn DMDGs.
Figures~\ref{fig:app_c_1} and \ref{fig:app_c_2} show the spatial distributions of these two galaxies and their nearby neighbors as a function of redshift.
The two galaxies encountered nearby galaxies and began to merge at $z<0.03$.
Their stellar masses are smaller than $5.21 \times 10^{8}\ \rm M_{\odot}$, and the stellar mass ratios with their counterparts are larger than 1:4, indicating major mergers.
These results suggest that the encounters occurred very recently, so the systems have not yet had sufficient time to relax.


\begin{sidewaysfigure}
\centering
\rotatebox{270}{\includegraphics[width=0.245\textwidth]{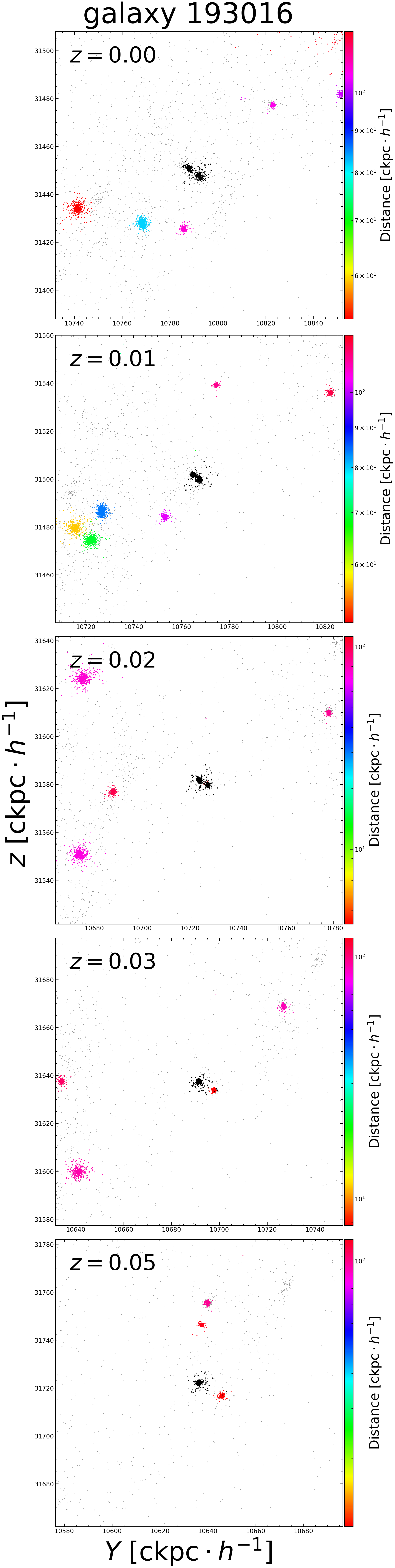}}
\caption{
Spatial distribution of stellar particles in galaxy 193016 and its nearby neighbors as a function of time.
The black points denote the target's members, while nearby neighbors are shown in different colors.
Redder colors indicate neighbors at smaller distances.
The gray points indicate unbound particles.
}
\label{fig:app_c_1}
\end{sidewaysfigure}

\begin{sidewaysfigure}
\centering
\rotatebox{270}{\includegraphics[width=0.245\textwidth]{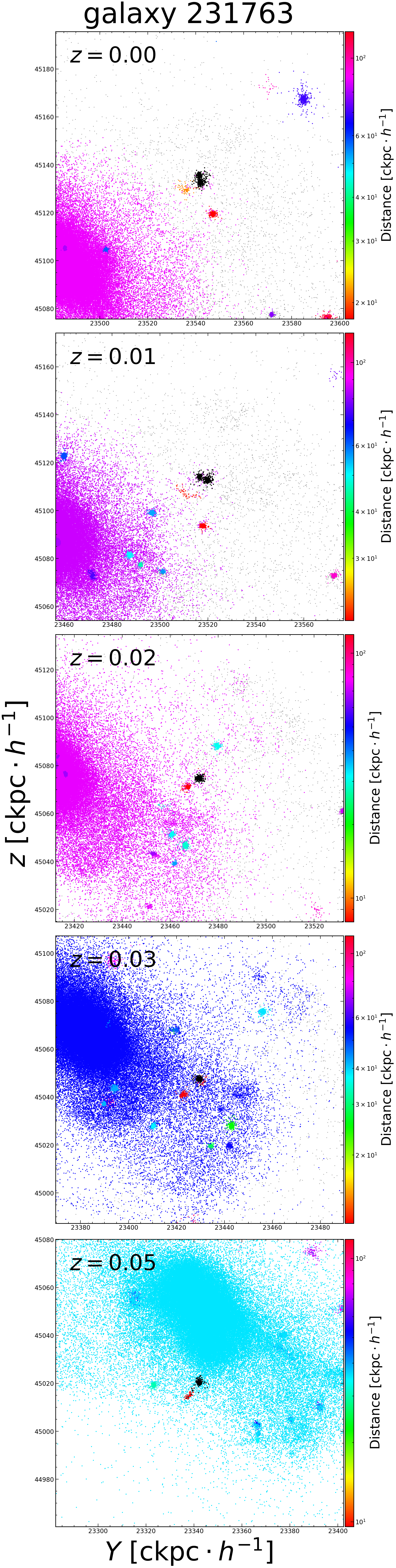}}
\caption{
Similar to Figure~\ref{fig:app_c_1}, but for galaxy 231763.
}
\label{fig:app_c_2}
\end{sidewaysfigure}

\bibliography{prophet}{}

@ARTICLE{Peebles1982,
       author = {{Peebles}, P.~J.~E.},
        title = "{Large-scale background temperature and mass fluctuations due to scale-invariant primeval perturbations}",
      journal = {\apjl},
         year = 1982,
        month = dec,
       volume = {263},
        pages = {L1-L5},
          doi = {10.1086/183911},
       adsurl = {https://ui.adsabs.harvard.edu/abs/1982ApJ...263L...1P}
}

@ARTICLE{WhiteRees1978,
       author = {{White}, S.~D.~M. and {Rees}, M.~J.},
        title = "{Core condensation in heavy halos: a two-stage theory for galaxy formation and clustering.}",
      journal = {\mnras},
         year = 1978,
        month = may,
       volume = {183},
        pages = {341-358},
          doi = {10.1093/mnras/183.3.341},
       adsurl = {https://ui.adsabs.harvard.edu/abs/1978MNRAS.183..341W}
}

@ARTICLE{Zwicky1956,
       author = {{Zwicky}, F.},
        title = "{Multiple Galaxies}",
      journal = {Ergebnisse der exakten Naturwissenschaften},
         year = 1956,
        month = jan,
       volume = {29},
        pages = {344-385},
       adsurl = {https://ui.adsabs.harvard.edu/abs/1956ErNW...29..344Z}
}

@ARTICLE{Mirabel1992,
       author = {{Mirabel}, I.~F. and {Dottori}, H. and {Lutz}, D.},
        title = "{Genesis of a dwarf galaxy from the debris of the Antennae.}",
      journal = {\aap},
         year = 1992,
        month = mar,
       volume = {256},
        pages = {L19-L22},
       adsurl = {https://ui.adsabs.harvard.edu/abs/1992A&A...256L..19M}
}

@ARTICLE{Duc2004,
       author = {{Duc}, P. -A. and {Bournaud}, F. and {Masset}, F.},
        title = "{A top-down scenario for the formation of massive Tidal Dwarf Galaxies}",
      journal = {\aap},
         year = 2004,
        month = dec,
       volume = {427},
        pages = {803-814},
          doi = {10.1051/0004-6361:20041410},
archivePrefix = {arXiv},
       eprint = {astro-ph/0408524},
 primaryClass = {astro-ph},
       adsurl = {https://ui.adsabs.harvard.edu/abs/2004A&A...427..803D}
}

@ARTICLE{Barnes1992,
       author = {{Barnes}, Joshua E. and {Hernquist}, Lars},
        title = "{Formation of dwarf galaxies in tidal tails}",
      journal = {\nat},
         year = 1992,
        month = dec,
       volume = {360},
       number = {6406},
        pages = {715-717},
          doi = {10.1038/360715a0},
       adsurl = {https://ui.adsabs.harvard.edu/abs/1992Natur.360..715B}
}

@ARTICLE{Ploeckinger2018,
       author = {{Ploeckinger}, Sylvia and {Sharma}, Kuldeep and {Schaye}, Joop and {Crain}, Robert A. and {Schaller}, Matthieu and {Barber}, Christopher},
        title = "{Tidal dwarf galaxies in cosmological simulations}",
      journal = {\mnras},
         year = 2018,
        month = feb,
       volume = {474},
       number = {1},
        pages = {580-596},
          doi = {10.1093/mnras/stx2787},
archivePrefix = {arXiv},
       eprint = {1710.09380},
 primaryClass = {astro-ph.GA},
       adsurl = {https://ui.adsabs.harvard.edu/abs/2018MNRAS.474..580P}
}

@ARTICLE{Braine2000,
       author = {{Braine}, Jonathan and {Lisenfeld}, Ute and {Due}, Pierre-Alain and {Leon}, St{\'e}phane},
        title = "{Formation of molecular gas in the tidal debris of violent galaxy-galaxy interactions}",
      journal = {\nat},
         year = 2000,
        month = feb,
       volume = {403},
       number = {6772},
        pages = {867-869},
          doi = {10.1038/35002521},
       adsurl = {https://ui.adsabs.harvard.edu/abs/2000Natur.403..867B}
}

@ARTICLE{Toomre1972,
       author = {{Toomre}, Alar and {Toomre}, Juri},
        title = "{Galactic Bridges and Tails}",
      journal = {\apj},
         year = 1972,
        month = dec,
       volume = {178},
        pages = {623-666},
          doi = {10.1086/151823},
       adsurl = {https://ui.adsabs.harvard.edu/abs/1972ApJ...178..623T}
}

@ARTICLE{vanDokkum2018,
       author = {{van Dokkum}, Pieter and {Danieli}, Shany and {Cohen}, Yotam and {Merritt}, Allison and {Romanowsky}, Aaron J. and {Abraham}, Roberto and {Brodie}, Jean and {Conroy}, Charlie and {Lokhorst}, Deborah and {Mowla}, Lamiya and {O'Sullivan}, Ewan and {Zhang}, Jielai},
        title = "{A galaxy lacking dark matter}",
      journal = {\nat},
         year = 2018,
        month = mar,
       volume = {555},
       number = {7698},
        pages = {629-632},
          doi = {10.1038/nature25767},
archivePrefix = {arXiv},
       eprint = {1803.10237},
 primaryClass = {astro-ph.GA},
       adsurl = {https://ui.adsabs.harvard.edu/abs/2018Natur.555..629V}
}

@ARTICLE{vanDokkum2019,
       author = {{van Dokkum}, Pieter and {Danieli}, Shany and {Abraham}, Roberto and {Conroy}, Charlie and {Romanowsky}, Aaron J.},
        title = "{A Second Galaxy Missing Dark Matter in the NGC 1052 Group}",
      journal = {\apjl},
         year = 2019,
        month = mar,
       volume = {874},
       number = {1},
          eid = {L5},
        pages = {L5},
          doi = {10.3847/2041-8213/ab0d92},
archivePrefix = {arXiv},
       eprint = {1901.05973},
 primaryClass = {astro-ph.GA},
       adsurl = {https://ui.adsabs.harvard.edu/abs/2019ApJ...874L...5V}
}

@ARTICLE{vanDokkum2022,
       author = {{van Dokkum}, Pieter and {Shen}, Zili and {Keim}, Michael A. and {Trujillo-Gomez}, Sebastian and {Danieli}, Shany and {Dutta Chowdhury}, Dhruba and {Abraham}, Roberto and {Conroy}, Charlie and {Kruijssen}, J.~M. Diederik and {Nagai}, Daisuke and {Romanowsky}, Aaron},
        title = "{A trail of dark-matter-free galaxies from a bullet-dwarf collision}",
      journal = {\nat},
         year = 2022,
        month = may,
       volume = {605},
       number = {7910},
        pages = {435-439},
          doi = {10.1038/s41586-022-04665-6},
archivePrefix = {arXiv},
       eprint = {2205.08552},
 primaryClass = {astro-ph.GA},
       adsurl = {https://ui.adsabs.harvard.edu/abs/2022Natur.605..435V}
}

@ARTICLE{Hunter2000,
       author = {{Hunter}, Deidre A. and {Hunsberger}, Sally D. and {Roye}, Erin W.},
        title = "{Identifying Old Tidal Dwarf Irregulars}",
      journal = {\apj},
         year = 2000,
        month = oct,
       volume = {542},
       number = {1},
        pages = {137-142},
          doi = {10.1086/309542},
archivePrefix = {arXiv},
       eprint = {astro-ph/0005257},
 primaryClass = {astro-ph},
       adsurl = {https://ui.adsabs.harvard.edu/abs/2000ApJ...542..137H}
}

@ARTICLE{Duc2007,
       author = {{Duc}, P. -A. and {Braine}, J. and {Lisenfeld}, U. and {Brinks}, E. and {Boquien}, M.},
        title = "{VCC 2062: an old tidal dwarf galaxy in the Virgo cluster?}",
      journal = {\aap},
         year = 2007,
        month = nov,
       volume = {475},
       number = {1},
        pages = {187-197},
          doi = {10.1051/0004-6361:20078335},
archivePrefix = {arXiv},
       eprint = {0709.2733},
 primaryClass = {astro-ph},
       adsurl = {https://ui.adsabs.harvard.edu/abs/2007A&A...475..187D}
}

@ARTICLE{Duc2014,
       author = {{Duc}, Pierre-Alain and {Paudel}, Sanjaya and {McDermid}, Richard M. and {Cuillandre}, Jean-Charles and {Serra}, Paolo and {Bournaud}, Fr{\'e}d{\'e}ric and {Cappellari}, Michele and {Emsellem}, Eric},
        title = "{Identification of old tidal dwarfs near early-type galaxies from deep imaging and H I observations}",
      journal = {\mnras},
         year = 2014,
        month = may,
       volume = {440},
       number = {2},
        pages = {1458-1469},
          doi = {10.1093/mnras/stu330},
archivePrefix = {arXiv},
       eprint = {1403.0626},
 primaryClass = {astro-ph.GA},
       adsurl = {https://ui.adsabs.harvard.edu/abs/2014MNRAS.440.1458D}
}

@ARTICLE{Sweet2014,
       author = {{Sweet}, Sarah M. and {Drinkwater}, Michael J. and {Meurer}, Gerhardt and {Bekki}, Kenji and {Dopita}, Michael A. and {Kilborn}, Virginia and {Nicholls}, David C.},
        title = "{Choirs H I Galaxy Groups: The Metallicity of Dwarf Galaxies}",
      journal = {\apj},
         year = 2014,
        month = feb,
       volume = {782},
       number = {1},
          eid = {35},
        pages = {35},
          doi = {10.1088/0004-637X/782/1/35},
archivePrefix = {arXiv},
       eprint = {1401.0955},
 primaryClass = {astro-ph.GA},
       adsurl = {https://ui.adsabs.harvard.edu/abs/2014ApJ...782...35S}
}

@ARTICLE{Buzzo2022,
       author = {{Buzzo}, Maria Luisa and {Forbes}, Duncan A. and {Brodie}, Jean P. and {Romanowsky}, Aaron J. and {Cluver}, Michelle E. and {Jarrett}, Thomas H. and {Laine}, Seppo and {Couch}, Warrick J. and {Gannon}, Jonah S. and {Ferr{\'e}-Mateu}, Anna and {Okabe}, Nobuhiro},
        title = "{The stellar populations of quiescent ultra-diffuse galaxies from optical to mid-infrared spectral energy distribution fitting}",
      journal = {\mnras},
         year = 2022,
        month = dec,
       volume = {517},
       number = {2},
        pages = {2231-2250},
          doi = {10.1093/mnras/stac2442},
archivePrefix = {arXiv},
       eprint = {2208.11819},
 primaryClass = {astro-ph.GA},
       adsurl = {https://ui.adsabs.harvard.edu/abs/2022MNRAS.517.2231B}
}

@ARTICLE{Fensch2019,
       author = {{Fensch}, J{\'e}r{\'e}my and {van der Burg}, Remco F.~J. and {Je{\v{r}}{\'a}bkov{\'a}}, Tereza and {Emsellem}, Eric and {Zanella}, Anita and {Agnello}, Adriano and {Hilker}, Michael and {M{\"u}ller}, Oliver and {Rejkuba}, Marina and {Duc}, Pierre-Alain and {Durrell}, Patrick and {Habas}, Rebecca and {Lim}, Sungsoon and {Marleau}, Francine R. and {Peng}, Eric W. and {S{\'a}nchez Janssen}, Rub{\'e}n},
        title = "{The ultra-diffuse galaxy NGC 1052-DF2 with MUSE. II. The population of DF2: stars, clusters, and planetary nebulae}",
      journal = {\aap},
         year = 2019,
        month = may,
       volume = {625},
          eid = {A77},
        pages = {A77},
          doi = {10.1051/0004-6361/201834911},
archivePrefix = {arXiv},
       eprint = {1812.07346},
 primaryClass = {astro-ph.GA},
       adsurl = {https://ui.adsabs.harvard.edu/abs/2019A&A...625A..77F}
}

@ARTICLE{vanDokkum2015,
       author = {{van Dokkum}, Pieter G. and {Abraham}, Roberto and {Merritt}, Allison and {Zhang}, Jielai and {Geha}, Marla and {Conroy}, Charlie},
        title = "{Forty-seven Milky Way-sized, Extremely Diffuse Galaxies in the Coma Cluster}",
      journal = {\apjl},
         year = 2015,
        month = jan,
       volume = {798},
       number = {2},
          eid = {L45},
        pages = {L45},
          doi = {10.1088/2041-8205/798/2/L45},
archivePrefix = {arXiv},
       eprint = {1410.8141},
 primaryClass = {astro-ph.GA},
       adsurl = {https://ui.adsabs.harvard.edu/abs/2015ApJ...798L..45V}
}

@ARTICLE{Shen2023,
       author = {{Shen}, Zili and {van Dokkum}, Pieter and {Danieli}, Shany},
        title = "{Confirmation of an Anomalously Low Dark Matter Content for the Galaxy NGC 1052-DF4 from Deep, High-resolution Continuum Spectroscopy}",
      journal = {\apj},
         year = 2023,
        month = nov,
       volume = {957},
       number = {1},
          eid = {6},
        pages = {6},
          doi = {10.3847/1538-4357/acfa70},
archivePrefix = {arXiv},
       eprint = {2309.08592},
 primaryClass = {astro-ph.GA},
       adsurl = {https://ui.adsabs.harvard.edu/abs/2023ApJ...957....6S}
}

@ARTICLE{Behroozi2010,
       author = {{Behroozi}, Peter S. and {Conroy}, Charlie and {Wechsler}, Risa H.},
        title = "{A Comprehensive Analysis of Uncertainties Affecting the Stellar Mass-Halo Mass Relation for 0 < z < 4}",
      journal = {\apj},
         year = 2010,
        month = jul,
       volume = {717},
       number = {1},
        pages = {379-403},
          doi = {10.1088/0004-637X/717/1/379},
archivePrefix = {arXiv},
       eprint = {1001.0015},
 primaryClass = {astro-ph.CO},
       adsurl = {https://ui.adsabs.harvard.edu/abs/2010ApJ...717..379B}
}

@ARTICLE{Moster2013,
       author = {{Moster}, Benjamin P. and {Naab}, Thorsten and {White}, Simon D.~M.},
        title = "{Galactic star formation and accretion histories from matching galaxies to dark matter haloes}",
      journal = {\mnras},
         year = 2013,
        month = feb,
       volume = {428},
       number = {4},
        pages = {3121-3138},
          doi = {10.1093/mnras/sts261},
archivePrefix = {arXiv},
       eprint = {1205.5807},
 primaryClass = {astro-ph.CO},
       adsurl = {https://ui.adsabs.harvard.edu/abs/2013MNRAS.428.3121M}
}

@ARTICLE{Martin2018,
       author = {{Martin}, Nicolas F. and {Collins}, Michelle L.~M. and {Longeard}, Nicolas and {Tollerud}, Erik},
        title = "{Current Velocity Data on Dwarf Galaxy NGC 1052-DF2 do not Constrain it to Lack Dark Matter}",
      journal = {\apjl},
         year = 2018,
        month = may,
       volume = {859},
       number = {1},
          eid = {L5},
        pages = {L5},
          doi = {10.3847/2041-8213/aac216},
archivePrefix = {arXiv},
       eprint = {1804.04136},
 primaryClass = {astro-ph.GA},
       adsurl = {https://ui.adsabs.harvard.edu/abs/2018ApJ...859L...5M}
}

@ARTICLE{Laporte2019,
       author = {{Laporte}, Chervin F.~P. and {Agnello}, Adriano and {Navarro}, Julio F.},
        title = "{Reconciling mass estimates of ultradiffuse galaxies}",
      journal = {\mnras},
         year = 2019,
        month = mar,
       volume = {484},
       number = {1},
        pages = {245-251},
          doi = {10.1093/mnras/sty2891},
archivePrefix = {arXiv},
       eprint = {1804.04139},
 primaryClass = {astro-ph.GA},
       adsurl = {https://ui.adsabs.harvard.edu/abs/2019MNRAS.484..245L}
}

@ARTICLE{Trujillo2019,
       author = {{Trujillo}, Ignacio and {Beasley}, Michael A. and {Borlaff}, Alejandro and {Carrasco}, Eleazar R. and {Di Cintio}, Arianna and {Filho}, Mercedes and {Monelli}, Matteo and {Montes}, Mireia and {Rom{\'a}n}, Javier and {Ruiz-Lara}, Tom{\'a}s and {S{\'a}nchez Almeida}, Jorge and {Valls-Gabaud}, David and {Vazdekis}, Alexandre},
        title = "{A distance of 13 Mpc resolves the claimed anomalies of the galaxy lacking dark matter}",
      journal = {\mnras},
         year = 2019,
        month = jun,
       volume = {486},
       number = {1},
        pages = {1192-1219},
          doi = {10.1093/mnras/stz771},
archivePrefix = {arXiv},
       eprint = {1806.10141},
 primaryClass = {astro-ph.GA},
       adsurl = {https://ui.adsabs.harvard.edu/abs/2019MNRAS.486.1192T}
}

@ARTICLE{Monelli2019ApJ,
       author = {{Monelli}, Matteo and {Trujillo}, Ignacio},
        title = "{The TRGB Distance to the Second Galaxy {\textquotedblleft}Missing Dark Matter{\textquotedblright}: Evidence for Two Groups of Galaxies at 13.5 and 19 Mpc in the Line of Sight of NGC 1052}",
      journal = {\apjl},
         year = 2019,
        month = jul,
       volume = {880},
       number = {1},
          eid = {L11},
        pages = {L11},
          doi = {10.3847/2041-8213/ab2fd2},
archivePrefix = {arXiv},
       eprint = {1907.03761},
 primaryClass = {astro-ph.GA},
       adsurl = {https://ui.adsabs.harvard.edu/abs/2019ApJ...880L..11M}
}

@ARTICLE{Danieli2020,
       author = {{Danieli}, Shany and {van Dokkum}, Pieter and {Abraham}, Roberto and {Conroy}, Charlie and {Dolphin}, Andrew E. and {Romanowsky}, Aaron J.},
        title = "{A Tip of the Red Giant Branch Distance to the Dark Matter Deficient Galaxy NGC 1052-DF4 from Deep Hubble Space Telescope Data}",
      journal = {\apjl},
         year = 2020,
        month = may,
       volume = {895},
       number = {1},
          eid = {L4},
        pages = {L4},
          doi = {10.3847/2041-8213/ab8dc4},
archivePrefix = {arXiv},
       eprint = {1910.07529},
 primaryClass = {astro-ph.GA},
       adsurl = {https://ui.adsabs.harvard.edu/abs/2020ApJ...895L...4D}
}

@ARTICLE{Shen2021,
       author = {{Shen}, Zili and {Danieli}, Shany and {van Dokkum}, Pieter and {Abraham}, Roberto and {Brodie}, Jean P. and {Conroy}, Charlie and {Dolphin}, Andrew E. and {Romanowsky}, Aaron J. and {Kruijssen}, J.~M. Diederik and {Dutta Chowdhury}, Dhruba},
        title = "{A Tip of the Red Giant Branch Distance of 22.1 {\ensuremath{\pm}} 1.2 Mpc to the Dark Matter Deficient Galaxy NGC 1052-DF2 from 40 Orbits of Hubble Space Telescope Imaging}",
      journal = {\apjl},
         year = 2021,
        month = jun,
       volume = {914},
       number = {1},
          eid = {L12},
        pages = {L12},
          doi = {10.3847/2041-8213/ac0335},
archivePrefix = {arXiv},
       eprint = {2104.03319},
 primaryClass = {astro-ph.GA},
       adsurl = {https://ui.adsabs.harvard.edu/abs/2021ApJ...914L..12S}
}

@ARTICLE{Guo2020,
       author = {{Guo}, Qi and {Hu}, Huijie and {Zheng}, Zheng and {Liao}, Shihong and {Du}, Wei and {Mao}, Shude and {Jiang}, Linhua and {Wang}, Jing and {Peng}, Yingjie and {Gao}, Liang and {Wang}, Jie and {Wu}, Hong},
        title = "{Further evidence for a population of dark-matter-deficient dwarf galaxies}",
      journal = {Nature Astronomy},
         year = 2020,
        month = jan,
       volume = {4},
        pages = {246-251},
          doi = {10.1038/s41550-019-0930-9},
archivePrefix = {arXiv},
       eprint = {1908.00046},
 primaryClass = {astro-ph.GA},
       adsurl = {https://ui.adsabs.harvard.edu/abs/2020NatAs...4..246G}
}

@ARTICLE{ManceraPina2022,
       author = {{Mancera Pi{\~n}a}, Pavel E. and {Fraternali}, Filippo and {Oosterloo}, Tom and {Adams}, Elizabeth A.~K. and {Oman}, Kyle A. and {Leisman}, Lukas},
        title = "{No need for dark matter: resolved kinematics of the ultra-diffuse galaxy AGC 114905}",
      journal = {\mnras},
         year = 2022,
        month = may,
       volume = {512},
       number = {3},
        pages = {3230-3242},
          doi = {10.1093/mnras/stab3491},
archivePrefix = {arXiv},
       eprint = {2112.00017},
 primaryClass = {astro-ph.GA},
       adsurl = {https://ui.adsabs.harvard.edu/abs/2022MNRAS.512.3230M}
}

@ARTICLE{Comeron2023,
       author = {{Comer{\'o}n}, S{\'e}bastien and {Trujillo}, Ignacio and {Cappellari}, Michele and {Buitrago}, Fernando and {Gardu{\~n}o}, Luis E. and {Zaragoza-Cardiel}, Javier and {Zinchenko}, Igor A. and {Lara-L{\'o}pez}, Maritza A. and {Ferr{\'e}-Mateu}, Anna and {Dib}, Sami},
        title = "{The massive relic galaxy NGC 1277 is dark matter deficient. From dynamical models of integral-field stellar kinematics out to five effective radii}",
      journal = {\aap},
         year = 2023,
        month = jul,
       volume = {675},
          eid = {A143},
        pages = {A143},
          doi = {10.1051/0004-6361/202346291},
archivePrefix = {arXiv},
       eprint = {2303.11360},
 primaryClass = {astro-ph.GA},
       adsurl = {https://ui.adsabs.harvard.edu/abs/2023A&A...675A.143C}
}

@ARTICLE{Buzzo2025,
       author = {{Buzzo}, Maria Luisa and {Forbes}, Duncan A. and {Romanowsky}, Aaron J. and {Haacke}, Lydia and {Gannon}, Jonah S. and {Tang}, Yimeng and {Hilker}, Michael and {Ferr{\'e}-Mateu}, Anna and {Janssens}, Steven R. and {Brodie}, Jean P. and {Valenzuela}, Lucas M.},
        title = "{A new class of dark matter-free dwarf galaxies?: I. Clues from FCC 224, NGC 1052-DF2, and NGC 1052-DF4}",
      journal = {\aap},
         year = 2025,
        month = mar,
       volume = {695},
          eid = {A124},
        pages = {A124},
          doi = {10.1051/0004-6361/202453522},
archivePrefix = {arXiv},
       eprint = {2502.05405},
 primaryClass = {astro-ph.GA},
       adsurl = {https://ui.adsabs.harvard.edu/abs/2025A&A...695A.124B}
}

@ARTICLE{Romanowsky2024,
       author = {{Romanowsky}, Aaron J. and {Cabrera}, Enrique and {Janssens}, Steven R.},
        title = "{A Candidate Dark Matter Deficient Dwarf Galaxy in the Fornax Cluster Identified through Overluminous Star Clusters}",
      journal = {Research Notes of the American Astronomical Society},
         year = 2024,
        month = aug,
       volume = {8},
       number = {8},
          eid = {202},
        pages = {202},
          doi = {10.3847/2515-5172/ad7112},
archivePrefix = {arXiv},
       eprint = {2408.12663},
 primaryClass = {astro-ph.GA},
       adsurl = {https://ui.adsabs.harvard.edu/abs/2024RNAAS...8..202R}
}

@ARTICLE{Tang2025,
       author = {{Tang}, Yimeng and {Romanowsky}, Aaron J. and {Gannon}, Jonah S. and {Janssens}, Steven R. and {Brodie}, Jean P. and {Bundy}, Kevin A. and {Buzzo}, Maria Luisa and {Cabrera}, Enrique A. and {Danieli}, Shany and {Ferr{\'e}-Mateu}, Anna and {Forbes}, Duncan A. and {van Dokkum}, Pieter G.},
        title = "{An Unexplained Origin for the Unusual Globular Cluster System in the Ultradiffuse Galaxy FCC 224}",
      journal = {\apj},
         year = 2025,
        month = mar,
       volume = {982},
       number = {1},
          eid = {1},
        pages = {1},
          doi = {10.3847/1538-4357/adae11},
archivePrefix = {arXiv},
       eprint = {2501.10665},
 primaryClass = {astro-ph.GA},
       adsurl = {https://ui.adsabs.harvard.edu/abs/2025ApJ...982....1T}
}

@ARTICLE{Ogiya2018,
       author = {{Ogiya}, Go},
        title = "{Tidal stripping as a possible origin of the ultra diffuse galaxy lacking dark matter}",
      journal = {\mnras},
         year = 2018,
        month = oct,
       volume = {480},
       number = {1},
        pages = {L106-L110},
          doi = {10.1093/mnrasl/sly138},
archivePrefix = {arXiv},
       eprint = {1804.06421},
 primaryClass = {astro-ph.GA},
       adsurl = {https://ui.adsabs.harvard.edu/abs/2018MNRAS.480L.106O}
}

@ARTICLE{Nusser2020,
       author = {{Nusser}, Adi},
        title = "{A Scenario for Ultradiffuse Satellite Galaxies with Low Velocity Dispersions: The Case of [KKS 2000]04}",
      journal = {\apj},
         year = 2020,
        month = apr,
       volume = {893},
       number = {1},
          eid = {66},
        pages = {66},
          doi = {10.3847/1538-4357/ab792c},
archivePrefix = {arXiv},
       eprint = {1907.08035},
 primaryClass = {astro-ph.GA},
       adsurl = {https://ui.adsabs.harvard.edu/abs/2020ApJ...893...66N}
}

@ARTICLE{Jing2019,
       author = {{Jing}, Yingjie and {Wang}, Chunxiang and {Li}, Ran and {Liao}, Shihong and {Wang}, Jie and {Guo}, Qi and {Gao}, Liang},
        title = "{Dark-matter-deficient galaxies in hydrodynamical simulations}",
      journal = {\mnras},
         year = 2019,
        month = sep,
       volume = {488},
       number = {3},
        pages = {3298-3307},
          doi = {10.1093/mnras/stz1839},
archivePrefix = {arXiv},
       eprint = {1811.09070},
 primaryClass = {astro-ph.GA},
       adsurl = {https://ui.adsabs.harvard.edu/abs/2019MNRAS.488.3298J}
}

@ARTICLE{Niemiec2019,
       author = {{Niemiec}, Anna and {Jullo}, Eric and {Giocoli}, Carlo and {Limousin}, Marceau and {Jauzac}, Mathilde},
        title = "{Dark matter stripping in galaxy clusters: a look at the stellar-to-halo mass relation in the Illustris simulation}",
      journal = {\mnras},
         year = 2019,
        month = jul,
       volume = {487},
       number = {1},
        pages = {653-666},
          doi = {10.1093/mnras/stz1318},
archivePrefix = {arXiv},
       eprint = {1811.04996},
 primaryClass = {astro-ph.GA},
       adsurl = {https://ui.adsabs.harvard.edu/abs/2019MNRAS.487..653N}
}

@ARTICLE{Maccio2021,
       author = {{Macci{\`o}}, Andrea V. and {Prats}, Daniel Huterer and {Dixon}, Keri L. and {Buck}, Tobias and {Waterval}, Stefan and {Arora}, Nikhil and {Courteau}, St{\'e}phane and {Kang}, Xi},
        title = "{Creating a galaxy lacking dark matter in a dark matter-dominated universe}",
      journal = {\mnras},
         year = 2021,
        month = feb,
       volume = {501},
       number = {1},
        pages = {693-700},
          doi = {10.1093/mnras/staa3716},
archivePrefix = {arXiv},
       eprint = {2010.02245},
 primaryClass = {astro-ph.GA},
       adsurl = {https://ui.adsabs.harvard.edu/abs/2021MNRAS.501..693M}
}

@ARTICLE{Ogiya2022,
       author = {{Ogiya}, Go and {van den Bosch}, Frank C. and {Burkert}, Andreas},
        title = "{On the tidal formation of dark matter-deficient galaxies}",
      journal = {\mnras},
         year = 2022,
        month = feb,
       volume = {510},
       number = {2},
        pages = {2724-2739},
          doi = {10.1093/mnras/stab3658},
archivePrefix = {arXiv},
       eprint = {2111.12104},
 primaryClass = {astro-ph.GA},
       adsurl = {https://ui.adsabs.harvard.edu/abs/2022MNRAS.510.2724O}
}

@ARTICLE{Jackson2021,
       author = {{Jackson}, R.~A. and {Kaviraj}, S. and {Martin}, G. and {Devriendt}, J.~E.~G. and {Slyz}, A. and {Silk}, J. and {Dubois}, Y. and {Yi}, S.~K. and {Pichon}, C. and {Volonteri}, M. and {Choi}, H. and {Kimm}, T. and {Kraljic}, K. and {Peirani}, S.},
        title = "{Dark matter-deficient dwarf galaxies form via tidal stripping of dark matter in interactions with massive companions}",
      journal = {\mnras},
         year = 2021,
        month = apr,
       volume = {502},
       number = {2},
        pages = {1785-1796},
          doi = {10.1093/mnras/stab093},
archivePrefix = {arXiv},
       eprint = {2010.02219},
 primaryClass = {astro-ph.GA},
       adsurl = {https://ui.adsabs.harvard.edu/abs/2021MNRAS.502.1785J}
}

@ARTICLE{Moreno2022,
       author = {{Moreno}, Jorge and {Danieli}, Shany and {Bullock}, James S. and {Feldmann}, Robert and {Hopkins}, Philip F. and {{\c{c}}atmabacak}, Onur and {Gurvich}, Alexander and {Lazar}, Alexandres and {Klein}, Courtney and {Hummels}, Cameron B. and {Hafen}, Zachary and {Mercado}, Francisco J. and {Yu}, Sijie and {Jiang}, Fangzhou and {Wheeler}, Coral and {Wetzel}, Andrew and {Angl{\'e}s-Alc{\'a}zar}, Daniel and {Boylan-Kolchin}, Michael and {Quataert}, Eliot and {Faucher-Gigu{\`e}re}, Claude-Andr{\'e} and {Kere{\v{s}}}, Du{\v{s}}an},
        title = "{Galaxies lacking dark matter produced by close encounters in a cosmological simulation}",
      journal = {Nature Astronomy},
         year = 2022,
        month = apr,
       volume = {6},
        pages = {496-502},
          doi = {10.1038/s41550-021-01598-4},
archivePrefix = {arXiv},
       eprint = {2202.05836},
 primaryClass = {astro-ph.GA},
       adsurl = {https://ui.adsabs.harvard.edu/abs/2022NatAs...6..496M}
}

@ARTICLE{Katayama2024,
       author = {{Katayama}, Ryosuke and {Nagamine}, Kentaro and {Kihara}, Kenji},
        title = "{Impact of dynamical friction on the tidal formation of NGC 1052-DF2}",
      journal = {\mnras},
         year = 2024,
        month = nov,
       volume = {535},
       number = {1},
        pages = {675-688},
          doi = {10.1093/mnras/stae2273},
archivePrefix = {arXiv},
       eprint = {2306.07756},
 primaryClass = {astro-ph.GA},
       adsurl = {https://ui.adsabs.harvard.edu/abs/2024MNRAS.535..675K}
}

@ARTICLE{Montero-Dorta2024,
       author = {{Montero-Dorta}, Antonio D. and {Rodriguez}, Facundo and {Artale}, M. Celeste and {Smith}, Rory and {Chaves-Montero}, Jon{\'a}s},
        title = "{Tracking the evolution of satellite galaxies: mass stripping and dark-matter deficient galaxies}",
      journal = {\mnras},
         year = 2024,
        month = jan,
       volume = {527},
       number = {3},
        pages = {5868-5885},
          doi = {10.1093/mnras/stad3268},
archivePrefix = {arXiv},
       eprint = {2212.12090},
 primaryClass = {astro-ph.GA},
       adsurl = {https://ui.adsabs.harvard.edu/abs/2024MNRAS.527.5868M}
}

@ARTICLE{Mitrasinovic2023,
       author = {{Mitra{\v{s}}inovi{\'c}}, Ana and {Smole}, Majda and {Micic}, Miroslav},
        title = "{The isolated dark matter-poor galaxy that ran away. An example from IllustrisTNG}",
      journal = {\aap},
         year = 2023,
        month = dec,
       volume = {680},
          eid = {L1},
        pages = {L1},
          doi = {10.1051/0004-6361/202348417},
archivePrefix = {arXiv},
       eprint = {2311.10643},
 primaryClass = {astro-ph.GA},
       adsurl = {https://ui.adsabs.harvard.edu/abs/2023A&A...680L...1M}
}

@ARTICLE{Silk2019,
       author = {{Silk}, Joseph},
        title = "{Ultra-diffuse galaxies without dark matter}",
      journal = {\mnras},
         year = 2019,
        month = sep,
       volume = {488},
       number = {1},
        pages = {L24-L28},
          doi = {10.1093/mnrasl/slz090},
archivePrefix = {arXiv},
       eprint = {1905.13235},
 primaryClass = {astro-ph.GA},
       adsurl = {https://ui.adsabs.harvard.edu/abs/2019MNRAS.488L..24S}
}

@ARTICLE{Shin2020,
       author = {{Shin}, Eun-jin and {Jung}, Minyong and {Kwon}, Goojin and {Kim}, Ji-hoon and {Lee}, Joohyun and {Jo}, Yongseok and {Oh}, Boon Kiat},
        title = "{Dark Matter Deficient Galaxies Produced via High-velocity Galaxy Collisions in High-resolution Numerical Simulations}",
      journal = {\apj},
         year = 2020,
        month = aug,
       volume = {899},
       number = {1},
          eid = {25},
        pages = {25},
          doi = {10.3847/1538-4357/aba434},
archivePrefix = {arXiv},
       eprint = {2007.09889},
 primaryClass = {astro-ph.GA},
       adsurl = {https://ui.adsabs.harvard.edu/abs/2020ApJ...899...25S}
}

@ARTICLE{LeeJoo2021,
       author = {{Lee}, Joohyun and {Shin}, Eun-jin and {Kim}, Ji-hoon},
        title = "{Dark Matter Deficient Galaxies and Their Member Star Clusters Form Simultaneously during High-velocity Galaxy Collisions in 1.25 pc Resolution Simulations}",
      journal = {\apjl},
         year = 2021,
        month = aug,
       volume = {917},
       number = {2},
          eid = {L15},
        pages = {L15},
          doi = {10.3847/2041-8213/ac16e0},
archivePrefix = {arXiv},
       eprint = {2108.01102},
 primaryClass = {astro-ph.GA},
       adsurl = {https://ui.adsabs.harvard.edu/abs/2021ApJ...917L..15L}
}

@ARTICLE{LeeJoo2024,
       author = {{Lee}, Joohyun and {Shin}, Eun-jin and {Kim}, Ji-hoon and {Shapiro}, Paul R. and {Chung}, Eunwoo},
        title = "{Multiple Beads on a String: Dark-matter-deficient Galaxy Formation in a Mini-Bullet Satellite{\textendash}Satellite Galaxy Collision}",
      journal = {\apj},
         year = 2024,
        month = may,
       volume = {966},
       number = {1},
          eid = {72},
        pages = {72},
          doi = {10.3847/1538-4357/ad2932},
archivePrefix = {arXiv},
       eprint = {2312.11350},
 primaryClass = {astro-ph.GA},
       adsurl = {https://ui.adsabs.harvard.edu/abs/2024ApJ...966...72L}
}

@ARTICLE{Otaki2023,
       author = {{Otaki}, Koki and {Mori}, Masao},
        title = "{Frequency of the dark matter subhalo collisions and bifurcation sequence arising formation of dwarf galaxies}",
      journal = {\mnras},
         year = 2023,
        month = oct,
       volume = {525},
       number = {2},
        pages = {2535-2552},
          doi = {10.1093/mnras/stad2432},
archivePrefix = {arXiv},
       eprint = {2308.03737},
 primaryClass = {astro-ph.GA},
       adsurl = {https://ui.adsabs.harvard.edu/abs/2023MNRAS.525.2535O}
}

@ARTICLE{Carleton2019,
       author = {{Carleton}, Timothy and {Errani}, Rapha{\"e}l and {Cooper}, Michael and {Kaplinghat}, Manoj and {Pe{\~n}arrubia}, Jorge and {Guo}, Yicheng},
        title = "{The formation of ultra-diffuse galaxies in cored dark matter haloes through tidal stripping and heating}",
      journal = {\mnras},
         year = 2019,
        month = may,
       volume = {485},
       number = {1},
        pages = {382-395},
          doi = {10.1093/mnras/stz383},
archivePrefix = {arXiv},
       eprint = {1805.06896},
 primaryClass = {astro-ph.GA},
       adsurl = {https://ui.adsabs.harvard.edu/abs/2019MNRAS.485..382C}
}

@ARTICLE{Saulder2020,
       author = {{Saulder}, Christoph and {Snaith}, Owain and {Park}, Changbom and {Laigle}, Clotilde},
        title = "{Isolated dark-matter-deprived galaxies in hydrodynamical simulations: real objects or artefacts?}",
      journal = {\mnras},
         year = 2020,
        month = jan,
       volume = {491},
       number = {1},
        pages = {1278-1286},
          doi = {10.1093/mnras/stz3058},
archivePrefix = {arXiv},
       eprint = {1907.03542},
 primaryClass = {astro-ph.GA},
       adsurl = {https://ui.adsabs.harvard.edu/abs/2020MNRAS.491.1278S}
}

@ARTICLE{Contreras-Santos2024,
       author = {{Contreras-Santos}, A. and {Buitrago}, F. and {Knebe}, A. and {Rasia}, E. and {Pearce}, F.~R. and {Cui}, W. and {Power}, C. and {Winstanley}, J.},
        title = "{The Three Hundred: The existence of massive dark matter-deficient satellite galaxies in cosmological simulations}",
      journal = {\aap},
         year = 2024,
        month = oct,
       volume = {690},
          eid = {A109},
        pages = {A109},
          doi = {10.1051/0004-6361/202451271},
archivePrefix = {arXiv},
       eprint = {2409.10356},
 primaryClass = {astro-ph.GA},
       adsurl = {https://ui.adsabs.harvard.edu/abs/2024A&A...690A.109C}
}

@ARTICLE{Nelson2019a,
       author = {{Nelson}, Dylan and {Springel}, Volker and {Pillepich}, Annalisa and {Rodriguez-Gomez}, Vicente and {Torrey}, Paul and {Genel}, Shy and {Vogelsberger}, Mark and {Pakmor}, Ruediger and {Marinacci}, Federico and {Weinberger}, Rainer and {Kelley}, Luke and {Lovell}, Mark and {Diemer}, Benedikt and {Hernquist}, Lars},
        title = "{The IllustrisTNG simulations: public data release}",
      journal = {Computational Astrophysics and Cosmology},
         year = 2019,
        month = may,
       volume = {6},
       number = {1},
          eid = {2},
        pages = {2},
          doi = {10.1186/s40668-019-0028-x},
archivePrefix = {arXiv},
       eprint = {1812.05609},
 primaryClass = {astro-ph.GA},
       adsurl = {https://ui.adsabs.harvard.edu/abs/2019ComAC...6....2N}
}

@ARTICLE{Pillepich2018a,
       author = {{Pillepich}, Annalisa and {Nelson}, Dylan and {Hernquist}, Lars and {Springel}, Volker and {Pakmor}, R{\"u}diger and {Torrey}, Paul and {Weinberger}, Rainer and {Genel}, Shy and {Naiman}, Jill P. and {Marinacci}, Federico and {Vogelsberger}, Mark},
        title = "{First results from the IllustrisTNG simulations: the stellar mass content of groups and clusters of galaxies}",
      journal = {\mnras},
         year = 2018,
        month = mar,
       volume = {475},
       number = {1},
        pages = {648-675},
          doi = {10.1093/mnras/stx3112},
archivePrefix = {arXiv},
       eprint = {1707.03406},
 primaryClass = {astro-ph.GA},
       adsurl = {https://ui.adsabs.harvard.edu/abs/2018MNRAS.475..648P}
}

@ARTICLE{Springel2018,
       author = {{Springel}, Volker and {Pakmor}, R{\"u}diger and {Pillepich}, Annalisa and {Weinberger}, Rainer and {Nelson}, Dylan and {Hernquist}, Lars and {Vogelsberger}, Mark and {Genel}, Shy and {Torrey}, Paul and {Marinacci}, Federico and {Naiman}, Jill},
        title = "{First results from the IllustrisTNG simulations: matter and galaxy clustering}",
      journal = {\mnras},
         year = 2018,
        month = mar,
       volume = {475},
       number = {1},
        pages = {676-698},
          doi = {10.1093/mnras/stx3304},
archivePrefix = {arXiv},
       eprint = {1707.03397},
 primaryClass = {astro-ph.GA},
       adsurl = {https://ui.adsabs.harvard.edu/abs/2018MNRAS.475..676S}
}

@ARTICLE{Nelson2018a,
       author = {{Nelson}, Dylan and {Pillepich}, Annalisa and {Springel}, Volker and {Weinberger}, Rainer and {Hernquist}, Lars and {Pakmor}, R{\"u}diger and {Genel}, Shy and {Torrey}, Paul and {Vogelsberger}, Mark and {Kauffmann}, Guinevere and {Marinacci}, Federico and {Naiman}, Jill},
        title = "{First results from the IllustrisTNG simulations: the galaxy colour bimodality}",
      journal = {\mnras},
         year = 2018,
        month = mar,
       volume = {475},
       number = {1},
        pages = {624-647},
          doi = {10.1093/mnras/stx3040},
archivePrefix = {arXiv},
       eprint = {1707.03395},
 primaryClass = {astro-ph.GA},
       adsurl = {https://ui.adsabs.harvard.edu/abs/2018MNRAS.475..624N}
}

@ARTICLE{Naiman2018,
       author = {{Naiman}, Jill P. and {Pillepich}, Annalisa and {Springel}, Volker and {Ramirez-Ruiz}, Enrico and {Torrey}, Paul and {Vogelsberger}, Mark and {Pakmor}, R{\"u}diger and {Nelson}, Dylan and {Marinacci}, Federico and {Hernquist}, Lars and {Weinberger}, Rainer and {Genel}, Shy},
        title = "{First results from the IllustrisTNG simulations: a tale of two elements - chemical evolution of magnesium and europium}",
      journal = {\mnras},
         year = 2018,
        month = jun,
       volume = {477},
       number = {1},
        pages = {1206-1224},
          doi = {10.1093/mnras/sty618},
archivePrefix = {arXiv},
       eprint = {1707.03401},
 primaryClass = {astro-ph.GA},
       adsurl = {https://ui.adsabs.harvard.edu/abs/2018MNRAS.477.1206N}
}

@ARTICLE{Marinacci2018,
       author = {{Marinacci}, Federico and {Vogelsberger}, Mark and {Pakmor}, R{\"u}diger and {Torrey}, Paul and {Springel}, Volker and {Hernquist}, Lars and {Nelson}, Dylan and {Weinberger}, Rainer and {Pillepich}, Annalisa and {Naiman}, Jill and {Genel}, Shy},
        title = "{First results from the IllustrisTNG simulations: radio haloes and magnetic fields}",
      journal = {\mnras},
         year = 2018,
        month = nov,
       volume = {480},
       number = {4},
        pages = {5113-5139},
          doi = {10.1093/mnras/sty2206},
archivePrefix = {arXiv},
       eprint = {1707.03396},
 primaryClass = {astro-ph.CO},
       adsurl = {https://ui.adsabs.harvard.edu/abs/2018MNRAS.480.5113M}
}

@ARTICLE{Weinberger2017,
       author = {{Weinberger}, Rainer and {Springel}, Volker and {Hernquist}, Lars and {Pillepich}, Annalisa and {Marinacci}, Federico and {Pakmor}, R{\"u}diger and {Nelson}, Dylan and {Genel}, Shy and {Vogelsberger}, Mark and {Naiman}, Jill and {Torrey}, Paul},
        title = "{Simulating galaxy formation with black hole driven thermal and kinetic feedback}",
      journal = {\mnras},
         year = 2017,
        month = mar,
       volume = {465},
       number = {3},
        pages = {3291-3308},
          doi = {10.1093/mnras/stw2944},
archivePrefix = {arXiv},
       eprint = {1607.03486},
 primaryClass = {astro-ph.GA},
       adsurl = {https://ui.adsabs.harvard.edu/abs/2017MNRAS.465.3291W}
}

@ARTICLE{Pillepich2018b,
       author = {{Pillepich}, Annalisa and {Springel}, Volker and {Nelson}, Dylan and {Genel}, Shy and {Naiman}, Jill and {Pakmor}, R{\"u}diger and {Hernquist}, Lars and {Torrey}, Paul and {Vogelsberger}, Mark and {Weinberger}, Rainer and {Marinacci}, Federico},
        title = "{Simulating galaxy formation with the IllustrisTNG model}",
      journal = {\mnras},
         year = 2018,
        month = jan,
       volume = {473},
       number = {3},
        pages = {4077-4106},
          doi = {10.1093/mnras/stx2656},
archivePrefix = {arXiv},
       eprint = {1703.02970},
 primaryClass = {astro-ph.GA},
       adsurl = {https://ui.adsabs.harvard.edu/abs/2018MNRAS.473.4077P}
}

@ARTICLE{PlanckColl2016,
       author = {{Planck Collaboration} and {Ade}, P.~A.~R. and {Aghanim}, N. and {Arnaud}, M. and {Ashdown}, M. and {Aumont}, J. and {Baccigalupi}, C. and {Banday}, A.~J. and {Barreiro}, R.~B. and {Bartlett}, J.~G. and {Bartolo}, N. and {Battaner}, E. and {Battye}, R. and {Benabed}, K. and {Beno{\^\i}t}, A. and {Benoit-L{\'e}vy}, A. and {Bernard}, J. -P. and {Bersanelli}, M. and {Bielewicz}, P. and {Bock}, J.~J. and {Bonaldi}, A. and {Bonavera}, L. and {Bond}, J.~R. and {Borrill}, J. and {Bouchet}, F.~R. and {Boulanger}, F. and {Bucher}, M. and {Burigana}, C. and {Butler}, R.~C. and {Calabrese}, E. and {Cardoso}, J. -F. and {Catalano}, A. and {Challinor}, A. and {Chamballu}, A. and {Chary}, R. -R. and {Chiang}, H.~C. and {Chluba}, J. and {Christensen}, P.~R. and {Church}, S. and {Clements}, D.~L. and {Colombi}, S. and {Colombo}, L.~P.~L. and {Combet}, C. and {Coulais}, A. and {Crill}, B.~P. and {Curto}, A. and {Cuttaia}, F. and {Danese}, L. and {Davies}, R.~D. and {Davis}, R.~J. and {de Bernardis}, P. and {de Rosa}, A. and {de Zotti}, G. and {Delabrouille}, J. and {D{\'e}sert}, F. -X. and {Di Valentino}, E. and {Dickinson}, C. and {Diego}, J.~M. and {Dolag}, K. and {Dole}, H. and {Donzelli}, S. and {Dor{\'e}}, O. and {Douspis}, M. and {Ducout}, A. and {Dunkley}, J. and {Dupac}, X. and {Efstathiou}, G. and {Elsner}, F. and {En{\ss}lin}, T.~A. and {Eriksen}, H.~K. and {Farhang}, M. and {Fergusson}, J. and {Finelli}, F. and {Forni}, O. and {Frailis}, M. and {Fraisse}, A.~A. and {Franceschi}, E. and {Frejsel}, A. and {Galeotta}, S. and {Galli}, S. and {Ganga}, K. and {Gauthier}, C. and {Gerbino}, M. and {Ghosh}, T. and {Giard}, M. and {Giraud-H{\'e}raud}, Y. and {Giusarma}, E. and {Gjerl{\o}w}, E. and {Gonz{\'a}lez-Nuevo}, J. and {G{\'o}rski}, K.~M. and {Gratton}, S. and {Gregorio}, A. and {Gruppuso}, A. and {Gudmundsson}, J.~E. and {Hamann}, J. and {Hansen}, F.~K. and {Hanson}, D. and {Harrison}, D.~L. and {Helou}, G. and {Henrot-Versill{\'e}}, S. and {Hern{\'a}ndez-Monteagudo}, C. and {Herranz}, D. and {Hildebrandt}, S.~R. and {Hivon}, E. and {Hobson}, M. and {Holmes}, W.~A. and {Hornstrup}, A. and {Hovest}, W. and {Huang}, Z. and {Huffenberger}, K.~M. and {Hurier}, G. and {Jaffe}, A.~H. and {Jaffe}, T.~R. and {Jones}, W.~C. and {Juvela}, M. and {Keih{\"a}nen}, E. and {Keskitalo}, R. and {Kisner}, T.~S. and {Kneissl}, R. and {Knoche}, J. and {Knox}, L. and {Kunz}, M. and {Kurki-Suonio}, H. and {Lagache}, G. and {L{\"a}hteenm{\"a}ki}, A. and {Lamarre}, J. -M. and {Lasenby}, A. and {Lattanzi}, M. and {Lawrence}, C.~R. and {Leahy}, J.~P. and {Leonardi}, R. and {Lesgourgues}, J. and {Levrier}, F. and {Lewis}, A. and {Liguori}, M. and {Lilje}, P.~B. and {Linden-V{\o}rnle}, M. and {L{\'o}pez-Caniego}, M. and {Lubin}, P.~M. and {Mac{\'\i}as-P{\'e}rez}, J.~F. and {Maggio}, G. and {Maino}, D. and {Mandolesi}, N. and {Mangilli}, A. and {Marchini}, A. and {Maris}, M. and {Martin}, P.~G. and {Martinelli}, M. and {Mart{\'\i}nez-Gonz{\'a}lez}, E. and {Masi}, S. and {Matarrese}, S. and {McGehee}, P. and {Meinhold}, P.~R. and {Melchiorri}, A. and {Melin}, J. -B. and {Mendes}, L. and {Mennella}, A. and {Migliaccio}, M. and {Millea}, M. and {Mitra}, S. and {Miville-Desch{\^e}nes}, M. -A. and {Moneti}, A. and {Montier}, L. and {Morgante}, G. and {Mortlock}, D. and {Moss}, A. and {Munshi}, D. and {Murphy}, J.~A. and {Naselsky}, P. and {Nati}, F. and {Natoli}, P. and {Netterfield}, C.~B. and {N{\o}rgaard-Nielsen}, H.~U. and {Noviello}, F. and {Novikov}, D. and {Novikov}, I. and {Oxborrow}, C.~A. and {Paci}, F. and {Pagano}, L. and {Pajot}, F. and {Paladini}, R. and {Paoletti}, D. and {Partridge}, B. and {Pasian}, F. and {Patanchon}, G. and {Pearson}, T.~J. and {Perdereau}, O. and {Perotto}, L. and {Perrotta}, F. and {Pettorino}, V. and {Piacentini}, F. and {Piat}, M. and {Pierpaoli}, E. and {Pietrobon}, D. and {Plaszczynski}, S. and {Pointecouteau}, E. and {Polenta}, G. and {Popa}, L. and {Pratt}, G.~W. and {Pr{\'e}zeau}, G.},
        title = "{Planck 2015 results. XIII. Cosmological parameters}",
      journal = {\aap},
         year = 2016,
        month = sep,
       volume = {594},
          eid = {A13},
        pages = {A13},
          doi = {10.1051/0004-6361/201525830},
archivePrefix = {arXiv},
       eprint = {1502.01589},
 primaryClass = {astro-ph.CO},
       adsurl = {https://ui.adsabs.harvard.edu/abs/2016A&A...594A..13P}
}

@ARTICLE{Davis1985,
       author = {{Davis}, M. and {Efstathiou}, G. and {Frenk}, C.~S. and {White}, S.~D.~M.},
        title = "{The evolution of large-scale structure in a universe dominated by cold dark matter}",
      journal = {\apj},
         year = 1985,
        month = may,
       volume = {292},
        pages = {371-394},
          doi = {10.1086/163168},
       adsurl = {https://ui.adsabs.harvard.edu/abs/1985ApJ...292..371D}
}

@ARTICLE{Audit1998,
       author = {{Audit}, Edouard and {Teyssier}, Romain and {Alimi}, Jean-Michel},
        title = "{Non-linear dynamics and mass function of cosmic structures. II. Numerical results}",
      journal = {\aap},
         year = 1998,
        month = may,
       volume = {333},
        pages = {779-789},
          doi = {10.48550/arXiv.astro-ph/9802287},
archivePrefix = {arXiv},
       eprint = {astro-ph/9802287},
 primaryClass = {astro-ph},
       adsurl = {https://ui.adsabs.harvard.edu/abs/1998A&A...333..779A}
}

@ARTICLE{Springel2001,
       author = {{Springel}, Volker and {White}, Simon D.~M. and {Tormen}, Giuseppe and {Kauffmann}, Guinevere},
        title = "{Populating a cluster of galaxies - I. Results at z=0}",
      journal = {\mnras},
         year = 2001,
        month = dec,
       volume = {328},
       number = {3},
        pages = {726-750},
          doi = {10.1046/j.1365-8711.2001.04912.x},
archivePrefix = {arXiv},
       eprint = {astro-ph/0012055},
 primaryClass = {astro-ph},
       adsurl = {https://ui.adsabs.harvard.edu/abs/2001MNRAS.328..726S}
}

@ARTICLE{Dolag2009,
       author = {{Dolag}, K. and {Borgani}, S. and {Murante}, G. and {Springel}, V.},
        title = "{Substructures in hydrodynamical cluster simulations}",
      journal = {\mnras},
         year = 2009,
        month = oct,
       volume = {399},
       number = {2},
        pages = {497-514},
          doi = {10.1111/j.1365-2966.2009.15034.x},
archivePrefix = {arXiv},
       eprint = {0808.3401},
 primaryClass = {astro-ph},
       adsurl = {https://ui.adsabs.harvard.edu/abs/2009MNRAS.399..497D}
}

@ARTICLE{Rodriguez-Gomez2015,
       author = {{Rodriguez-Gomez}, Vicente and {Genel}, Shy and {Vogelsberger}, Mark and {Sijacki}, Debora and {Pillepich}, Annalisa and {Sales}, Laura V. and {Torrey}, Paul and {Snyder}, Greg and {Nelson}, Dylan and {Springel}, Volker and {Ma}, Chung-Pei and {Hernquist}, Lars},
        title = "{The merger rate of galaxies in the Illustris simulation: a comparison with observations and semi-empirical models}",
      journal = {\mnras},
         year = 2015,
        month = may,
       volume = {449},
       number = {1},
        pages = {49-64},
          doi = {10.1093/mnras/stv264},
archivePrefix = {arXiv},
       eprint = {1502.01339},
 primaryClass = {astro-ph.GA},
       adsurl = {https://ui.adsabs.harvard.edu/abs/2015MNRAS.449...49R}
}

@ARTICLE{Dressler1980,
       author = {{Dressler}, A.},
        title = "{Galaxy morphology in rich clusters: implications for the formation and evolution of galaxies.}",
      journal = {\apj},
         year = 1980,
        month = mar,
       volume = {236},
        pages = {351-365},
          doi = {10.1086/157753},
       adsurl = {https://ui.adsabs.harvard.edu/abs/1980ApJ...236..351D}
}

@ARTICLE{Postman1984,
       author = {{Postman}, M. and {Geller}, M.~J.},
        title = "{The morphology-density relation - The group connection.}",
      journal = {\apj},
         year = 1984,
        month = jun,
       volume = {281},
        pages = {95-99},
          doi = {10.1086/162078},
       adsurl = {https://ui.adsabs.harvard.edu/abs/1984ApJ...281...95P}
}

@ARTICLE{Postman2005,
       author = {{Postman}, M. and {Franx}, M. and {Cross}, N.~J.~G. and {Holden}, B. and {Ford}, H.~C. and {Illingworth}, G.~D. and {Goto}, T. and {Demarco}, R. and {Rosati}, P. and {Blakeslee}, J.~P. and {Tran}, K. -V. and {Ben{\'\i}tez}, N. and {Clampin}, M. and {Hartig}, G.~F. and {Homeier}, N. and {Ardila}, D.~R. and {Bartko}, F. and {Bouwens}, R.~J. and {Bradley}, L.~D. and {Broadhurst}, T.~J. and {Brown}, R.~A. and {Burrows}, C.~J. and {Cheng}, E.~S. and {Feldman}, P.~D. and {Golimowski}, D.~A. and {Gronwall}, C. and {Infante}, L. and {Kimble}, R.~A. and {Krist}, J.~E. and {Lesser}, M.~P. and {Martel}, A.~R. and {Mei}, S. and {Menanteau}, F. and {Meurer}, G.~R. and {Miley}, G.~K. and {Motta}, V. and {Sirianni}, M. and {Sparks}, W.~B. and {Tran}, H.~D. and {Tsvetanov}, Z.~I. and {White}, R.~L. and {Zheng}, W.},
        title = "{The Morphology-Density Relation in z \raisebox{-0.5ex}\textasciitilde 1 Clusters}",
      journal = {\apj},
         year = 2005,
        month = apr,
       volume = {623},
       number = {2},
        pages = {721-741},
          doi = {10.1086/428881},
archivePrefix = {arXiv},
       eprint = {astro-ph/0501224},
 primaryClass = {astro-ph},
       adsurl = {https://ui.adsabs.harvard.edu/abs/2005ApJ...623..721P}
}

@ARTICLE{Park2007,
       author = {{Park}, Changbom and {Choi}, Yun-Young and {Vogeley}, Michael S. and {Gott}, III, J. Richard and {Blanton}, Michael R. and {SDSS Collaboration}},
        title = "{Environmental Dependence of Properties of Galaxies in the Sloan Digital Sky Survey}",
      journal = {\apj},
         year = 2007,
        month = apr,
       volume = {658},
       number = {2},
        pages = {898-916},
          doi = {10.1086/511059},
archivePrefix = {arXiv},
       eprint = {astro-ph/0611610},
 primaryClass = {astro-ph},
       adsurl = {https://ui.adsabs.harvard.edu/abs/2007ApJ...658..898P}
}

@ARTICLE{Park2009,
       author = {{Park}, Changbom and {Hwang}, Ho Seong},
        title = "{Interactions of Galaxies in the Galaxy Cluster Environment}",
      journal = {\apj},
         year = 2009,
        month = jul,
       volume = {699},
       number = {2},
        pages = {1595-1609},
          doi = {10.1088/0004-637X/699/2/1595},
archivePrefix = {arXiv},
       eprint = {0812.2088},
 primaryClass = {astro-ph},
       adsurl = {https://ui.adsabs.harvard.edu/abs/2009ApJ...699.1595P}
}

@ARTICLE{Thomas2010,
       author = {{Thomas}, Daniel and {Maraston}, Claudia and {Schawinski}, Kevin and {Sarzi}, Marc and {Silk}, Joseph},
        title = "{Environment and self-regulation in galaxy formation}",
      journal = {\mnras},
         year = 2010,
        month = jun,
       volume = {404},
       number = {4},
        pages = {1775-1789},
          doi = {10.1111/j.1365-2966.2010.16427.x},
archivePrefix = {arXiv},
       eprint = {0912.0259},
 primaryClass = {astro-ph.CO},
       adsurl = {https://ui.adsabs.harvard.edu/abs/2010MNRAS.404.1775T}
}

@ARTICLE{Smith2005,
       author = {{Smith}, Graham P. and {Treu}, Tommaso and {Ellis}, Richard S. and {Moran}, Sean M. and {Dressler}, Alan},
        title = "{Evolution since z = 1 of the Morphology-Density Relation for Galaxies}",
      journal = {\apj},
         year = 2005,
        month = feb,
       volume = {620},
       number = {1},
        pages = {78-87},
          doi = {10.1086/426930},
archivePrefix = {arXiv},
       eprint = {astro-ph/0403455},
 primaryClass = {astro-ph},
       adsurl = {https://ui.adsabs.harvard.edu/abs/2005ApJ...620...78S}
}

@ARTICLE{Hwang2009,
       author = {{Hwang}, Ho Seong and {Park}, Changbom},
        title = "{Evidence for Morphology and Luminosity Transformation of Galaxies at High Redshifts}",
      journal = {\apj},
         year = 2009,
        month = jul,
       volume = {700},
       number = {1},
        pages = {791-798},
          doi = {10.1088/0004-637X/700/1/791},
archivePrefix = {arXiv},
       eprint = {0905.2586},
 primaryClass = {astro-ph.CO},
       adsurl = {https://ui.adsabs.harvard.edu/abs/2009ApJ...700..791H}
}

@ARTICLE{Muldrew2012,
       author = {{Muldrew}, Stuart I. and {Croton}, Darren J. and {Skibba}, Ramin A. and {Pearce}, Frazer R. and {Ann}, Hong Bae and {Baldry}, Ivan K. and {Brough}, Sarah and {Choi}, Yun-Young and {Conselice}, Christopher J. and {Cowan}, Nicolas B. and {Gallazzi}, Anna and {Gray}, Meghan E. and {Gr{\"u}tzbauch}, Ruth and {Li}, I. -Hui and {Park}, Changbom and {Pilipenko}, Sergey V. and {Podgorzec}, Bret J. and {Robotham}, Aaron S.~G. and {Wilman}, David J. and {Yang}, Xiaohu and {Zhang}, Youcai and {Zibetti}, Stefano},
        title = "{Measures of galaxy environment - I. What is 'environment'?}",
      journal = {\mnras},
         year = 2012,
        month = jan,
       volume = {419},
       number = {3},
        pages = {2670-2682},
          doi = {10.1111/j.1365-2966.2011.19922.x},
archivePrefix = {arXiv},
       eprint = {1109.6328},
 primaryClass = {astro-ph.CO},
       adsurl = {https://ui.adsabs.harvard.edu/abs/2012MNRAS.419.2670M}
}

@ARTICLE{Park2008,
       author = {{Park}, Changbom and {Gott}, III, J. Richard and {Choi}, Yun-Young},
        title = "{Transformation of Morphology and Luminosity Classes of the SDSS Galaxies}",
      journal = {\apj},
         year = 2008,
        month = feb,
       volume = {674},
       number = {2},
        pages = {784-796},
          doi = {10.1086/524192},
archivePrefix = {arXiv},
       eprint = {0708.4118},
 primaryClass = {astro-ph},
       adsurl = {https://ui.adsabs.harvard.edu/abs/2008ApJ...674..784P}
}

@ARTICLE{Park2009b,
       author = {{Park}, Changbom and {Choi}, Yun-Young},
        title = "{Combined Effects of Galaxy Interactions and Large-Scale Environment on Galaxy Properties}",
      journal = {\apj},
         year = 2009,
        month = feb,
       volume = {691},
       number = {2},
        pages = {1828-1845},
          doi = {10.1088/0004-637X/691/2/1828},
archivePrefix = {arXiv},
       eprint = {0809.2156},
 primaryClass = {astro-ph},
       adsurl = {https://ui.adsabs.harvard.edu/abs/2009ApJ...691.1828P}
}

@ARTICLE{Jeong2025,
       author = {{Jeong}, Daeun and {Hwang}, Ho Seong and {Chung}, Haeun and {Yoon}, Yongmin},
        title = "{Diverse Rotation Curves of Galaxies in a Simulated Universe: The Observed Dependence on Stellar Mass and Morphology Reproduced}",
      journal = {\apj},
         year = 2025,
        month = mar,
       volume = {982},
       number = {1},
          eid = {11},
        pages = {11},
          doi = {10.3847/1538-4357/adb1be},
archivePrefix = {arXiv},
       eprint = {2502.01625},
 primaryClass = {astro-ph.GA},
       adsurl = {https://ui.adsabs.harvard.edu/abs/2025ApJ...982...11J}
}

@ARTICLE{GLee2024,
       author = {{Lee}, Gain and {Hwang}, Ho Seong and {Lee}, Jaehyun and {Shin}, Jihye and {Song}, Hyunmi},
        title = "{Understanding the Formation and Evolution of Dark Galaxies in a Simulated Universe}",
      journal = {\apj},
         year = 2024,
        month = feb,
       volume = {962},
       number = {2},
          eid = {129},
        pages = {129},
          doi = {10.3847/1538-4357/ad1e5d},
archivePrefix = {arXiv},
       eprint = {2401.07007},
 primaryClass = {astro-ph.GA},
       adsurl = {https://ui.adsabs.harvard.edu/abs/2024ApJ...962..129L}
}

@ARTICLE{Haslbauer2019,
       author = {{Haslbauer}, M. and {Dabringhausen}, J. and {Kroupa}, P. and {Javanmardi}, B. and {Banik}, I.},
        title = "{Galaxies lacking dark matter in the Illustris simulation}",
      journal = {\aap},
         year = 2019,
        month = jun,
       volume = {626},
          eid = {A47},
        pages = {A47},
          doi = {10.1051/0004-6361/201833771},
archivePrefix = {arXiv},
       eprint = {1905.03258},
 primaryClass = {astro-ph.GA},
       adsurl = {https://ui.adsabs.harvard.edu/abs/2019A&A...626A..47H}
}

@ARTICLE{Zaragoza-Cardiel2024,
       author = {{Zaragoza-Cardiel}, Javier and {Smith}, Beverly J. and {Jones}, Mark G. and {Giroux}, Mark L. and {Toner}, Shawn and {Alzate}, Jairo A. and {Fern{\'a}ndez-Arenas}, David and {Mayya}, Divakara and {Ortiz-Le{\'o}n}, Gisela and {Portilla}, Mauricio},
        title = "{Detection and characterization of detached tidal dwarf galaxies}",
      journal = {\aap},
         year = 2024,
        month = sep,
       volume = {689},
          eid = {A206},
        pages = {A206},
          doi = {10.1051/0004-6361/202450349},
archivePrefix = {arXiv},
       eprint = {2406.05179},
 primaryClass = {astro-ph.GA},
       adsurl = {https://ui.adsabs.harvard.edu/abs/2024A&A...689A.206Z}
}

@ARTICLE{Lovell2025,
       author = {{Lovell}, Mark R. and {Pillepich}, Annalisa and {Engler}, Christoph and {Nelson}, Dylan and {Ramesh}, Rahul and {Springel}, Volker and {Hernquist}, Lars},
        title = "{Numerical effects on the stripping of dark matter and stars in IllustrisTNG galaxy groups and clusters}",
      journal = {\mnras},
         year = 2025,
        month = dec,
       volume = {544},
       number = {4},
        pages = {4367-4389},
          doi = {10.1093/mnras/staf2012},
archivePrefix = {arXiv},
       eprint = {2509.07078},
 primaryClass = {astro-ph.GA},
       adsurl = {https://ui.adsabs.harvard.edu/abs/2025MNRAS.544.4367L}
}

@ARTICLE{Springel2021,
       author = {{Springel}, Volker and {Pakmor}, R{\"u}diger and {Zier}, Oliver and {Reinecke}, Martin},
        title = "{Simulating cosmic structure formation with the GADGET-4 code}",
      journal = {\mnras},
         year = 2021,
        month = sep,
       volume = {506},
       number = {2},
        pages = {2871-2949},
          doi = {10.1093/mnras/stab1855},
archivePrefix = {arXiv},
       eprint = {2010.03567},
 primaryClass = {astro-ph.IM},
       adsurl = {https://ui.adsabs.harvard.edu/abs/2021MNRAS.506.2871S}
}

@ARTICLE{Yoon2021,
       author = {{Yoon}, Yongmin and {Park}, Changbom and {Chung}, Haeun and {Zhang}, Kai},
        title = "{Rotation Curves of Galaxies and Their Dependence on Morphology and Stellar Mass}",
      journal = {\apj},
         year = 2021,
        month = dec,
       volume = {922},
       number = {2},
          eid = {249},
        pages = {249},
          doi = {10.3847/1538-4357/ac2302},
archivePrefix = {arXiv},
       eprint = {2110.06033},
 primaryClass = {astro-ph.GA},
       adsurl = {https://ui.adsabs.harvard.edu/abs/2021ApJ...922..249Y}
}

@ARTICLE{Ubler2021,
       author = {{{\"U}bler}, Hannah and {Genel}, Shy and {Sternberg}, Amiel and {Genzel}, Reinhard and {Price}, Sedona H. and {F{\"o}rster Schreiber}, Natascha M. and {Shimizu}, Taro T. and {Pillepich}, Annalisa and {Nelson}, Dylan and {Burkert}, Andreas and {Davies}, Ric and {Hernquist}, Lars and {Lang}, Philipp and {Lutz}, Dieter and {Pakmor}, R{\"u}diger and {Tacconi}, Linda J.},
        title = "{The kinematics and dark matter fractions of TNG50 galaxies at z = 2 from an observational perspective}",
      journal = {\mnras},
         year = 2021,
        month = jan,
       volume = {500},
       number = {4},
        pages = {4597-4619},
          doi = {10.1093/mnras/staa3464},
archivePrefix = {arXiv},
       eprint = {2008.05486},
 primaryClass = {astro-ph.GA},
       adsurl = {https://ui.adsabs.harvard.edu/abs/2021MNRAS.500.4597U}
}

@ARTICLE{Farahi2022,
       author = {{Farahi}, Arya and {Nagai}, Daisuke and {Anbajagane}, Dhayaa},
        title = "{Correlations of Dark Matter, Gas, and Stellar Profiles in Dark Matter Halos}",
      journal = {\apj},
         year = 2022,
        month = jul,
       volume = {933},
       number = {1},
          eid = {48},
        pages = {48},
          doi = {10.3847/1538-4357/ac721e},
archivePrefix = {arXiv},
       eprint = {2204.13578},
 primaryClass = {astro-ph.CO},
       adsurl = {https://ui.adsabs.harvard.edu/abs/2022ApJ...933...48F}
}

@ARTICLE{Puglisi2023,
       author = {{Puglisi}, Annagrazia and {Dudzevi{\v{c}}i{\={u}}t{\.{e}}}, Ugn{\.{e}} and {Swinbank}, Mark and {Gillman}, Steven and {Tiley}, Alfred L. and {Bower}, Richard G. and {Cirasuolo}, Michele and {Cortese}, Luca and {Glazebrook}, Karl and {Harrison}, Chris and {Ibar}, Edo and {Molina}, Juan and {Obreschkow}, Danail and {Oman}, Kyle A. and {Schaller}, Matthieu and {Shankar}, Francesco and {Sharples}, Ray M.},
        title = "{KURVS: the outer rotation curve shapes and dark matter fractions of z   1.5 star-forming galaxies}",
      journal = {\mnras},
         year = 2023,
        month = sep,
       volume = {524},
       number = {2},
        pages = {2814-2835},
          doi = {10.1093/mnras/stad1966},
archivePrefix = {arXiv},
       eprint = {2305.04382},
 primaryClass = {astro-ph.GA},
       adsurl = {https://ui.adsabs.harvard.edu/abs/2023MNRAS.524.2814P}
}

@ARTICLE{Shankar2025,
       author = {{Shankar}, Rohan and {Prakash}, Adithya and {Mehta}, Aarya},
        title = "{Estimations of dark matter fractions for ETGs using the broken-power-law model and machine learning techniques}",
      journal = {\mnras},
         year = 2025,
        month = jul,
       volume = {540},
       number = {3},
        pages = {2269-2278},
          doi = {10.1093/mnras/staf811},
       adsurl = {https://ui.adsabs.harvard.edu/abs/2025MNRAS.540.2269S}
}

@ARTICLE{Hwang2010,
       author = {{Hwang}, H.~S. and {Elbaz}, D. and {Lee}, J.~C. and {Jeong}, W.-S. and {Park}, C. and {Lee}, M.~G. and {Lee}, H.~M.},
        title = "{Environmental dependence of local luminous infrared galaxies}",
      journal = {\aap},
         year = 2010,
        month = nov,
       volume = {522},
          eid = {A33},
        pages = {A33},
          doi = {10.1051/0004-6361/201014807},
archivePrefix = {arXiv},
       eprint = {1006.2166},
 primaryClass = {astro-ph.CO},
       adsurl = {https://ui.adsabs.harvard.edu/abs/2010A&A...522A..33H}
}

@ARTICLE{Hwang2012,
       author = {{Hwang}, H.~S. and {Park}, C. and {Elbaz}, D. and {Choi}, Y.-Y.},
        title = "{Activity in galactic nuclei of cluster and field galaxies in the local universe}",
      journal = {\aap},
         year = 2012,
        month = feb,
       volume = {538},
          eid = {A15},
        pages = {A15},
          doi = {10.1051/0004-6361/201117351},
archivePrefix = {arXiv},
       eprint = {1111.1973},
 primaryClass = {astro-ph.CO},
       adsurl = {https://ui.adsabs.harvard.edu/abs/2012A&A...538A..15H}
}

@ARTICLE{Kwon2025,
       author = {{Kwon}, Minseong and {Hwang}, Ho Seong and {Kent}, Brian R. and {Yoon}, Ilsang and {Lee}, Gain and {Yoon}, Hyein},
        title = "{Searching for Dark Galaxies with H I Detection from the Arecibo Legacy Fast ALFA (ALFALFA) Survey}",
      journal = {\apjs},
         year = 2025,
        month = aug,
       volume = {279},
       number = {2},
          eid = {38},
        pages = {38},
          doi = {10.3847/1538-4365/ade0b8},
archivePrefix = {arXiv},
       eprint = {2506.03678},
 primaryClass = {astro-ph.GA},
       adsurl = {https://ui.adsabs.harvard.edu/abs/2025ApJS..279...38K}
}

@ARTICLE{Keim2026,
       author = {{Keim}, Michael A. and {van Dokkum}, Pieter and {Shen}, Zili and {Danieli}, Shany and {Pasha}, Imad},
        title = "{A Third Galaxy Missing Dark Matter along a Trail of Galaxies in the NGC 1052 Field}",
      journal = {arXiv e-prints},
         year = 2026,
        month = mar,
          eid = {arXiv:2603.15860},
        pages = {arXiv:2603.15860},
          doi = {10.48550/arXiv.2603.15860},
archivePrefix = {arXiv},
       eprint = {2603.15860},
 primaryClass = {astro-ph.GA},
       adsurl = {https://ui.adsabs.harvard.edu/abs/2026arXiv260315860K}
}

@ARTICLE{vanderWel2014,
       author = {{van der Wel}, A. and {Franx}, M. and {van Dokkum}, P.~G. and {Skelton}, R.~E. and {Momcheva}, I.~G. and {Whitaker}, K.~E. and {Brammer}, G.~B. and {Bell}, E.~F. and {Rix}, H.-W. and {Wuyts}, S. and {Ferguson}, H.~C. and {Holden}, B.~P. and {Barro}, G. and {Koekemoer}, A.~M. and {Chang}, Yu-Yen and {McGrath}, E.~J. and {H{\"a}ussler}, B. and {Dekel}, A. and {Behroozi}, P. and {Fumagalli}, M. and {Leja}, J. and {Lundgren}, B.~F. and {Maseda}, M.~V. and {Nelson}, E.~J. and {Wake}, D.~A. and {Patel}, S.~G. and {Labb{\'e}}, I. and {Faber}, S.~M. and {Grogin}, N.~A. and {Kocevski}, D.~D.},
        title = "{3D-HST+CANDELS: The Evolution of the Galaxy Size-Mass Distribution since z = 3}",
      journal = {\apj},
         year = 2014,
        month = jun,
       volume = {788},
       number = {1},
          eid = {28},
        pages = {28},
          doi = {10.1088/0004-637X/788/1/28},
archivePrefix = {arXiv},
       eprint = {1404.2844},
 primaryClass = {astro-ph.GA},
       adsurl = {https://ui.adsabs.harvard.edu/abs/2014ApJ...788...28V}
}

@ARTICLE{vanDokkum2015b,
       author = {{van Dokkum}, Pieter G. and {Nelson}, Erica June and {Franx}, Marijn and {Oesch}, Pascal and {Momcheva}, Ivelina and {Brammer}, Gabriel and {F{\"o}rster Schreiber}, Natascha M. and {Skelton}, Rosalind E. and {Whitaker}, Katherine E. and {van der Wel}, Arjen and {Bezanson}, Rachel and {Fumagalli}, Mattia and {Illingworth}, Garth D. and {Kriek}, Mariska and {Leja}, Joel and {Wuyts}, Stijn},
        title = "{Forming Compact Massive Galaxies}",
      journal = {\apj},
         year = 2015,
        month = nov,
       volume = {813},
       number = {1},
          eid = {23},
        pages = {23},
          doi = {10.1088/0004-637X/813/1/23},
archivePrefix = {arXiv},
       eprint = {1506.03085},
 primaryClass = {astro-ph.GA},
       adsurl = {https://ui.adsabs.harvard.edu/abs/2015ApJ...813...23V}
}

@ARTICLE{Cappellari2013,
       author = {{Cappellari}, Michele and {McDermid}, Richard M. and {Alatalo}, Katherine and {Blitz}, Leo and {Bois}, Maxime and {Bournaud}, Fr{\'e}d{\'e}ric and {Bureau}, M. and {Crocker}, Alison F. and {Davies}, Roger L. and {Davis}, Timothy A. and {de Zeeuw}, P.~T. and {Duc}, Pierre-Alain and {Emsellem}, Eric and {Khochfar}, Sadegh and {Krajnovi{\'c}}, Davor and {Kuntschner}, Harald and {Morganti}, Raffaella and {Naab}, Thorsten and {Oosterloo}, Tom and {Sarzi}, Marc and {Scott}, Nicholas and {Serra}, Paolo and {Weijmans}, Anne-Marie and {Young}, Lisa M.},
        title = "{The ATLAS$^{3D}$ project - XX. Mass-size and mass-{\ensuremath{\sigma}} distributions of early-type galaxies: bulge fraction drives kinematics, mass-to-light ratio, molecular gas fraction and stellar initial mass function}",
      journal = {\mnras},
         year = 2013,
        month = jul,
       volume = {432},
       number = {3},
        pages = {1862-1893},
          doi = {10.1093/mnras/stt644},
archivePrefix = {arXiv},
       eprint = {1208.3523},
 primaryClass = {astro-ph.CO},
       adsurl = {https://ui.adsabs.harvard.edu/abs/2013MNRAS.432.1862C}
}

@ARTICLE{Cappellari2016,
       author = {{Cappellari}, Michele},
        title = "{Structure and Kinematics of Early-Type Galaxies from Integral Field Spectroscopy}",
      journal = {\araa},
         year = 2016,
        month = sep,
       volume = {54},
        pages = {597-665},
          doi = {10.1146/annurev-astro-082214-122432},
archivePrefix = {arXiv},
       eprint = {1602.04267},
 primaryClass = {astro-ph.GA},
       adsurl = {https://ui.adsabs.harvard.edu/abs/2016ARA&A..54..597C}
}

@ARTICLE{Gannon2023,
       author = {{Gannon}, Jonah S. and {Buzzo}, Maria Luisa and {Ferr{\'e}-Mateu}, Anna and {Forbes}, Duncan A. and {Brodie}, Jean P. and {Romanowsky}, Aaron J.},
        title = "{Keck spectroscopy of NGC 1052-DF9: stellar populations in the context of the NGC 1052 group}",
      journal = {\mnras},
         year = 2023,
        month = sep,
       volume = {524},
       number = {2},
        pages = {2624-2629},
          doi = {10.1093/mnras/stad1883},
archivePrefix = {arXiv},
       eprint = {2306.08540},
 primaryClass = {astro-ph.GA},
       adsurl = {https://ui.adsabs.harvard.edu/abs/2023MNRAS.524.2624G}
}

@ARTICLE{Buzzo2026,
       author = {{Buzzo}, Maria Luisa and {van Dokkum}, Pieter and {Hilker}, Michael and {Forbes}, Duncan A. and {Romanowsky}, Aaron J. and {Tang}, Yimeng},
        title = "{Dark matter-deficient twins: FCC 224 and FCC 240 as possible analogues of NGC 1052-DF2 and DF4}",
      journal = {arXiv e-prints},
         year = 2026,
        month = may,
          eid = {arXiv:2605.24099},
        pages = {arXiv:2605.24099},
          doi = {10.48550/arXiv.2605.24099},
archivePrefix = {arXiv},
       eprint = {2605.24099},
 primaryClass = {astro-ph.GA},
       adsurl = {https://ui.adsabs.harvard.edu/abs/2026arXiv260524099B}
}

@ARTICLE{Mitrasinovic2025,
       author = {{Mitra{\v{s}}inovi{\'c}}, Ana},
        title = "{Living on the edge: A quantitative warning on boundary artifacts in the IllustrisTNG}",
      journal = {\aap},
         year = 2025,
        month = nov,
       volume = {703},
          eid = {L16},
        pages = {L16},
          doi = {10.1051/0004-6361/202557557},
archivePrefix = {arXiv},
       eprint = {2511.03562},
 primaryClass = {astro-ph.CO},
       adsurl = {https://ui.adsabs.harvard.edu/abs/2025A&A...703L..16M}
}

@ARTICLE{Makarova2002,
       author = {{Makarova}, L.~N. and {Grebel}, E.~K. and {Karachentsev}, I.~D. and {Dolphin}, A.~E. and {Karachentseva}, V.~E. and {Sharina}, M.~E. and {Geisler}, D. and {Guhathakurta}, P. and {Hodge}, P.~W. and {Sarajedini}, A. and {Seitzer}, P.},
        title = "{Tidal dwarfs in the M81 group: The second generation?}",
      journal = {\aap},
         year = 2002,
        month = dec,
       volume = {396},
        pages = {473-487},
          doi = {10.1051/0004-6361:20021426},
       adsurl = {https://ui.adsabs.harvard.edu/abs/2002A&A...396..473M}
}

@ARTICLE{Recchi2007,
       author = {{Recchi}, S. and {Theis}, C. and {Kroupa}, P. and {Hensler}, G.},
        title = "{The early evolution of tidal dwarf galaxies}",
      journal = {\aap},
         year = 2007,
        month = jul,
       volume = {470},
       number = {1},
        pages = {L5-L8},
          doi = {10.1051/0004-6361:20077264},
archivePrefix = {arXiv},
       eprint = {0706.0093},
 primaryClass = {astro-ph},
       adsurl = {https://ui.adsabs.harvard.edu/abs/2007A&A...470L...5R}
}

@ARTICLE{Ploeckinger2015,
       author = {{Ploeckinger}, S. and {Recchi}, S. and {Hensler}, G. and {Kroupa}, P.},
        title = "{Chemodynamical evolution of tidal dwarf galaxies - II. The long-term evolution and influence of a tidal field}",
      journal = {\mnras},
         year = 2015,
        month = mar,
       volume = {447},
       number = {3},
        pages = {2512-2525},
          doi = {10.1093/mnras/stu2629},
archivePrefix = {arXiv},
       eprint = {1412.3450},
 primaryClass = {astro-ph.GA},
       adsurl = {https://ui.adsabs.harvard.edu/abs/2015MNRAS.447.2512P}
}

@ARTICLE{Weilbacher2003,
       author = {{Weilbacher}, P.~M. and {Duc}, P.-A. and {Fritze-v. Alvensleben}, U.},
        title = "{Tidal dwarf candidates in a sample of interacting galaxies. II. Properties and kinematics of the ionized gas}",
      journal = {\aap},
         year = 2003,
        month = jan,
       volume = {397},
        pages = {545-555},
          doi = {10.1051/0004-6361:20021522},
archivePrefix = {arXiv},
       eprint = {astro-ph/0210393},
 primaryClass = {astro-ph},
       adsurl = {https://ui.adsabs.harvard.edu/abs/2003A&A...397..545W}
}

@INPROCEEDINGS{Duc2012,
       author = {{Duc}, Pierre-Alain},
        title = "{Birth, Life and Survival of Tidal Dwarf Galaxies}",
    booktitle = {Dwarf Galaxies: Keys to Galaxy Formation and Evolution},
         year = 2012,
       editor = {{Papaderos}, Polychronis and {Recchi}, Simone and {Hensler}, Gerhard},
       series = {Astrophysics and Space Science Proceedings},
       volume = {28},
        month = jan,
        pages = {305},
          doi = {10.1007/978-3-642-22018-0_37},
archivePrefix = {arXiv},
       eprint = {1101.4834},
 primaryClass = {astro-ph.CO},
       adsurl = {https://ui.adsabs.harvard.edu/abs/2012ASSP...28..305D}
}

@ARTICLE{Fensch2019b,
       author = {{Fensch}, J{\'e}r{\'e}my and {Duc}, Pierre-Alain and {Boquien}, M{\'e}d{\'e}ric and {Elmegreen}, Debra M. and {Elmegreen}, Bruce G. and {Bournaud}, Fr{\'e}d{\'e}ric and {Brinks}, Elias and {de Grijs}, Richard and {Lelli}, Federico and {Renaud}, Florent and {Weilbacher}, Peter M.},
        title = "{Massive star cluster formation and evolution in tidal dwarf galaxies}",
      journal = {\aap},
         year = 2019,
        month = aug,
       volume = {628},
          eid = {A60},
        pages = {A60},
          doi = {10.1051/0004-6361/201834403},
archivePrefix = {arXiv},
       eprint = {1903.10789},
 primaryClass = {astro-ph.GA},
       adsurl = {https://ui.adsabs.harvard.edu/abs/2019A&A...628A..60F}
}

@ARTICLE{Simha2014,
       author = {{Simha}, Vimal and {Weinberg}, David H. and {Conroy}, Charlie and {Dave}, Romeel and {Fardal}, Mark and {Katz}, Neal and {Oppenheimer}, Benjamin D.},
        title = "{Parametrising Star Formation Histories}",
      journal = {arXiv e-prints},
         year = 2014,
        month = apr,
          eid = {arXiv:1404.0402},
        pages = {arXiv:1404.0402},
          doi = {10.48550/arXiv.1404.0402},
archivePrefix = {arXiv},
       eprint = {1404.0402},
 primaryClass = {astro-ph.GA},
       adsurl = {https://ui.adsabs.harvard.edu/abs/2014arXiv1404.0402S}
}

@ARTICLE{Diemer2017,
       author = {{Diemer}, Benedikt and {Sparre}, Martin and {Abramson}, Louis E. and {Torrey}, Paul},
        title = "{Log-normal Star Formation Histories in Simulated and Observed Galaxies}",
      journal = {\apj},
         year = 2017,
        month = apr,
       volume = {839},
       number = {1},
          eid = {26},
        pages = {26},
          doi = {10.3847/1538-4357/aa68e5},
archivePrefix = {arXiv},
       eprint = {1701.02308},
 primaryClass = {astro-ph.GA},
       adsurl = {https://ui.adsabs.harvard.edu/abs/2017ApJ...839...26D}
}

@ARTICLE{Joshi2021,
       author = {{Joshi}, Gandhali D. and {Pillepich}, Annalisa and {Nelson}, Dylan and {Zinger}, Elad and {Marinacci}, Federico and {Springel}, Volker and {Vogelsberger}, Mark and {Hernquist}, Lars},
        title = "{The cumulative star formation histories of dwarf galaxies with TNG50. I: environment-driven diversity and connection to quenching}",
      journal = {\mnras},
         year = 2021,
        month = dec,
       volume = {508},
       number = {2},
        pages = {1652-1674},
          doi = {10.1093/mnras/stab2573},
archivePrefix = {arXiv},
       eprint = {2101.12226},
 primaryClass = {astro-ph.GA},
       adsurl = {https://ui.adsabs.harvard.edu/abs/2021MNRAS.508.1652J}
}

@ARTICLE{Rodriguez-Gomez2016,
       author = {{Rodriguez-Gomez}, Vicente and {Pillepich}, Annalisa and {Sales}, Laura V. and {Genel}, Shy and {Vogelsberger}, Mark and {Zhu}, Qirong and {Wellons}, Sarah and {Nelson}, Dylan and {Torrey}, Paul and {Springel}, Volker and {Ma}, Chung-Pei and {Hernquist}, Lars},
        title = "{The stellar mass assembly of galaxies in the Illustris simulation: growth by mergers and the spatial distribution of accreted stars}",
      journal = {\mnras},
         year = 2016,
        month = may,
       volume = {458},
       number = {3},
        pages = {2371-2390},
          doi = {10.1093/mnras/stw456},
archivePrefix = {arXiv},
       eprint = {1511.08804},
 primaryClass = {astro-ph.GA},
       adsurl = {https://ui.adsabs.harvard.edu/abs/2016MNRAS.458.2371R}
}

@ARTICLE{Rodriguez-Gomez2017,
       author = {{Rodriguez-Gomez}, Vicente and {Sales}, Laura V. and {Genel}, Shy and {Pillepich}, Annalisa and {Zjupa}, Jolanta and {Nelson}, Dylan and {Griffen}, Brendan and {Torrey}, Paul and {Snyder}, Gregory F. and {Vogelsberger}, Mark and {Springel}, Volker and {Ma}, Chung-Pei and {Hernquist}, Lars},
        title = "{The role of mergers and halo spin in shaping galaxy morphology}",
      journal = {\mnras},
         year = 2017,
        month = may,
       volume = {467},
       number = {3},
        pages = {3083-3098},
          doi = {10.1093/mnras/stx305},
archivePrefix = {arXiv},
       eprint = {1609.09498},
 primaryClass = {astro-ph.GA},
       adsurl = {https://ui.adsabs.harvard.edu/abs/2017MNRAS.467.3083R}
}

@ARTICLE{Martizzi2019,
       author = {{Martizzi}, Davide and {Vogelsberger}, Mark and {Artale}, Maria Celeste and {Haider}, Markus and {Torrey}, Paul and {Marinacci}, Federico and {Nelson}, Dylan and {Pillepich}, Annalisa and {Weinberger}, Rainer and {Hernquist}, Lars and {Naiman}, Jill and {Springel}, Volker},
        title = "{Baryons in the Cosmic Web of IllustrisTNG - I: gas in knots, filaments, sheets, and voids}",
      journal = {\mnras},
         year = 2019,
        month = jul,
       volume = {486},
       number = {3},
        pages = {3766-3787},
          doi = {10.1093/mnras/stz1106},
archivePrefix = {arXiv},
       eprint = {1810.01883},
 primaryClass = {astro-ph.CO},
       adsurl = {https://ui.adsabs.harvard.edu/abs/2019MNRAS.486.3766M}
}

@ARTICLE{Hui1997,
       author = {{Hui}, Lam and {Gnedin}, Nickolay Y.},
        title = "{Equation of state of the photoionized intergalactic medium}",
      journal = {\mnras},
         year = 1997,
        month = nov,
       volume = {292},
       number = {1},
        pages = {27-42},
          doi = {10.1093/mnras/292.1.27},
archivePrefix = {arXiv},
       eprint = {astro-ph/9612232},
 primaryClass = {astro-ph},
       adsurl = {https://ui.adsabs.harvard.edu/abs/1997MNRAS.292...27H}
}

@ARTICLE{Wiersma2009,
       author = {{Wiersma}, Robert P.~C. and {Schaye}, Joop and {Smith}, Britton D.},
        title = "{The effect of photoionization on the cooling rates of enriched, astrophysical plasmas}",
      journal = {\mnras},
         year = 2009,
        month = feb,
       volume = {393},
       number = {1},
        pages = {99-107},
          doi = {10.1111/j.1365-2966.2008.14191.x},
archivePrefix = {arXiv},
       eprint = {0807.3748},
 primaryClass = {astro-ph},
       adsurl = {https://ui.adsabs.harvard.edu/abs/2009MNRAS.393...99W}
}

@ARTICLE{Springel2010,
       author = {{Springel}, Volker},
        title = "{E pur si muove: Galilean-invariant cosmological hydrodynamical simulations on a moving mesh}",
      journal = {\mnras},
         year = 2010,
        month = jan,
       volume = {401},
       number = {2},
        pages = {791-851},
          doi = {10.1111/j.1365-2966.2009.15715.x},
archivePrefix = {arXiv},
       eprint = {0901.4107},
 primaryClass = {astro-ph.CO},
       adsurl = {https://ui.adsabs.harvard.edu/abs/2010MNRAS.401..791S}
}

@ARTICLE{Genel2013,
       author = {{Genel}, Shy and {Vogelsberger}, Mark and {Nelson}, Dylan and {Sijacki}, Debora and {Springel}, Volker and {Hernquist}, Lars},
        title = "{Following the flow: tracer particles in astrophysical fluid simulations}",
      journal = {\mnras},
         year = 2013,
        month = oct,
       volume = {435},
       number = {2},
        pages = {1426-1442},
          doi = {10.1093/mnras/stt1383},
archivePrefix = {arXiv},
       eprint = {1305.2195},
 primaryClass = {astro-ph.IM},
       adsurl = {https://ui.adsabs.harvard.edu/abs/2013MNRAS.435.1426G}
}

@ARTICLE{deAlmeida2026,
       author = {{de Almeida}, Abhner P. and {Mamon}, Gary A. and {Lima Neto}, Gast{\~a}o B.},
        title = "{Satellite compaction pathways: environmental drivers shaping dwarf galaxy corpulence in the TNG50 simulation}",
      journal = {\mnras},
         year = 2026,
        month = jul,
       volume = {550},
       number = {1},
          eid = {stag1105},
        pages = {stag1105},
          doi = {10.1093/mnras/stag1105},
archivePrefix = {arXiv},
       eprint = {2606.09817},
 primaryClass = {astro-ph.GA},
       adsurl = {https://ui.adsabs.harvard.edu/abs/2026MNRAS.550g1105D}
}

@ARTICLE{Hammer2026,
       author = {{Hammer}, Francois and {Yang}, Yanbin},
        title = "{Tidal origin of dark-matter free dwarf galaxies in the NGC 1052 group}",
      journal = {arXiv e-prints},
         year = 2026,
        month = jun,
          eid = {arXiv:2606.30718},
        pages = {arXiv:2606.30718},
          doi = {10.48550/arXiv.2606.30718},
archivePrefix = {arXiv},
       eprint = {2606.30718},
 primaryClass = {astro-ph.GA},
       adsurl = {https://ui.adsabs.harvard.edu/abs/2026arXiv260630718H}
}

@ARTICLE{Springel2003,
       author = {{Springel}, Volker and {Hernquist}, Lars},
        title = "{Cosmological smoothed particle hydrodynamics simulations: a hybrid multiphase model for star formation}",
      journal = {\mnras},
         year = 2003,
        month = feb,
       volume = {339},
       number = {2},
        pages = {289-311},
          doi = {10.1046/j.1365-8711.2003.06206.x},
archivePrefix = {arXiv},
       eprint = {astro-ph/0206393},
 primaryClass = {astro-ph},
       adsurl = {https://ui.adsabs.harvard.edu/abs/2003MNRAS.339..289S}
}

@ARTICLE{Sheen2009,
       author = {{Sheen}, Yun-Kyeong and {Jeong}, Hyunjin and {Yi}, Sukyoung K. and {Ferreras}, Ignacio and {Lotz}, Jennifer M. and {Olsen}, Knut A.~G. and {Dickinson}, Mark and {Barnes}, Sydney and {Park}, Jang-Hyun and {Ree}, Chang H. and {Madore}, Barry F. and {Barlow}, Tom A. and {Conrow}, Tim and {Foster}, Karl and {Friendman}, Peter G. and {Lee}, Young-Wook and {Martin}, D. Christopher and {Morrissey}, Patrick and {Neff}, Susan G. and {Schiminovich}, David and {Seibert}, Mark and {Small}, Todd and {Wyder}, Ted K.},
        title = "{Tidal Dwarf Galaxies Around a Post-merger Galaxy, NGC 4922}",
      journal = {\aj},
         year = 2009,
        month = dec,
       volume = {138},
       number = {6},
        pages = {1911-1916},
          doi = {10.1088/0004-6256/138/6/1911},
archivePrefix = {arXiv},
       eprint = {0910.2351},
 primaryClass = {astro-ph.CO},
       adsurl = {https://ui.adsabs.harvard.edu/abs/2009AJ....138.1911S}
}

@ARTICLE{Wen2012,
       author = {{Wen}, Zhang-Zheng and {Zheng}, Xian-Zhong and {Zhao}, Ying-He and {Gao}, Yu},
        title = "{Tidal dwarf galaxies at intermediate redshifts}",
      journal = {\apss},
         year = 2012,
        month = feb,
       volume = {337},
       number = {2},
        pages = {729-737},
          doi = {10.1007/s10509-011-0882-8},
archivePrefix = {arXiv},
       eprint = {1103.2546},
 primaryClass = {astro-ph.CO},
       adsurl = {https://ui.adsabs.harvard.edu/abs/2012Ap&SS.337..729W}
}
\bibliographystyle{aasjournalv7}



\end{document}